\documentclass[10pt,journal,compsoc]{IEEEtran}
\ifCLASSOPTIONcompsoc
\usepackage[nocompress]{cite}
\else
\usepackage{cite}
\fi

\usepackage{epsfig}
\usepackage{graphicx}
\usepackage{amsmath}
\usepackage{amssymb}
\usepackage{array}
\usepackage{bm}
\usepackage{color}
\usepackage[table]{xcolor}
\usepackage{arydshln, collcell}
\usepackage{algorithm}
\usepackage{algorithmic}
\usepackage[labelformat=simple]{subcaption}
\usepackage{multirow}
\usepackage{booktabs}
\usepackage{graphbox} 
\usepackage{url}

\newcolumntype{?}[1]{!{\vrule width #1}}

\newcolumntype{C}[1]{>{\centering\arraybackslash}p{#1}}

\newcommand{\eg}{\textit{e.g., }}
\newcommand{\ie}{\textit{i.e., }}

\newcommand{\etc}{\textit{etc.}}

\newenvironment{packed_enum}{
	\begin{enumerate}
		\setlength{\itemsep}{1pt}
		\setlength{\parskip}{0pt}
		\setlength{\parsep}{0pt}
	}{\end{enumerate}}

\begin{document}
%
\title{Physics-based Noise Modeling for \\ Extreme Low-light Photography}
%
%
%
%


\author{Kaixuan~Wei, 
	Ying~Fu,~\IEEEmembership{Member,~IEEE,}
	~Yinqiang~Zheng,~\IEEEmembership{Member,~IEEE}
	and~Jiaolong~Yang,~\IEEEmembership{Member,~IEEE}
	
	\IEEEcompsocitemizethanks{\IEEEcompsocthanksitem Kaixuan Wei and Ying Fu are with School of Computer Science and Technology, Beijing Institute of Technology,
		Beijing, 100081, China.
		\IEEEcompsocthanksitem Yinqiang Zheng is with the Next Generation Artiﬁcial Intelligence Research Center, The University of Tokyo, Tokyo, 113-8656, Japan.
		\IEEEcompsocthanksitem Jiaolong Yang is with Microsoft Research Asia, Beijing, 100080, China. Major work done by Feb. 2021
		\IEEEcompsocthanksitem Corresponding author: Ying Fu (e-mail: fuying@bit.edu.cn)
		\IEEEcompsocthanksitem This research was supported by the National Natural Science Foundation of China under Grants No. 61827901 and No. 62088101.		
		}

}

%
%

\markboth{IEEE TRANSACTIONS ON PATTERN ANALYSIS AND MACHINE INTELLIGENCE}%
{Shell \MakeLowercase{\textit{et al.}}: Bare Demo of IEEEtran.cls for Computer Society Journals}
%


\IEEEtitleabstractindextext{%
\begin{abstract}
Enhancing the visibility in extreme low-light environments is a challenging task. Under nearly lightless condition, existing image denoising methods could easily break down due to significantly low SNR. 
In this paper, we systematically study the noise statistics in the imaging pipeline of CMOS photosensors, and formulate a comprehensive noise model that can accurately characterize the real noise structures.
Our novel model considers the noise sources caused by digital camera electronics which are largely overlooked by existing methods yet have significant influence on raw measurement in the dark. It provides a way to decouple the intricate noise structure into different statistical distributions with physical interpretations. Moreover, our noise model can be used to synthesize realistic training data for learning-based low-light denoising algorithms. In this regard, although promising results have been shown recently with deep convolutional neural networks, the success heavily depends on abundant noisy-clean image pairs for training, which are tremendously difficult to obtain in practice. Generalizing their trained models to images from new devices is also problematic. Extensive experiments on multiple low-light denoising datasets -- including a newly collected one in this work covering various devices -- show that a deep neural network trained with our proposed noise formation model can reach surprisingly-high accuracy. The results are on par with or sometimes even outperform training with paired real data, opening a new door to real-world extreme low-light photography. 

\end{abstract}

}

\maketitle
\IEEEdisplaynontitleabstractindextext

\ifCLASSOPTIONpeerreview
\begin{center} \bfseries EDICS Category: 3-BBND \end{center}
\fi
%
\IEEEpeerreviewmaketitle

\IEEEraisesectionheading{\section{Introduction}\label{sec:introduction}}

\IEEEPARstart{L}{ight} is of paramount importance to photography. 
Night and low light place very demanding
constraints on photography due to very limited photon count and inescapable noise.
One natural reaction is to gather more light by, \eg enlarging aperture
setting, lengthening exposure time and opening flashlight. However, each method brings a trade-off -- large aperture incurs small depth of field, and is usually unavailable in smartphone cameras; long exposure can induce blur due to scene variations or camera motions; flash can cause color aberrations and is useful only for nearby objects. 

Instead of attempting to gather more light during capturing, another way for low-light photography is to make use of a small number of photon detection to recover the desired scene information in the dark. 
This can be achieved either by hardware-based approach using advanced low-light imagers \cite{dussault2004noise,torr1986intensified,guerrieri2009fast,bruschini2019single,kirmani2014first,morris2015imaging,gariepy2015single,shin2016photon,heide2018sub,OToole_2017_CVPR} or computation-based approach using modern denoising algorithms \cite{dabov2007BM3D,Gu2014Weighted,Gu_2019_ICCV,Xu_2017_ICCV,xu2018trilateral,mao2016image,zhang2017beyond,tai2017memnet,Chen_2018_ECCV,Shi2018Toward,Anwar_2019_ICCV,Chen_2018_CVPR,Liu2014Fast,Mildenhall_2018_CVPR,Hasinoff2016Burst,liba2019handheld,Xia_2020_CVPR}. 
Our interest in this paper lies in computation-based denoising, which could be applied on commodity cameras without need for dedicated imaging hardware. 
Along this line of research, significant progress has been made in recent years. 
Representative works include the burst processing scheme \cite{Hasinoff2016Burst,liba2019handheld} which aligns and fuses a burst of photos to increase the signal-to-noise ratio (SNR), 
as well as the deep learning approach \cite{Chen_2018_CVPR,Chen_2019_ICCV} that automatically learns the mapping from a low-light noisy image to its long-exposure clean counterpart. 
Nevertheless, the existing methods still suffer from several limitations and the problem is still far from beings solved. For example, burst processing may generate unwanted ghosting effect \cite{Hasinoff2016Burst} when capturing dynamic scenes in the presence of vehicles, humans, \etc; the deep learning approach, on the other hand, requires a large volume of densely-labeled real training data which is tremendously difficult to acquire. 

\begin{figure*}[t]
	\centering
	\setlength\tabcolsep{1pt}
	\begin{tabular}{cccccccc}
		\rotatebox[origin=c]{90}{SID \cite{Chen_2018_CVPR}}  
		& \includegraphics[align=c,width=0.135\linewidth]{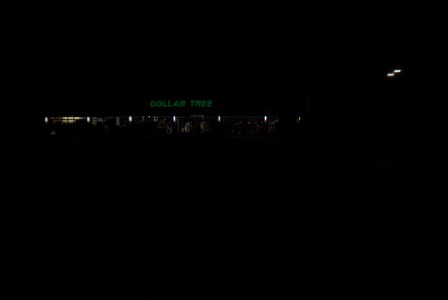}         	 	
		& \includegraphics[align=c,width=0.135\linewidth]{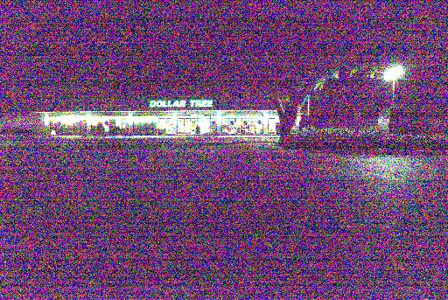}         
		& \includegraphics[align=c,width=0.135\linewidth]{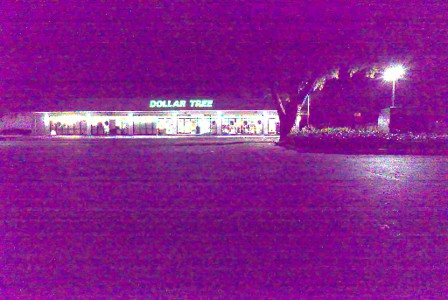} 
		& \includegraphics[align=c,width=0.135\linewidth]{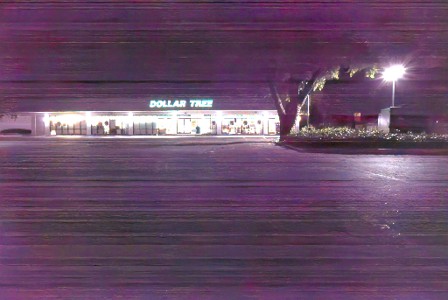} 
		& \includegraphics[align=c,width=0.135\linewidth]{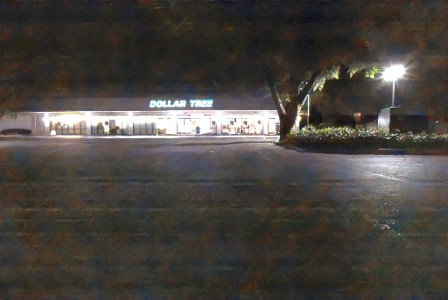} 
		& \includegraphics[align=c,width=0.135\linewidth]{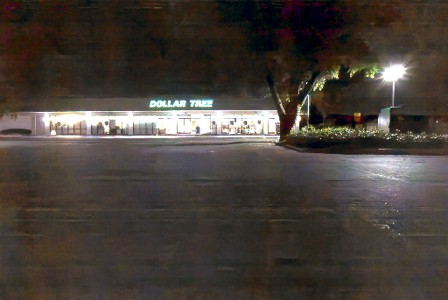} 
		& \includegraphics[align=c,width=0.135\linewidth]{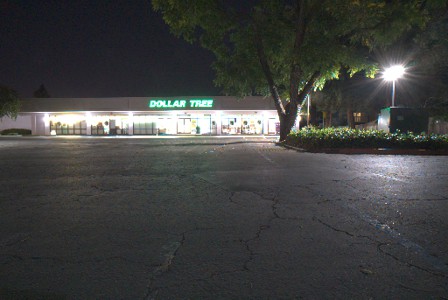} \\
		\addlinespace[1.5pt]
		\rotatebox[origin=c]{90}{DRV \cite{Chen_2019_ICCV}}  
		& \includegraphics[align=c,width=0.135\linewidth]{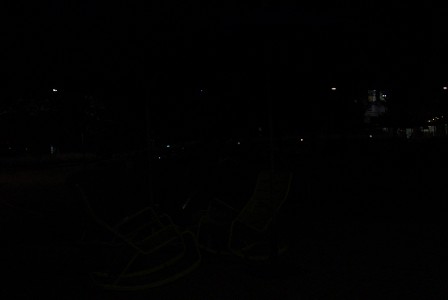}   		
		& \includegraphics[align=c,width=0.135\linewidth]{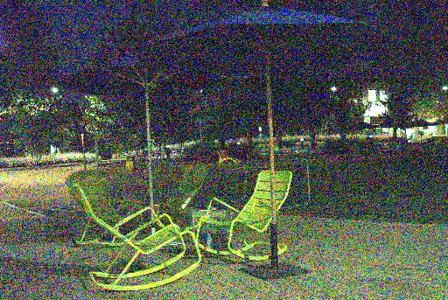}        
		& \includegraphics[align=c,width=0.135\linewidth]{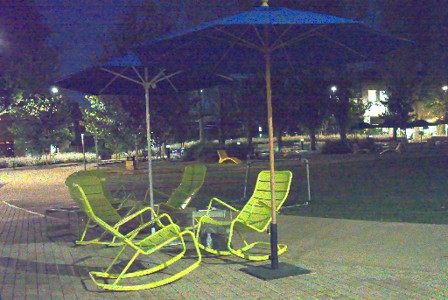} 
		& \includegraphics[align=c,width=0.135\linewidth]{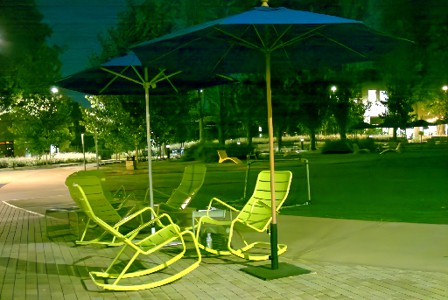} 
		& \includegraphics[align=c,width=0.135\linewidth]{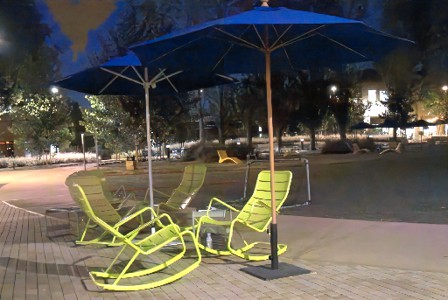} 
		& \includegraphics[align=c,width=0.135\linewidth]{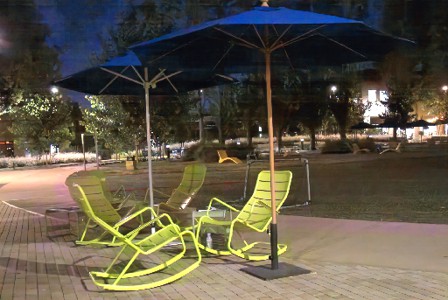} 
		& \includegraphics[align=c,width=0.135\linewidth]{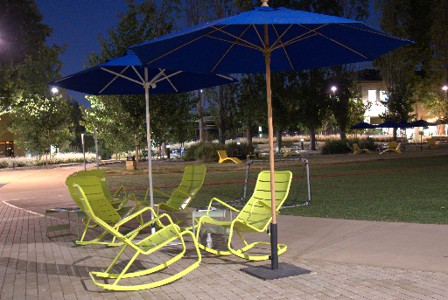} \\
		\addlinespace[1.5pt]
		\rotatebox[origin=c]{90}{ELD}  
		& \includegraphics[align=c,width=0.135\linewidth]{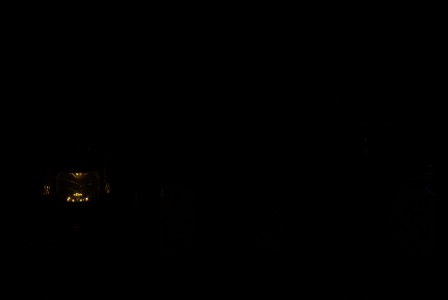}      		
		& \includegraphics[align=c,width=0.135\linewidth]{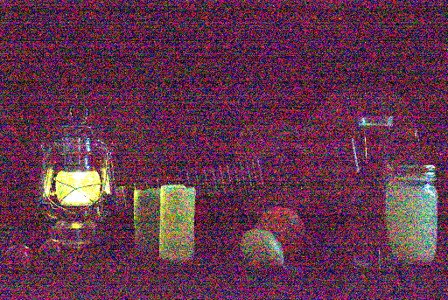}        
		& \includegraphics[align=c,width=0.135\linewidth]{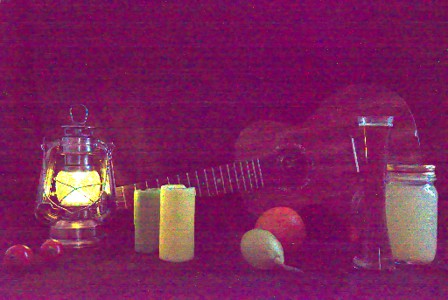} 
		& \includegraphics[align=c,width=0.135\linewidth]{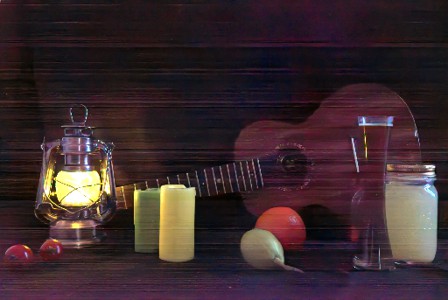} 
		& \includegraphics[align=c,width=0.135\linewidth]{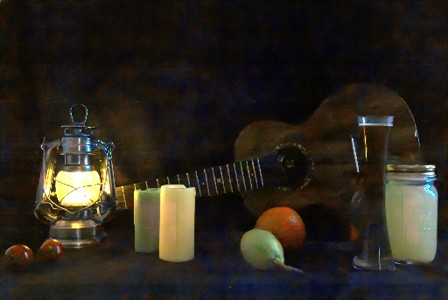} 
		& \includegraphics[align=c,width=0.135\linewidth]{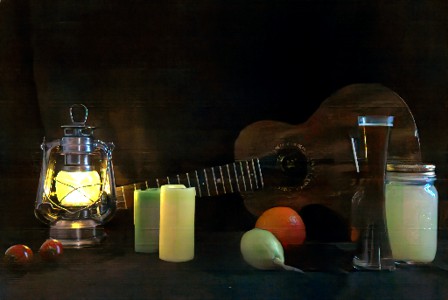} 
		& \includegraphics[align=c,width=0.135\linewidth]{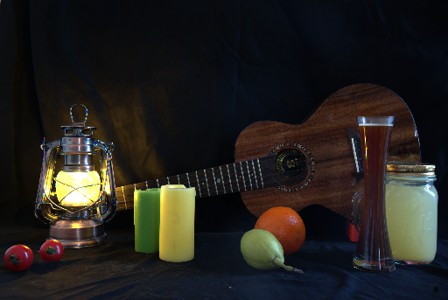} \\ 		
		\addlinespace[1.5pt]
		& (a) \footnotesize Short-exposure & (b) \footnotesize Amplified input & (c) \footnotesize BM3D \cite{dabov2007BM3D} & (d) \footnotesize $G$+$P$ \cite{Foi2008Practical} & (e) \footnotesize Paired data \cite{Chen_2018_CVPR} & (f) \footnotesize Ours & (g) \footnotesize Long-exposure \\	
	\end{tabular}
	\caption{Raw denoising results of images from three real-world datasets, \ie the See-in-the-Dark (SID) dataset \cite{Chen_2018_CVPR}, the Dark Raw Video (DRV) dataset \cite{Chen_2019_ICCV} and our Extreme Low-light Denoising (ELD) dataset, where we present (a) the short-exposure low-light image; (b) amplified noisy input image from (a); (c) the denoised output by the well-known denoising method BM3D \cite{dabov2007BM3D}; (g) the long-exposure reference image;  (d-f) the outputs of UNets \cite{ronneberger2015u} trained with (d) synthetic data generated by the signal-dependent heteroscedastic Gaussian noise model (G+P) \cite{Foi2008Practical},  (e)  paired real data of \cite{Chen_2018_CVPR,Chen_2019_ICCV}, and (f) synthetic data generated by our proposed noise model respectively. 
		\emph{All images were converted from raw Bayer space to sRGB for visualization; similarly hereinafter.} \textbf{(Best viewed with zoom)} }
	\label{fig:example}
\end{figure*}

This paper approaches computational low-light imaging from a fundamental perspective -- directly modeling the noise in the imaging procedure of photosensors. Our first observation is that the successful design of denoising algorithms is highly contingent upon the accuracy of the adopted noise model \cite{Xu_2017_ICCV,xu2018trilateral,Shi2018Toward,Brooks2018Unprocessing}. 
A precise noise model can not only motivate the design of suitable regularizers for optimization-based methods, but also benefit  learning-based algorithms by synthesizing rich realistic training data. Our second observation is the noise pattern are complex and cannot be accurately modeled by existing methods. Even the state-of-the-art heteroscedastic Gaussian noise model \cite{Foi2008Practical} cannot delineate the full picture of sensor noise under severely low illuminance. An illustrative example is shown in Figure~\ref{fig:example}, where the objectionable banding pattern artifacts, an unmodeled noise component that is exacerbated in dim environments, become clearly noticeable by human eyes.

\begin{figure}[!t]
	\centering
	\includegraphics[width=.98\linewidth,clip,keepaspectratio]{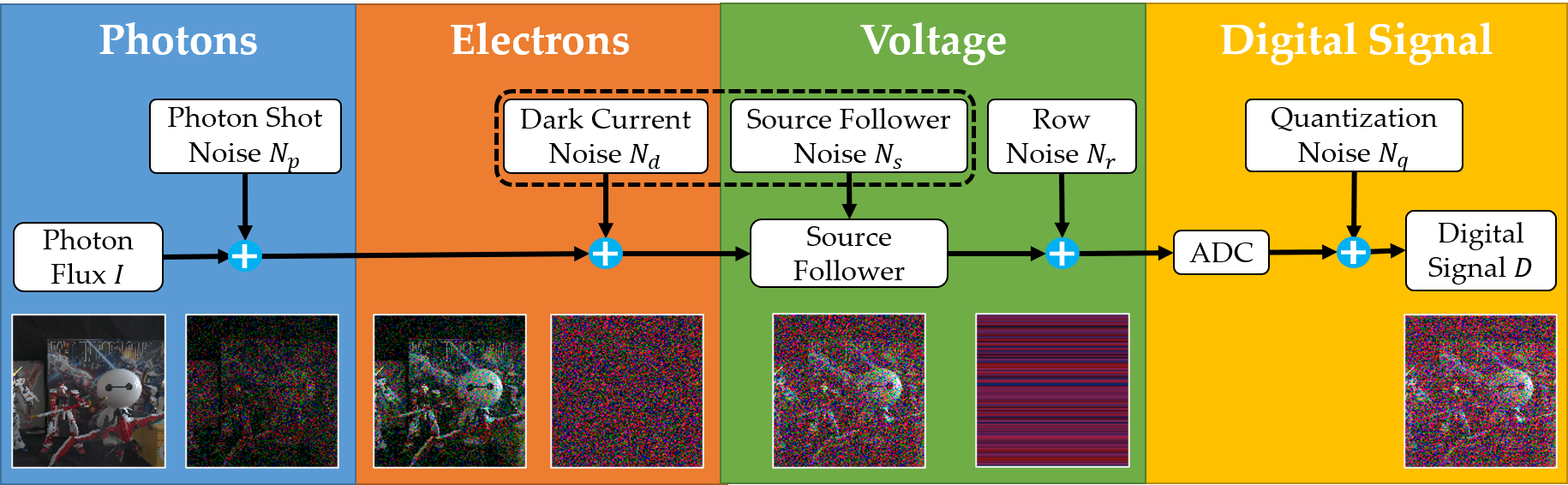}
	\caption{Overview of electronic imaging pipeline and visualization of noise source and resulting image at each stage. }
	\label{fig:photosensor}
\end{figure}

In this work, we mainly focus on the noise formation model for \emph{raw} images to avoid the impact on noise model from the image processing pipeline (ISP) when converting raw data to sRGB (\eg gamma correction, gamut mapping, and tone mapping). 
We propose a physics-based noise formation model for extreme low-light imaging, which explicitly leverages the characteristics of CMOS photosensors
to better match the physics of noise formation. As shown in Figure~\ref{fig:photosensor}, our proposed synthetic pipeline derives from the inherent
process of electronic imaging by considering how photons go through several
stages, and model sensor noise in a fine-grained manner that includes many noise sources, \eg photon shot noise, dark current noise, and pixel circuit noise. This yields a comprehensive statistical noise model that can accurately represent various elusive noise-driven phenomena (\eg the banding pattern artifacts and the color bias issue) in very low light. 
It also provides a way to decouple the complicated real noise structure into different statistical distributions with clear physical interpretations, which facilitates the understanding of the real noise occurred in extreme low-light conditions. 

Furthermore, we devise a method to calibrate the noise parameters from available digital cameras, by first estimating the noise parameters at various ISO settings, then modeling the joint distributions of noise parameters. With this calibration technique, our approach can be easily adapted into any new camera devices at little cost on recording calibration data, thus bypassing the time-consuming paired real training data collection. In order to systematically investigate the generality of our noise model, we additionally introduce
an Extreme Low-light Denoising (ELD) dataset taken by various camera devices.

We evaluate our approach on a wide spectrum of low-light imaging applications, including extreme low-light raw denoising, extreme low-light image processing, extreme low-light video denoising as well as several downstream vision applications (\ie depth estimation, optical flow, object detection/recognition) in the dark. For these tasks, we use the proposed noise model to synthesize training data for deep neural networks. The experiments collectively show that 
the network trained only with the synthetic data from our noise model can generate highly-accurate denoising results, which are on par with or sometimes even outperform training with real labeled data.



Our main contributions are summarized as follows: 
\begin{itemize}
	\item We formulate a comprehensive noise model that can accurately characterize the real noise structure in the dark. This not only enables us to synthesize realistic noisy images that can match the quality of real data, but also facilitates the understanding of the complicated real noise structure. We demonstrate the usefulness of our noise model on a wide range of low-light imaging applications and  show that training deep neural network with it achieves state-of-the-art denoising performance.
	\item We propose a noise parameter calibration method that can fit our noise model into any given camera devices at various ISO settings. It allows the fast adaptation and deployment of our noise model into any new devices at minimal extra cost, thereby circumventing the labor-intensive paired real data acquisition. 
	\item We collect an extreme low-light denoising dataset ELD which contains images captured by various camera devices under different scenes to verify the effectiveness and generality of the proposed noise model and compare different methods.
\end{itemize}



The remainder of this paper is organized as follows. 
In Section \ref{sec:related-work}, we review related low-light imaging methods including both hardware-based approach and computation-based approach. 
Section \ref{sec:noise-model} introduces our physics-based noise formation model and the noise parameter calibration method for extreme low-light imaging.  Both quantitative and qualitative experimental analysis of our approach are provided in Section \ref{sec:experiments}, followed by discussions of applicability scope in Section \ref{sec:discussion-scope}.  
Conclusions and discussions of open problems are drawn in Section \ref{sec:conclusion}. A preliminary version of this work was presented as a conference paper~\cite{wei2020physics}.

\section{Related Work} \label{sec:related-work}
The challenge of imaging in low light is well-known in the computational photography/imaging community. In this section, we provide an overview of low-light imaging and denoising techniques.

\subsection{Hardware-based Approach}

Conventional photon detectors \ie CMOS and CCD photosensors, generally require collecting a large number of photons ($\sim10^{3}$  photons per pixel) 
to make a photograph \cite{holst1998ccd,llull2013coded}, which is largely infeasible in many scientific applications, \eg non-stellar observation in astronomy \cite{mclean2008electronic}, remote sensing of human activity from space \cite{levin2020remote} as well as imaging of delicate biological samples in microscopy \cite{joens2013helium}. Therefore, many advanced imagers with superior light sensitivity, \eg EMCCD, ICCD, and sCMOS imagers, have been designed and fabricated to address these imaging challenges \cite{torr1986intensified,dussault2004noise}.
All these imagers could be operated at very low-photon-flux regimes, where moonlight ($\sim$ 10 photons per pixel) could provide sufficient illumination.
To further push the limit of low-light imaging, recent years have witnessed significant advancements on ultralow-photon imaging techniques powered by single-photon avalanche diode (SPAD) imagers  \cite{guerrieri2009fast,bruschini2019single,kirmani2014first,morris2015imaging,gariepy2015single,shin2016photon,heide2018sub,OToole_2017_CVPR}, making imaging under even starlight ($\sim$ 1 photon per pixel)  possible. 

However, the complex physical mechanism of these hardware-based techniques renders the whole imaging system sophisticated and expensive. 
Moreover, their applications are often restricted to the laboratory environment, thus are not suitable for consumer-level applications in the wild, \eg night vision for autonomous vehicles. In this work, we mainly focus on computational low-light imaging using a conventional camera.
We demonstrate our method can considerably reduce the number of photons needed by the conventional camera by one to two orders of the magnitude,  
which even approaches the low-light sensitivity of advanced low-light imagers.

\subsection{Computation-based Approach}
In contrast to the hardware-based approach, computation-based approaches typically rely on image priors and noise model to reconstruct a high-SNR image from its low-SNR observation acquired by camera. 
Crafting
an analytical regularizer associated with image priors (\eg smoothness,
sparsity, self-similarity, low rank), therefore, plays a critical role in the design pipeline of traditional denoising algorithms \cite{Rudin1992Nonlinear,osher2005iterative,elad2006image,dong2011sparsity,mairal2008sparse,dabov2007BM3D,Buades2005A,Gu2014Weighted}.
In the modern era, most image denoising algorithms are entirely
data-driven, which rely on deep neural networks that implicitly learn the
statistical regularities to infer clean images from their noisy counterparts
\cite{Schmidt_2014_CVPR,Chen_2015_CVPR,mao2016image,zhang2017beyond,Gharbi:2016:DJD:2980179.2982399,tai2017memnet,Chen_2018_ECCV,Shi2018Toward}. Although simple and powerful, these
learning-based approaches are often trained on
synthetic image data due to practical constraints. The most widely-used additive, white, Gaussian noise model
deviates strongly from realistic evaluation scenarios, resulting in significant
performance declines on photographs with real noise \cite{Plotz_2017_CVPR,Abdelhamed_2018_CVPR}.

To step aside the domain gap between synthetic images and real photographs, some
works have resorted to collecting paired real data not just for evaluation but
for training \cite{Abdelhamed_2018_CVPR,Chen_2018_CVPR,Schwartz2018DeepISP,Chen_2019_ICCV,Jiang_2019_ICCV}.
Notwithstanding the promising results, collecting sufficient real data with ground-truth labels to prevent overfitting is exceedingly expensive and
time-consuming. 
Recent works exploit the use of paired (Noise2Noise
\cite{pmlr-v80-lehtinen18a}) or single (Noise2Void
\cite{Krull_2019_CVPR}) noisy images as training data instead
of paired noisy and noise-free images. Still, they can not substantially ease the
burden of labor requirements for capturing a massive amount of real-world
training data.

Another line of research has focused on improving the realism of synthetic
training data to circumvent the difficulties in acquiring abundant real data. By considering both photon arrival statistics (``shot" noise) and sensor
readout effects (``read" noise), the works of \cite{Mildenhall_2018_CVPR,Brooks2018Unprocessing} employed a signal-dependent heteroscedastic Gaussian
model \cite{Foi2008Practical} to characterize the noise properties in raw sensor
data. Most recently, \cite{Wang_2019_ICCV} proposes a noise model which considers the dynamic streak noise, color channel heterogeneous and clipping effect, to simulate the high-sensitivity noise on real low-light color images.  Concurrently, a flow-based generative model \cite{Abdelhamed_2019_ICCV} is proposed to formulate the distribution of real noise using latent variables with tractable density, and 
a GAN-based model is presented to learn camera-aware noise model \cite{chang2020learning}. 
However, these approaches 
oversimplify the modern sensor imaging pipeline, especially the noise sources caused by camera electronics, which have been extensively studied in the electronic
imaging community \cite{Mikhail2014Highlevel,Healey1994Radiometric,Gow2007A,Baer2006A,El2005CMOS,Farrell2008Sensor,Irie2008tech,Boie1992An,Wach2004Noise,Costantini2004Virtual}. 

In this work, we propose a physics-based comprehensive noise formation model stemming from the essential process of electronic imaging to synthesize the noisy-image dataset. We show sizeable improvements of denoising performance on real data, particularly under extremely low illuminance.

\section{Physics-based Noise Formation Model} \label{sec:noise-model}
The creation of a digital sensor raw image $D$ can be generally formulated by a
linear model
\begin{equation}
\label{eq:formation}
D = K I + N,
\end{equation}
where $I$ is the number of photoelectrons that is proportional to the scene
irradiation, $K$ represents the overall system gain composed by analog and
digital gains, and $N$ denotes the summation of all noise sources physically caused
by light or camera.
Under extreme low light, the characteristics of $N$ are formated in terms of the sensor
physical process, 
which is beyond the existing noise models.
In the following, we first describe the detailed procedures of the physical formation of a sensor raw image as well as the noise sources
introduced during the whole process. 
An overview of this process is shown in
Figure~\ref{fig:photosensor}.

\subsection{Sensor Raw Image Formation} \label{sec: raw formation}
Our photosensor model is primarily based upon the CMOS sensor, which is the dominating imaging sensor nowadays \cite{grandviewresearch}. 
We consider the electronic imaging pipeline of how incident light is converted from
photons to electrons, from electrons to voltage, and finally from voltage to
digital numbers, to model noise.

\subsubsection{From Photon to Electrons}
During exposure, incident lights in the form of photons hit the photosensor pixel area, which liberates photon-generated electrons
(photoelectrons) proportional to the light intensity given the photoelectric effect. 

\vspace{6pt}
\noindent\textbf{Photon shot noise.~} Due to the quantum nature of light, there exists an inevitable
uncertainty in the number of electrons collected. 
Such uncertainty imposes a Poisson distribution over this number of electrons, which follows
\begin{align}
(I + N_p) \sim \mathcal{P} \left( I \right),
\label{eq: noise-poisson}
\end{align}
where $N_p$ is termed as the photon shot noise and $\mathcal{P}$ denotes the
Poisson distribution. This type of noise depends on the signal (\ie light intensity). Shot noise is a fundamental limitation and cannot be avoided even
for a perfect sensor.

There are some other noise sources introduced during the
photon-to-electron stage, such as photo response non-uniformity and dark current
noise, as reported in the previous literature \cite{Healey1994Radiometric,Gow2007A,Wach2004Noise,Baer2006A}. Over the last decade, technical advancements
in CMOS sensor design and fabrication, \eg on-sensor dark current suppression,
have led to a new generation of digital single lens reflex (DSLR) cameras with lower dark current and better
photo response uniformity \cite{Fossum2014A,lin2016high}. Therefore, we assume a constant
photo response and absorb the effect of dark current noise $N_d$ into read noise $N_{read}$, which will be presented next.

\subsubsection{From Electrons to Voltage} \label{sec:from-electrons}
After electrons are collected at each site, they are typically integrated, amplified
and read out as measurable charge or voltage at the end of exposure time.
Noise present during the electrons-to-voltage stage depends on the circuit design and processing technology used, and thus is referred to as pixel circuit noise~\cite{Gow2007A}. It includes thermal noise, reset noise~\cite{Mikhail2014Highlevel}, source follower noise~\cite{Leyris2005Trap} and banding pattern noise~\cite{Gow2007A}. The physical origin of these noise components can be found in the electronic imaging literature \cite{Mikhail2014Highlevel,Gow2007A,Wach2004Noise,Leyris2005Trap}.
For instance,  source follower noise is attributed to the action of traps in silicon lattice which randomly capture and emit carriers; banding pattern noise, consisting of row and column artifacts, is associated with the CMOS circuit readout pattern and the amplifier gain mismatch.
By leveraging this knowledge,  we consider the thermal noise $N_{t}$, source follower noise $N_s$ and banding
pattern noise $N_{b}$ in our model. 

\vspace{6pt}
\noindent\textbf{Read noise.~} To simplify analysis, we absorb multiple noise sources including dark current noise $N_d$, thermal noise $N_t$ and source follower noise $N_s$
into a unified term, \ie read noise:
\begin{align}
N_{read} = N_{d} + N_{t} +  N_s. 
\end{align}
Banding pattern noise $N_{b}$ will be considered later in this section. Read noise is generally assumed to follow a Gaussian distribution, but the analysis of
noise data (in Section \ref{sec:noise-param}) tells a long-tailed nature of its
shape. This can be attributed by the flicker and random telegraph signal components of source
follower noise \cite{Gow2007A}, or the dark spikes raised by dark current
\cite{Mikhail2014Highlevel}. 
Therefore, we propose using a statistical distribution that can better characterize the long-tail shape.
Specifically, we model the read
noise by a Tukey lambda distribution ($TL$)~\cite{Joiner1971Some}, which is a
distributional family that can approximate a number of common distributions (\eg
a heavy-tailed Cauchy distribution):
\begin{align}
N_{read} \sim \mathop{TL} \left( \lambda; 0, \sigma_{TL} \right),
\label{eq: noise-tl}
\end{align}
where $\lambda$ and $\sigma_{TL}$ indicate the shape and scale parameters
respectively, while the location parameter is set to be zero with zero-mean noise assumption.

\vspace{6pt}
\noindent\textbf{Color-biased read noise.~} 
Although zero-mean noise assumption is generally applicable in most situations, 
we find it breaks down under extreme low-light settings, because of the non-negligible direct current (DC) (\ie zero-frequency) noise component (see Section~\ref{sec:noise-param}).
This component originates from the dark current noise, \ie the averaged number of thermally generated electrons, which renders the noise distribution no longer zero-centered. 
Though this component has been largely eliminated by the modern on-sensor dark current suppression techniques, \eg pinned photodiode architecture \cite{darkcurrent_Clark}, the residual component is still impactful due to the large system gain $K$ applied to amplify the signal as well as the noise. Even worse, the evaluation (in Section 
\ref{sec:noise-param}) reveals this DC noise component varies across color channels, potentially owing to the stack photodiodes design of modern CMOS sensors \cite{gilblom2004real}.
Such a color heterogeneous DC component is  the culprit that leads to the color bias phenomenon frequently observed under extreme low illuminance (\emph{c.f.} Figure~\ref{fig:example}). 

To make our noise model simple yet compact, we model the DC noise component (\ie color bias) as a mean value of the read noise model. As a result, we modify the Equation~\eqref{eq: noise-tl} to
\begin{align}
N_{read} \sim \mathop{TL} \left( \lambda; \mu_c, \sigma_{TL} \right),
\label{eq: noise-tl-bias}
\end{align}
where $\mu_c$ denotes the color-wise DC noise component. The separated analysis of DC component and long-tailed $TL$ distribution will be performed in Section \ref{sec:noise-ablation}.

\vspace{6pt}
\noindent\textbf{Row noise.~} We introduce row noise $N_r$ to account for banding pattern noise $N_b$. Though $N_b$ may appear in images as horizontal or vertical lines, we only consider the row-wise component (horizontal stripes) in our model, as the column-wise counterpart is generally negligible
when measuring the noise data (Section \ref{sec:noise-param}). We simulate the
$N_r$ by sampling a value from zero-mean Gaussian distribution $\mathcal{N} \left( 0, \sigma_r \right)$ with scale parameter $\sigma_r$ for each row, \ie
\begin{align}
N_r \sim \mathcal{N} \left( 0, \sigma_r \right),
\label{eq: noise-row}
\end{align}
then adding it as an offset to all the pixels within that row.

\subsubsection{From Voltage to Digital Numbers}
To generate an image that can be stored in a digital storage medium, the
analog voltage signal read out during last stage is quantized into discrete
codes using an analog-to-digital converter (ADC), which introduces quantization noise.

\vspace{6pt}
\noindent\textbf{Quantization noise.~}  
Quantization noise $N_q$ is a rounding error between the analog input voltage to the ADC and the output digitized
value, which can be assumed to follow a uniform distribution, \ie 
\begin{align}
N_q \sim U \left( -1/2 q , 1/2 q \right),
\end{align}
where $U (\cdot, \cdot)$ denotes the uniform distribution over the range
$[-1/2 q , 1/2 q]$ and $q$ is the quantization step.

\vspace{6pt}
\textbf{To summarize}, our noise formation model consists of four major noise components:
\begin{align}
N = K N_p + N_{read} + N_r + N_q,
\label{eq: noise-formation}
\end{align}
where $K$, $N_p$, $N_{read}$, $N_r$ and $N_q$ denotes the overall system gain, photon shot noise, read noise, row noise and quantization noise, respectively. 

\vspace{-6pt}
\subsection{Sensor Noise Evaluation} \label{sec:noise-param}

\begin{figure}[!t]
	\centering
	\begin{subfigure}[b]{.318\linewidth}
		\centering
		\includegraphics[width=1\linewidth,clip,keepaspectratio]{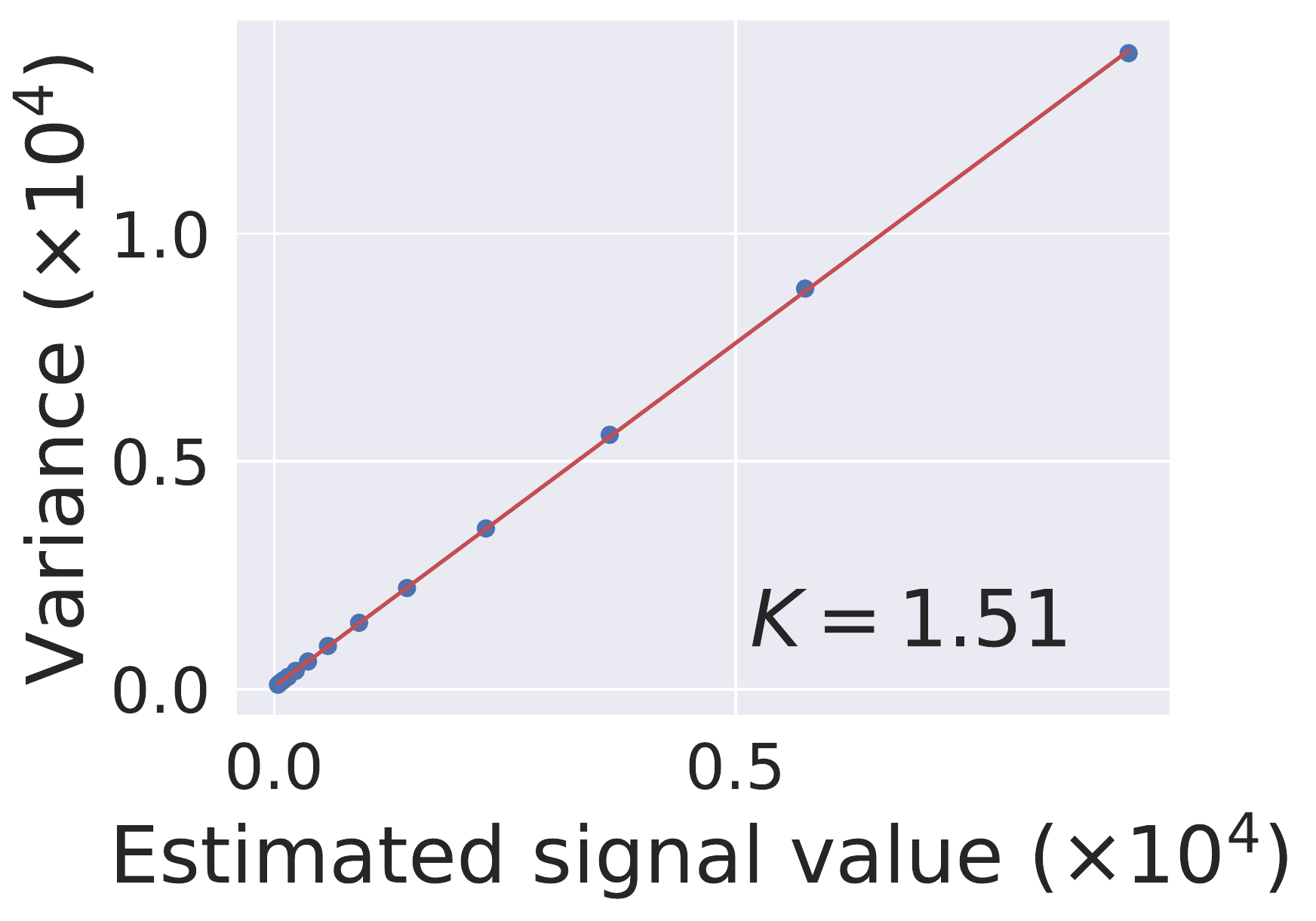}
		\caption{\footnotesize SonyA7S2}
	\end{subfigure}
	\begin{subfigure}[b]{.329\linewidth}
		\centering
		\includegraphics[width=1\linewidth,clip,keepaspectratio]{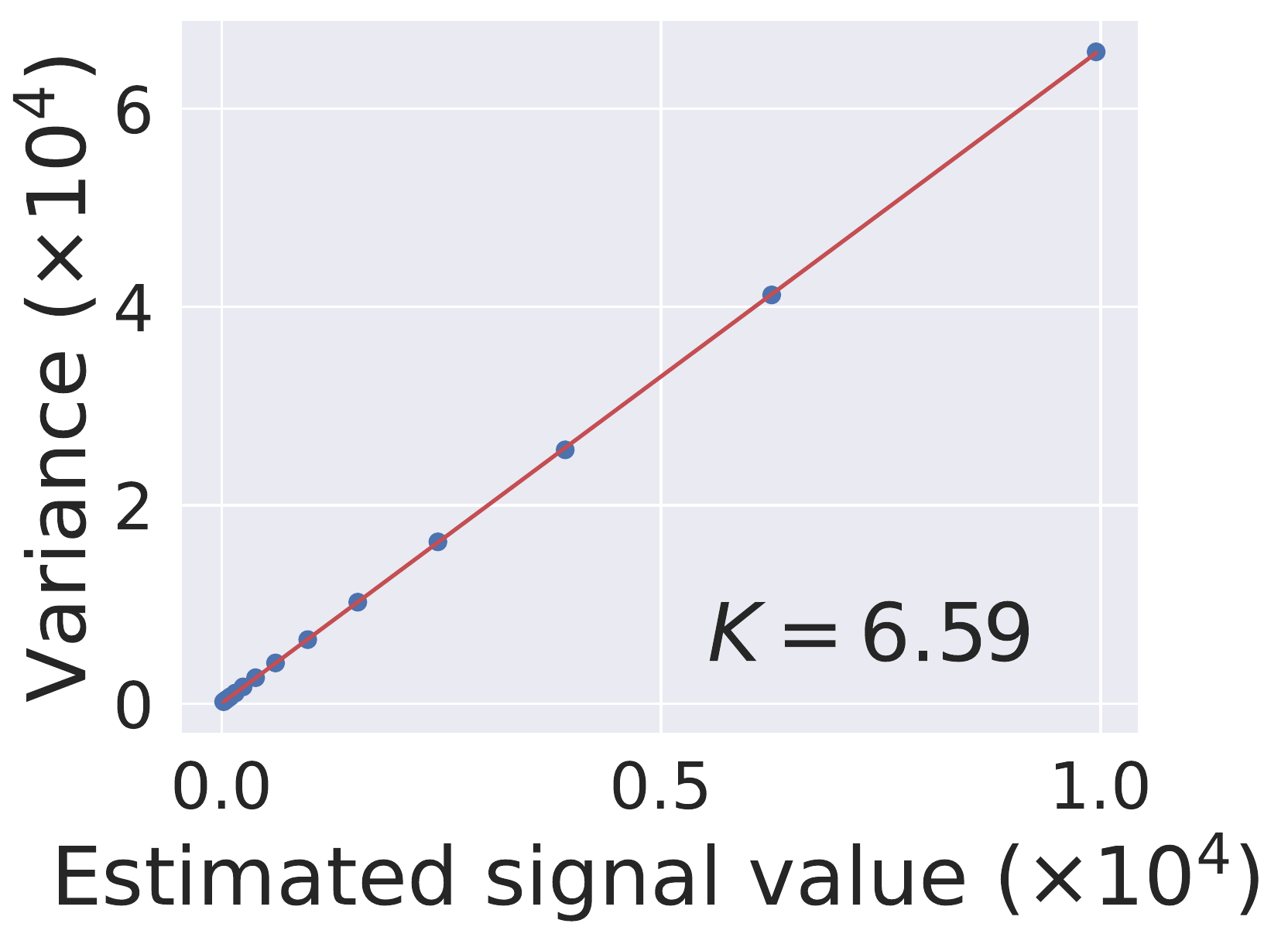}
		\caption{\footnotesize NikonD850}
	\end{subfigure}
	\begin{subfigure}[b]{.329\linewidth}
		\centering
		\includegraphics[width=1\linewidth,clip,keepaspectratio]{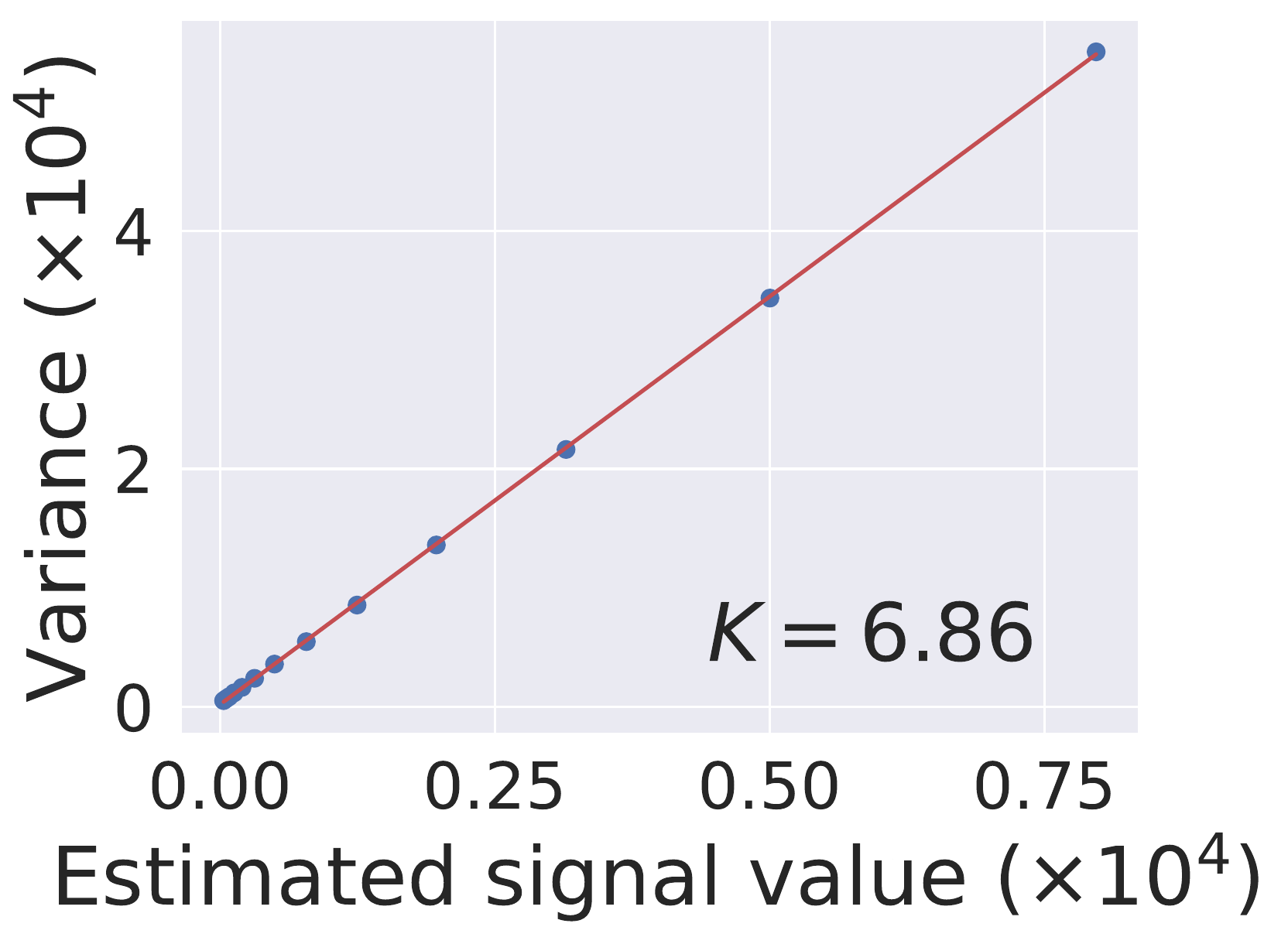}
		\caption{\footnotesize CanonEOS70D}
	\end{subfigure}
	\caption{Estimation of the overall system gain $K$ for the three cameras. The noisy signal variance ($y$-axis) and the underlying true signal value ($x$-axis) satisfy a linear function whose slope implies $K$ at the measured ISO (1600) setting. }
	\label{fig:photon-transfer}
\end{figure}

\begin{figure}[!t]
	\centering
	\begin{tabular}{cc}
		\rotatebox[origin=c]{90}{\footnotesize SonyA7S2}\!\!\!\! &
		\includegraphics[align=c,width=.9\linewidth,clip,keepaspectratio]{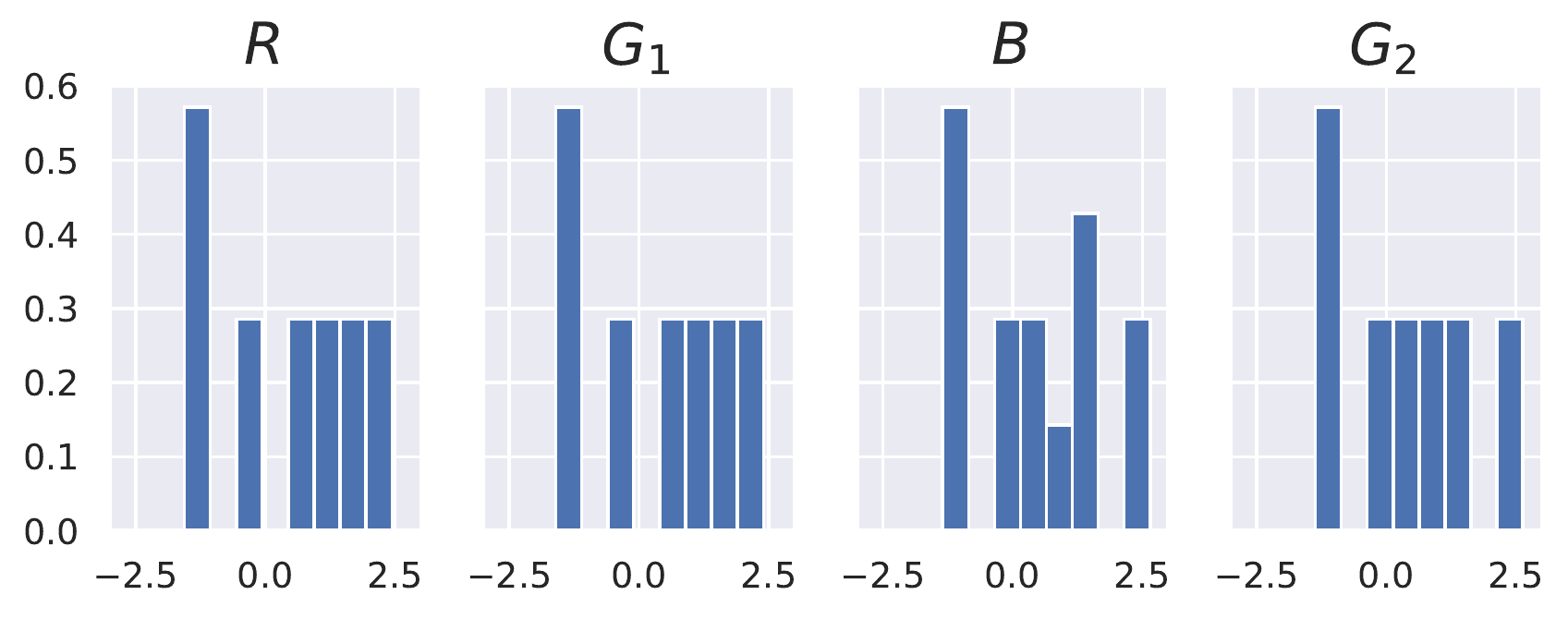} \\
		\rotatebox[origin=c]{90}{\footnotesize NikonD850}\!\!\!\! &
		\includegraphics[align=c,width=.9\linewidth,clip,keepaspectratio]{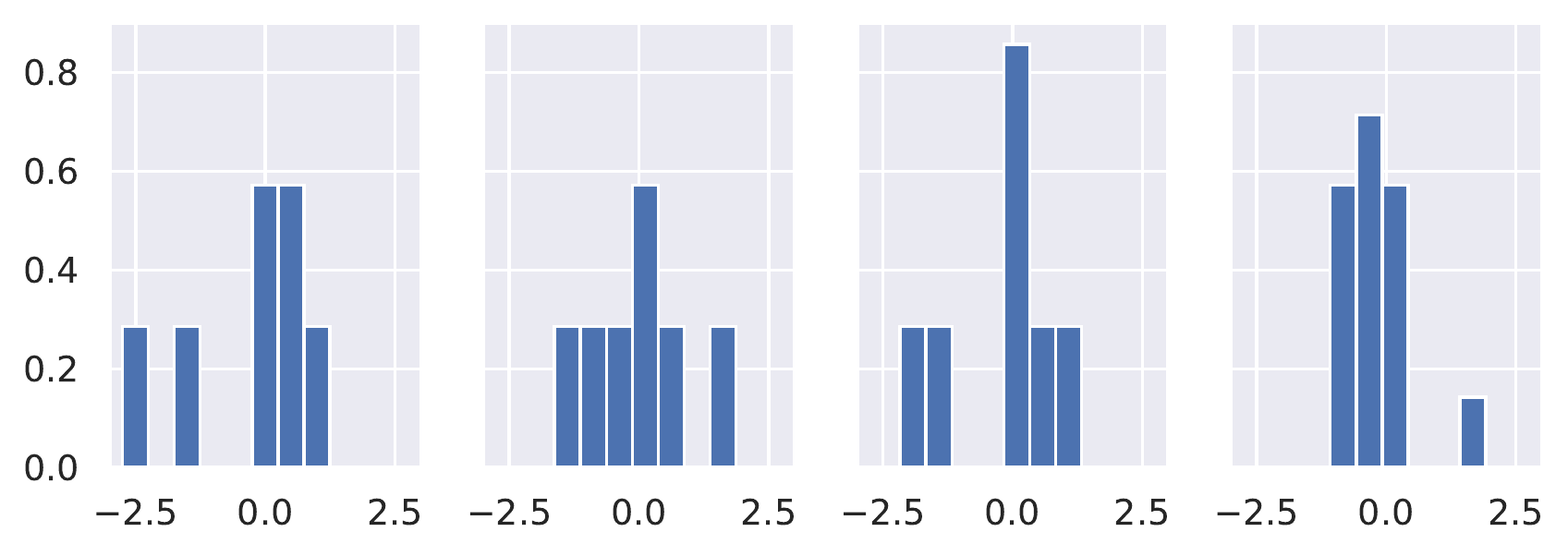} \\
		\rotatebox[origin=c]{90}{\footnotesize CanonEOS70D}\!\!\!\! &
		\includegraphics[align=c,width=.9\linewidth,clip,keepaspectratio]{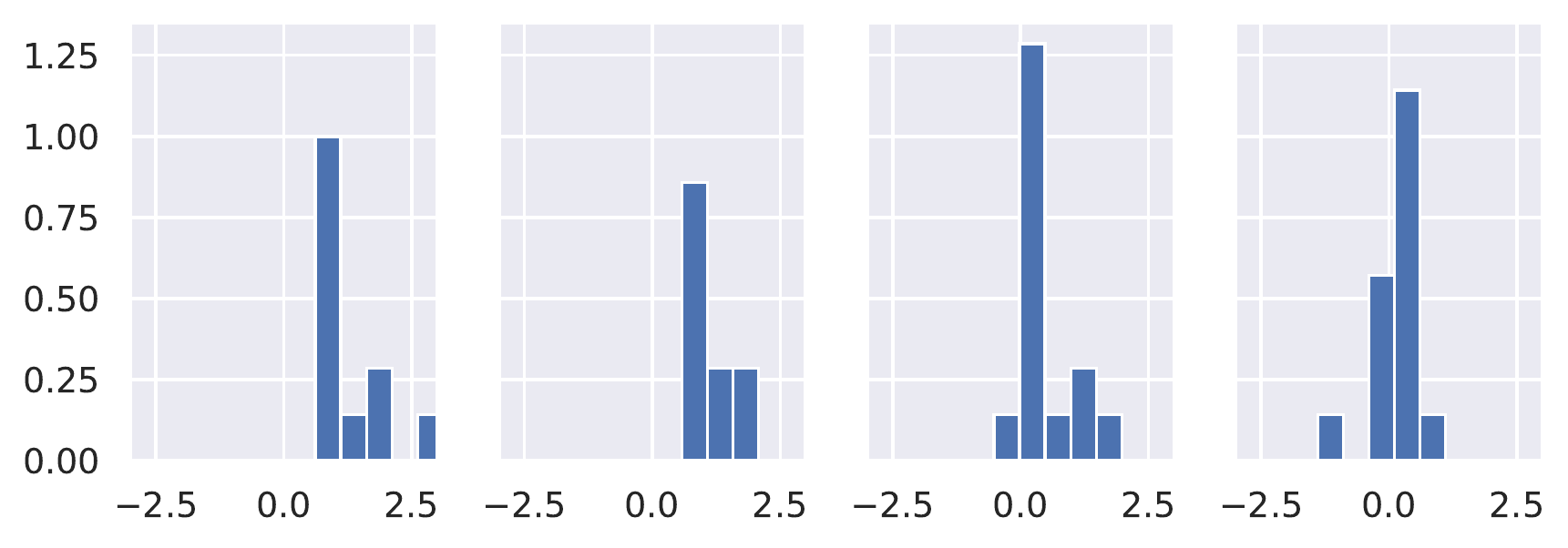} \\
	\end{tabular}
	\vspace{-4pt}
	\caption{Normalized density histogram (with bin size set to 0.5) of the color-wise DC noise component $\mu_c$ from different cameras (14-bit depth). }
	\vspace{-4pt}
	\label{fig:color-bias}
\end{figure}

In this section, we present a noise parameter calibration method attached to our proposed noise formation model. 
Table~\ref{tb:noise-model} summarizes the necessary parameters in our noise model, which include
overall system gain $K$ for photon shot noise $N_p$, shape and scale parameters
($\lambda$ and $\sigma_{TL}$) for read noise $N_{read}$, $\mu_c$ for DC noise component (color bias), and scale parameter $\sigma_r$ for row noise $N_r$.
Given a camera device,  our noise calibration method consists of two main procedures: 1) estimating noise parameters at various ISO settings\footnote{Noise parameters are generally stationary at a fixed ISO.}, and 2)  modeling joint distributions of noise parameters. 

\begin{table}[!t]
	\centering
	\caption{Summary of our noise model}
	\footnotesize
	\setlength{\tabcolsep}{1mm}{
		\begin{tabular}{C{.33\linewidth}|C{.39\linewidth}|C{.2\linewidth}}
			\hline
			\cellcolor[HTML]{EFEFEF} Noise type & \cellcolor[HTML]{EFEFEF} Formulation & \cellcolor[HTML]{EFEFEF} Parameters \\ \hline
			\multirow{2}{*}{Photon shot noise $N_p$} &  Poisson distribution  & \multirow{2}{*}{System gain $K$} \\ 
			& $(I + N_p) \sim \mathcal{P} \left( I \right)$ &  \\\hline
			\multirow{3}{*}{Read noise $N_{read}$} 
			& \multirow{3}{*}{\shortstack[l]{Tukey lambda distribution\\ $N_{read} \sim \mathop{TL} \left( \lambda; \mu_c, \sigma_{TL} \right)$}}  & Shape $\lambda$ \\
			&   & Color bias $\mu_c$ \\
			& & Scale $\sigma_{TL}$       \\ \hline
			\multirow{2}{*}{Row noise $N_r$} & Gaussian distribution  & \multirow{2}{*}{Scale $\sigma_r$} \\   		& $N_r \sim \mathcal{N} \left( 0, \sigma_r \right)$ &   \\ \hline
			\multirow{2}{*}{Quantization noise $N_q$} & Uniform distribution  & \multirow{2}{*}{None} \\ 
			& $N_q \sim U\left( -1/2 q , 1/2 q \right)$ & \\\hline						
			
	\end{tabular}}
	\label{tb:noise-model}
\end{table}

\vspace{-6pt}
\subsubsection{Estimating Noise Parameters} 
Our calibration method makes use of two types of raw images captured at specialized settings, \ie \emph{flat-field frames} and \emph{bias frames} to estimate noise parameters. 
Flat-field frames are the images captured when sensor is uniformly illuminated. 
We take them of a white paper on a uniformly-lit wall. The camera is mounted on a tripod close to the paper, and the lens is focused on infinity to diminish non-uniformity.
Bias frames are the images captured under a lightless environment with the
shortest exposure time. We take them at a dark room and the camera lens is capped-on. 
Flat-field frames characterize the light-dependent photon shot noise, thus can be used to estimate the related parameter $K$, while bias frames delineate the dark noise picture independent of light, which can be used to derive other noise parameters $\mu_c$, $\sigma_r$, $\lambda$, $\sigma_{TL}$ sequentially.

\vspace{6pt}
\noindent\textbf{Estimate $K$ for photon shot noise.~}
According to Equation~\eqref{eq:formation} \eqref{eq: noise-poisson} \eqref{eq: noise-formation}, a noisy sensor raw data $D$ can be expressed by 
\begin{align}
D = K (I + N_p) + N_{o}
\label{eq:raw}
\end{align}
where $N_{o} = N_{read} + N_r + N_q$ accounts for other noise sources independent of light.  Following \eqref{eq:raw}, the variance of noisy raw data  is given by
\begin{align}
Var(D) = K^2 Var(I+N_p) +   Var(N_o)
\end{align}
where $Var(\cdot)$ denotes the variance operator that calculates the variance of a given random variable. 
Given that $(I+N_p)$ follows a Possion distribution, whose variance equals to its mean,  we have
\begin{align}
Var(D) &= K^2 I +   Var(N_o) \nonumber\\
&= K (KI) + Var(N_o)  \label{eq:photon-transfer}
\end{align}
where $KI$ is the underlying true signal and represented by digital numbers. 

Equation~\eqref{eq:photon-transfer} reveals a linear relationship between the signal variance $Var(D)$ and the underlying clean digital signal $KI$. Determining $K$, therefore,  simply requires fitting a line,  if a set of observations ($Var(D)$, $KI$) are accessible to us.  Signal variance is easy to calculate, but the true signal value is generally unavailable. By utilizing the flat-field frame, the true signal value $KI$ can be approximated by the median statistics owing to the uniformity. 
Now we can plot the estimated signal intensity against the variance of the noisy image, and calculate a linear least square regression for these two sets of measurements. As shown in Figure~\ref{fig:photon-transfer}, these data points almost perfectly lie in a straight line, whose slope characterizes the overall system  gain $K$  at the measured ISO setting of the camera.

Given estimated $K$, we can simulate realistic photon shot noise by firstly convert a raw digital signal $D$ into the number of photoelectrons $I$, then impose a Poisson distribution on it, and finally revert it to $D$.

\begin{figure}[!t]
	\centering
	\begin{subfigure}[b]{.3\linewidth}
		\centering
		\includegraphics[width=1\linewidth,clip,keepaspectratio]{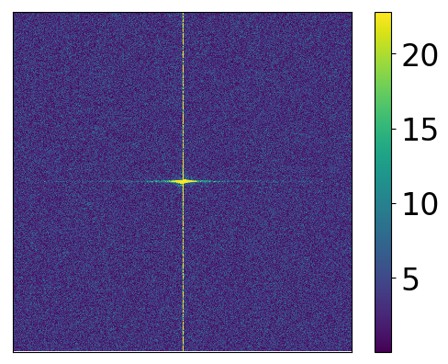}
		\caption{\footnotesize SonyA7S2}
	\end{subfigure}
	\begin{subfigure}[b]{.3\linewidth}
		\centering
		\includegraphics[width=1\linewidth,clip,keepaspectratio]{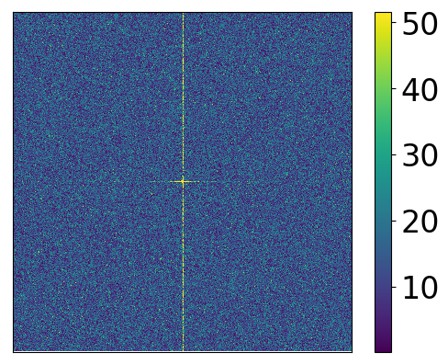}
		\caption{\footnotesize NikonD850}
	\end{subfigure}
	\begin{subfigure}[b]{.3\linewidth}
		\centering
		\includegraphics[width=1\linewidth,clip,keepaspectratio]{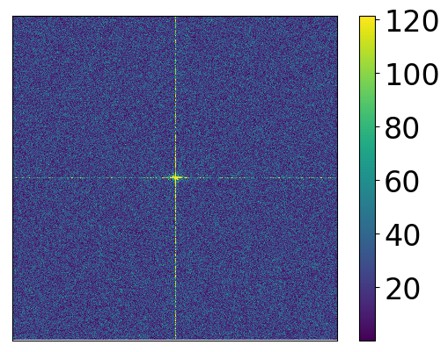}
		\caption{\footnotesize CanonEOS70D}
	\end{subfigure}
	\vspace{-4pt}
	\caption{Centralized Fourier spectrum of the bias frames captured by three cameras.}
	\vspace{-4pt}
	\label{fig:fft_bf}
\end{figure}

\begin{figure}[!t]
	\centering
	\begin{subfigure}[b]{.325\linewidth}
		\centering
		\includegraphics[width=1\linewidth,clip,keepaspectratio]{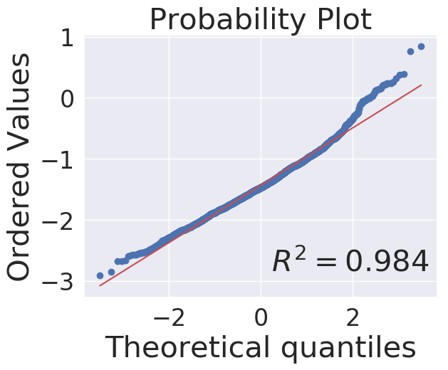}
		\caption{\footnotesize SonyA7S2}
	\end{subfigure}
	\begin{subfigure}[b]{.325\linewidth}
		\centering
		\includegraphics[width=1\linewidth,clip,keepaspectratio]{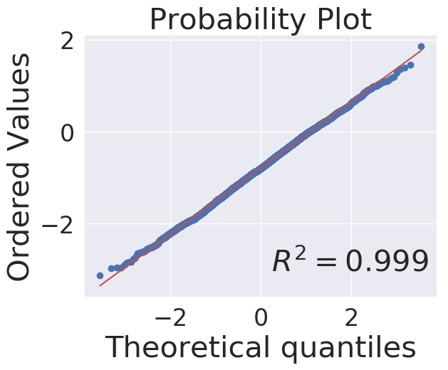}
		\caption{\footnotesize NikonD850}
	\end{subfigure}
	\begin{subfigure}[b]{.325\linewidth}
		\centering
		\includegraphics[width=1\linewidth,clip,keepaspectratio]{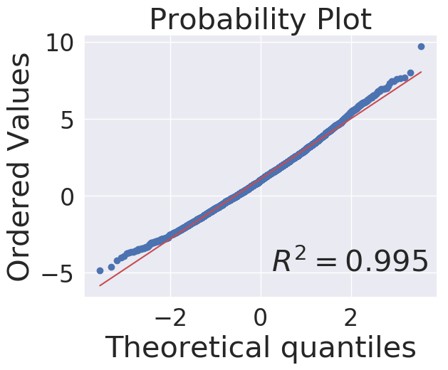}
		\caption{\footnotesize CanonEOS70D}
	\end{subfigure}
	\caption{Normal probability plots of row noise data for three cameras. The resulting image looks close to a straight line if the data are approximately normally distributed.}
	\label{fig:qqplot-row}
\end{figure}

\vspace{6pt}
\noindent\textbf{Estimate $\mu_c$ for color bias.~} Given a bias frame, 
the DC noise component can be examined by averaging all pixel values within each color channel of the bias frame. If the noise distribution is zero-centered without DC component, the resulting color-wise values should almost equal to the black level\footnote{\emph{Black level} denotes the ideal noise-free readout value when no light hits the pixel array. } (per channel) recorded in the metadata of raw images. However, we find these values are severely biased from the recorded black level, which cannot be explained as  random fluctuations caused by other zero-mean noise - this discloses the existence of DC noise component. Figure~\ref{fig:color-bias} shows the density histogram of these biases $\mu_c$ for each color channel calculated from a number of bias frames. 
Obvious differences on histogram can be observed among color channels, which demonstrate the varied statistics of DC noise component over color channels.  

Note this observation and the new model are very important to extreme low-light denoising, since small bias could lead to severe color shift in extreme low light due to the large digital gain (e.g., $\times$100) applied to signal as well as noise. It challenges the commonly used zero-mean noise assumption, and significantly improves the low-light denoising performance (see Section \ref{sec:noise-ablation}).


To exclude the influence of DC noise in the following evaluation of other noise components, the mean values of each color channel are subtracted from the bias frames.

\vspace{6pt}
\noindent\textbf{Estimate $\sigma_r$ for row noise.~}
The banding pattern noise can be tested via performing discrete Fourier transform on the bias frame.  In Figure~\ref{fig:fft_bf}, the highlighted vertical pattern in the centralized Fourier spectrum reveals the existence of row noise component. 
To analyze the distribution of row noise, we extract the
mean values of each row from raw data. These values, therefore, serve as good
estimates to the underlying row noise intensities, given the zero-mean nature of other remaining noise sources.  
A normal probability plot \cite{Wilk1968Probability} is drawn in Figure~\ref{fig:qqplot-row}, to compare the empirical
distribution of row noise with a normal distribution. 
The normality of the row noise data is also tested by a Shapiro-Wilk test \cite{Shapiro1975An}: the resulting $p$-value is higher than $0.05$, suggesting the null hypothesis that the data are normally distributed cannot be rejected. 
The related scale parameter $\sigma_r$ can be easily
estimated by maximizing the log-likelihood.
The estimated row noise are then removed from the bias frames for the following calibration process.

\begin{figure}[!t]
	\centering
	\begin{tabular}{cccc}
		\!\!\!\rotatebox[origin=c]{90}{\footnotesize SonyA7S2}\!\!\!\!\! &
		\includegraphics[align=c,width=.293\linewidth,clip,keepaspectratio]{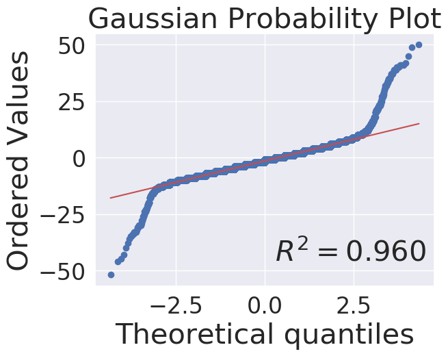}\!\!\! &
		\includegraphics[align=c,width=.293\linewidth,clip,keepaspectratio]{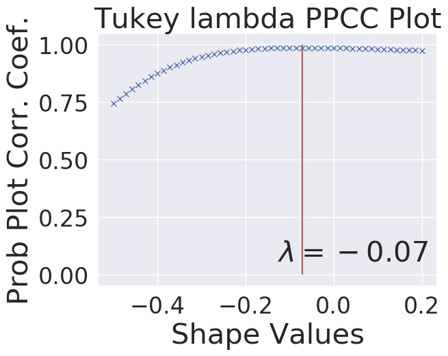}\!\!\! &
		\includegraphics[align=c,width=.293\linewidth,clip,keepaspectratio]{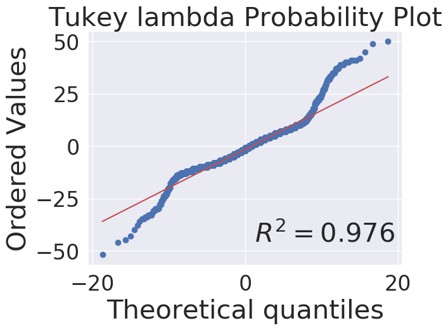} \\
		\!\!\!\rotatebox[origin=c]{90}{\footnotesize NikonD850}\!\!\!\!\! &
		\includegraphics[align=c,width=.293\linewidth,clip,keepaspectratio]{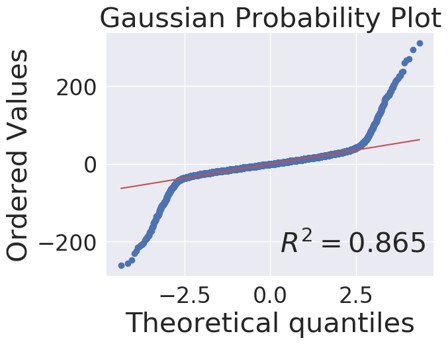}\!\!\! &
		\includegraphics[align=c,width=.293\linewidth,clip,keepaspectratio]{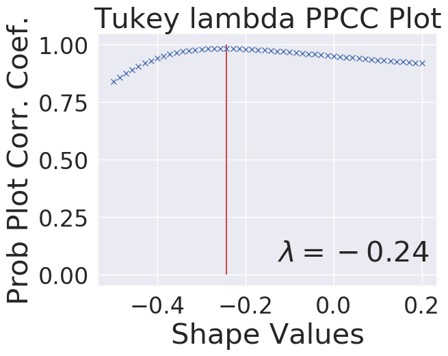}\!\!\! &
		\includegraphics[align=c,width=.293\linewidth,clip,keepaspectratio]{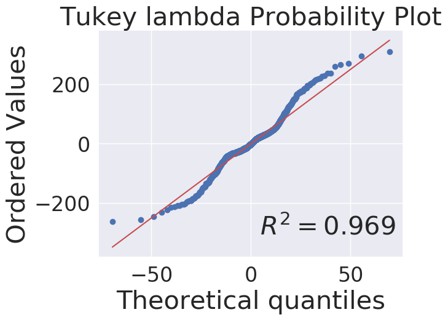} \\
		\!\!\!\rotatebox[origin=c]{90}{\footnotesize CanonEOS70D}\!\!\!\!\! &
		\includegraphics[align=c,width=.293\linewidth,clip,keepaspectratio]{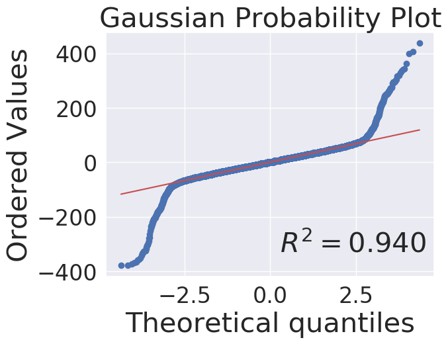}\!\!\! &
		\includegraphics[align=c,width=.293\linewidth,clip,keepaspectratio]{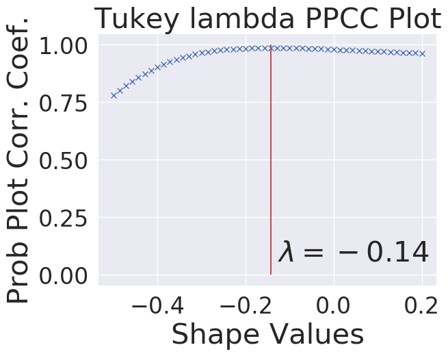}\!\!\! &
		\includegraphics[align=c,width=.293\linewidth,clip,keepaspectratio]{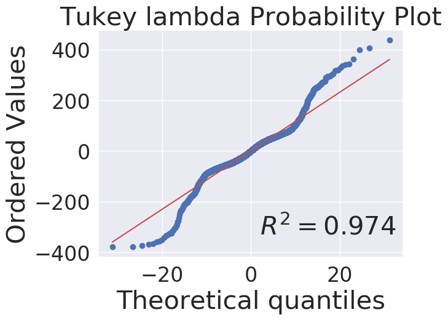} \\
	\end{tabular}
	\vspace{-4pt}
	\caption{Distribution fitting of read noise for three cameras. \textit{Left:} probability plot against the Gaussian distribution.
		\textit{Middle:} Tukey lambda PPCC plot that determines the optimal $\lambda$ (shown in red line). \textit{Right:} probability plot against the Tukey Lambda distribution. A higher $R^2$ indicates a better fit. }
	\vspace{-4pt}
	\label{fig:TL-PPCC}
\end{figure}

\vspace{6pt}
\noindent\textbf{Estimate $\lambda$ and $\sigma_{TL}$ for read noise.~} Statistical models can be used to fit the empirical distribution of the residual read noise. A preliminary diagnosis (Figure~\ref{fig:TL-PPCC} Left) shows the main
body of the data may follow a Gaussian distribution, but it also
unveils the long-tail nature of the underlying distribution. 
In contrast to regarding extreme values as outliers, 
we observe an appropriate long-tail statistical distribution can
characterize the noise data better. 

We generate a probability plot correlation
coefficient (PPCC) plot~\cite{Filliben1975The} to identify a statistical model
from a Tukey lambda distributional family~\cite{Joiner1971Some} that best
describes the data. The Tukey lambda distribution is a family of distributions
that can approximate many distributions by varying its shape parameter
$\lambda$. It can approximate a Gaussian distribution if $\lambda = 0.14$,  or derive a
heavy-tailed distribution if $\lambda < 0.14$.  
The PPCC plot (Figure~\ref{fig:TL-PPCC} Middle) is used to find a good value of $\lambda$. The
probability plot~\cite{Wilk1968Probability} (Figure~\ref{fig:TL-PPCC} Right) is then employed to estimate the scale parameter $\sigma_{TL}$. 
The goodness-of-fit can be evaluated by $R^2$ -- the
coefficient of determination \emph{w.r.t.} the resulting probability plot~\cite{MORGAN201115}. 
The $R^2$ of the fitted Tukey
Lambda distribution is much higher than the Gaussian distribution (\eg $0.972$ \emph{vs.}
$0.886$), indicating a much better fit to the empirical data.

It should be noted the Poisson mixture model was used in \cite{zhang2017improved} for heavy tail modeling, which utilized two Poisson components to capture the long-tailed behavior of sensor noise. It implicitly imposes a signal/light-dependent nature on noise due to the Poisson model. However, as we found in Figure~\ref{fig:TL-PPCC}, the long-tailed characteristics is virtually a feature of read noise, which is the noise present at bias frames captured under lightless conditions. It violates the underlying assumption of Poisson mixture model. Consequently, we still adopt the Tukey lambda distribution for long-tailed read noise modeling. 

\begin{figure}[t]
	\centering
	\setlength\tabcolsep{1pt}
	\renewcommand\arraystretch{1}	
	\begin{tabular}{cccc}
		& \footnotesize Real Bias Frame & \footnotesize Gaussian Model & \footnotesize Ours \\
		\rotatebox[origin=c]{90}{\footnotesize SonyA7S2}	&
		\includegraphics[align=c,width=.3\linewidth,clip,keepaspectratio]{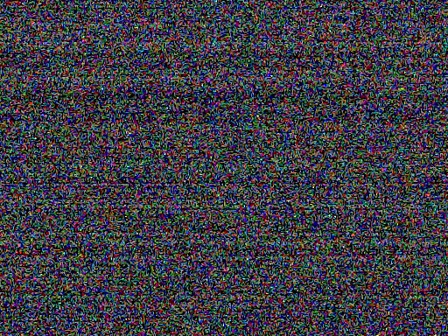} &
		\includegraphics[align=c,width=.3\linewidth,clip,keepaspectratio]{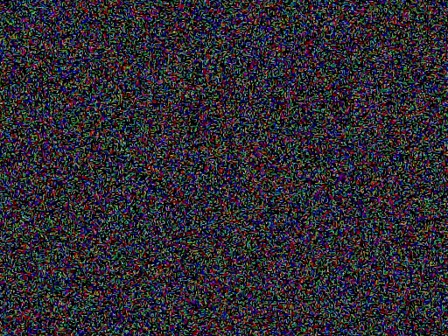} &
		\includegraphics[align=c,width=.3\linewidth,clip,keepaspectratio]{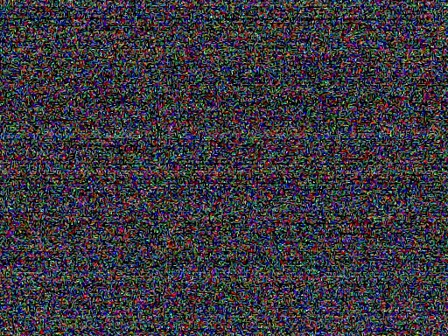} \\
		&  \footnotesize ($R^2$) & \footnotesize (0.961) & \footnotesize (0.978)\\
		\rotatebox[origin=c]{90}{\footnotesize NikonD850}	&
		\includegraphics[align=c,width=.3\linewidth,clip,keepaspectratio]{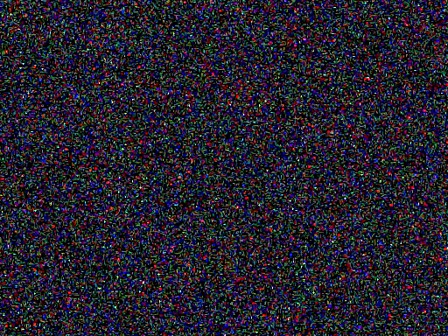} &
		\includegraphics[align=c,width=.3\linewidth,clip,keepaspectratio]{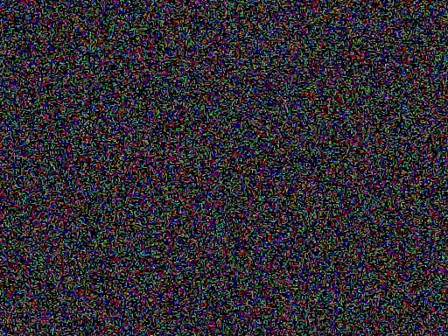} &
		\includegraphics[align=c,width=.3\linewidth,clip,keepaspectratio]{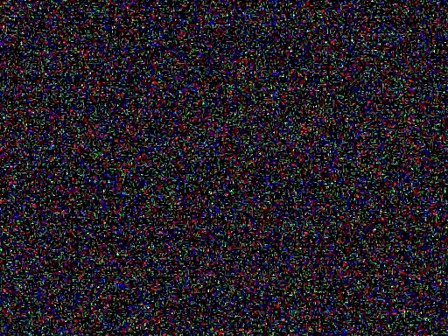} \\
		& \footnotesize ($R^2$) & \footnotesize (0.880) & \footnotesize (0.972) \\
		\rotatebox[origin=c]{90}{\footnotesize CanonEOS70D}	&
		\includegraphics[align=c,width=.3\linewidth,clip,keepaspectratio]{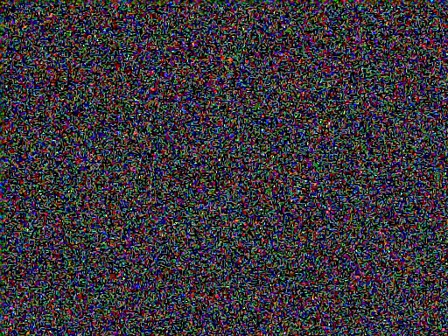} &
		\includegraphics[align=c,width=.3\linewidth,clip,keepaspectratio]{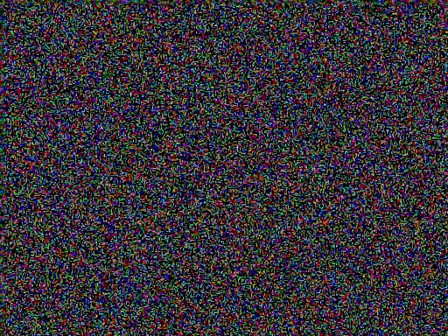} &
		\includegraphics[align=c,width=.3\linewidth,clip,keepaspectratio]{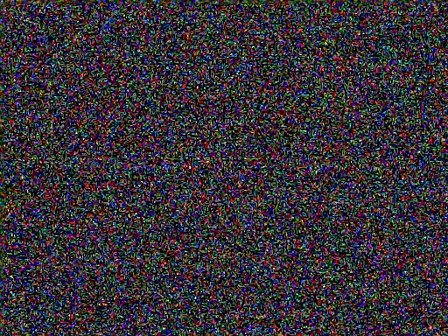} \\
		& \footnotesize ($R^2$) & \footnotesize (0.915) & \footnotesize (0.962) \\
	\end{tabular}
	\vspace{-4pt}
	\caption{Simulated and real bias frames of three cameras; A higher $R^2$ indicates a better fit. \textbf{(Best viewed with zoom)}}
	\vspace{-4pt}	
	\label{fig:noise_comparision}
\end{figure}

Although we use a unified noise model for different cameras,  
the noise parameters estimated from different cameras are highly diverse.  Figure~\ref{fig:TL-PPCC} shows the selected optimal shape parameter $\lambda$ differs camera by camera, implying distributions with varying degree of heavy tails across cameras.  The visual comparisons of real and simulated bias frames are presented in Figure~\ref{fig:noise_comparision}. Our model is capable of synthesizing realistic noise across various cameras, which outperforms the Gaussian noise model both in terms of the goodness-of-fit measure (\ie $R^2$) and the visual similarity to real noise.

\begin{figure}[!t]
	\centering
	\setlength\tabcolsep{1pt}
	\renewcommand\arraystretch{1}	
	\begin{tabular}{ccc}
		\includegraphics[width=.325\linewidth,clip,keepaspectratio]{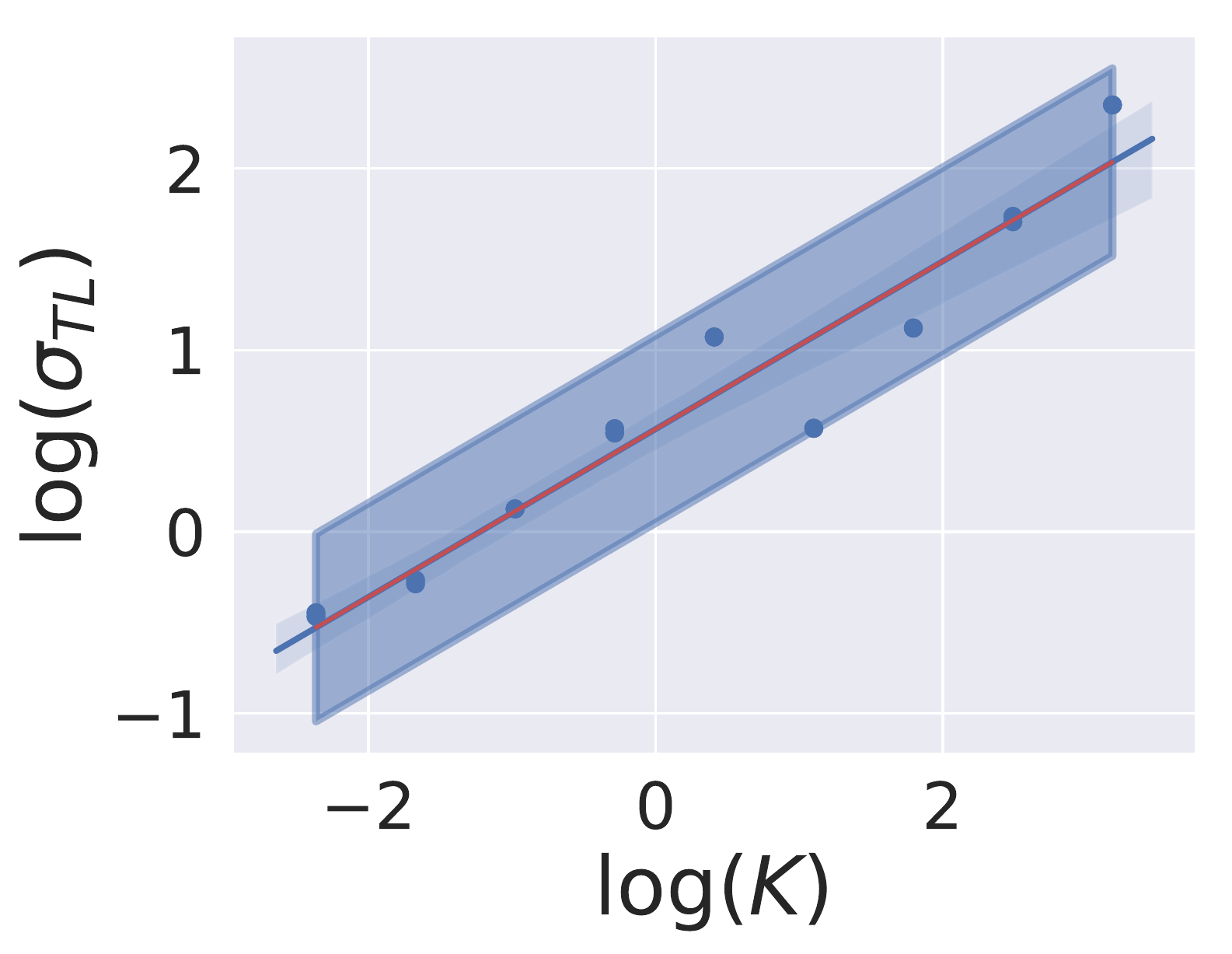} &
		\includegraphics[width=.325\linewidth,clip,keepaspectratio]{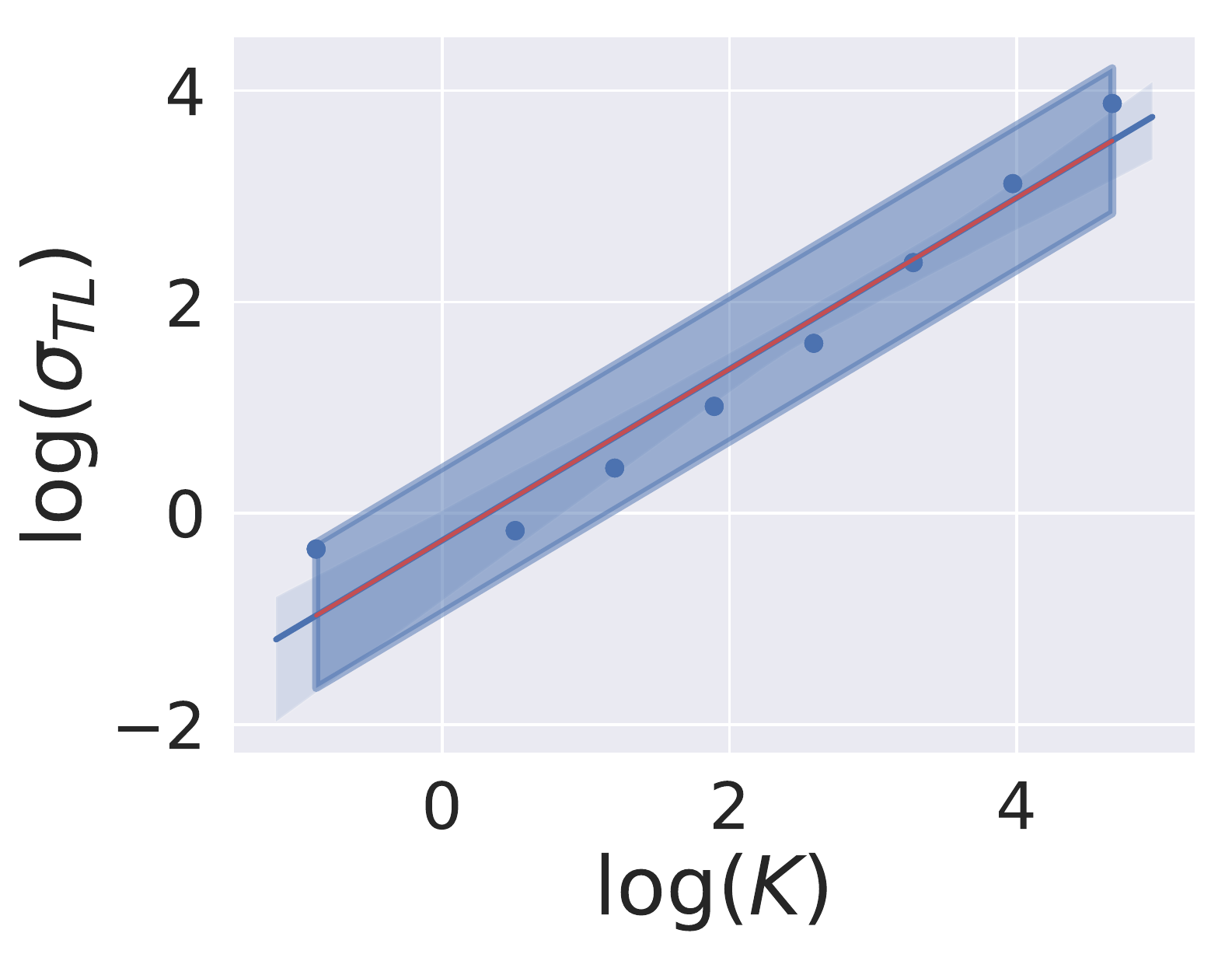} &
		\includegraphics[width=.325\linewidth,clip,keepaspectratio]{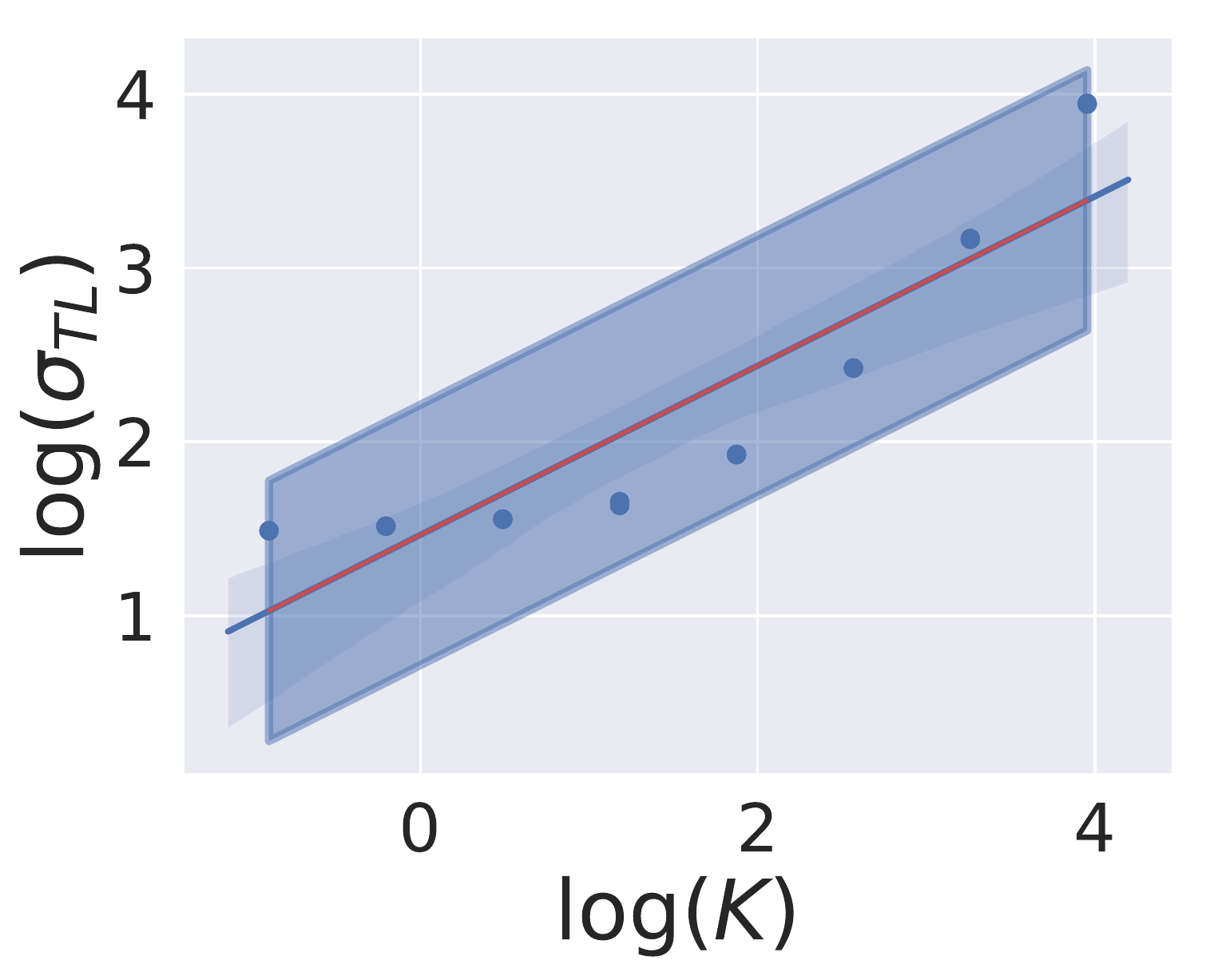} \\
		\includegraphics[width=.325\linewidth,clip,keepaspectratio]{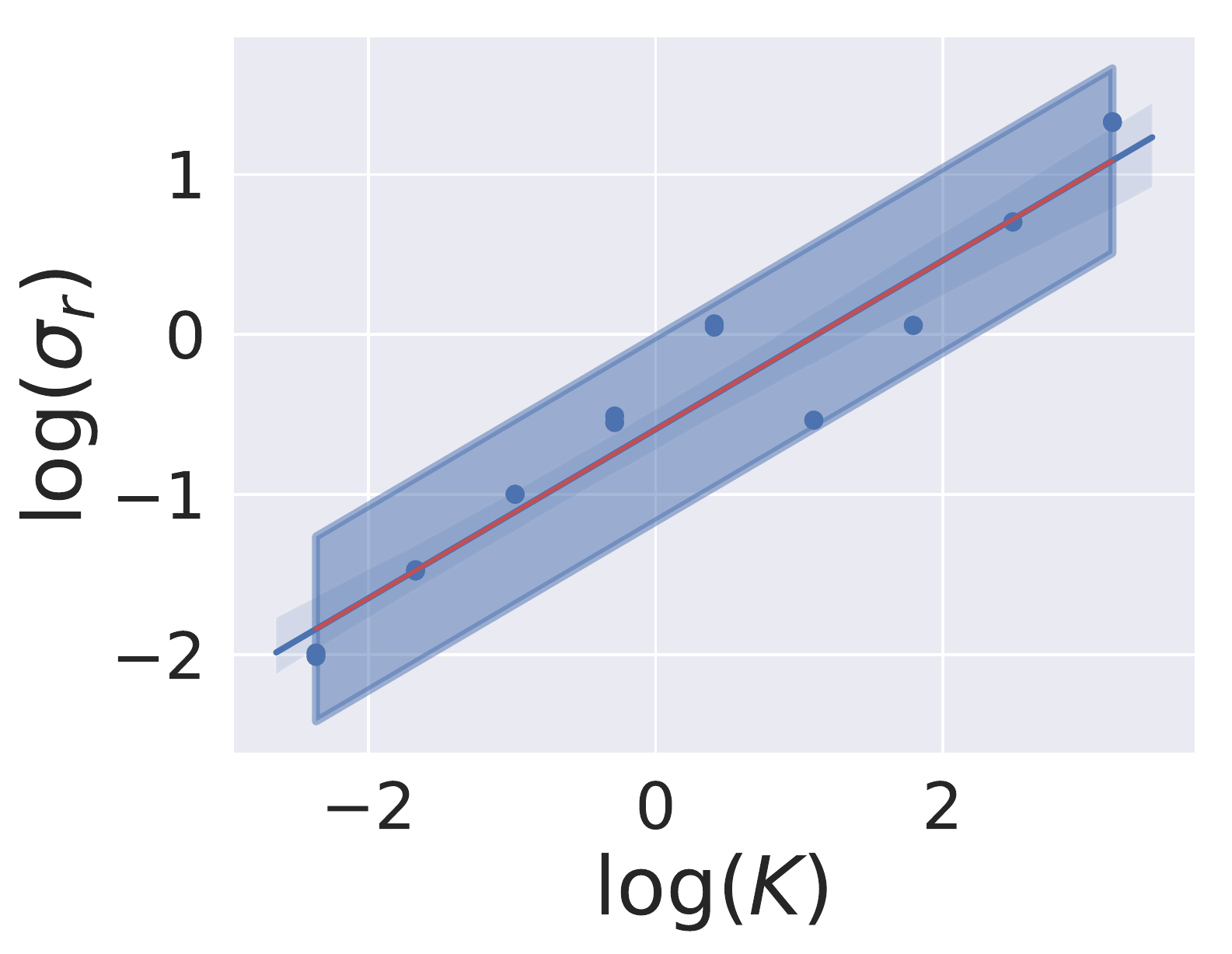} &
		\includegraphics[width=.325\linewidth,clip,keepaspectratio]{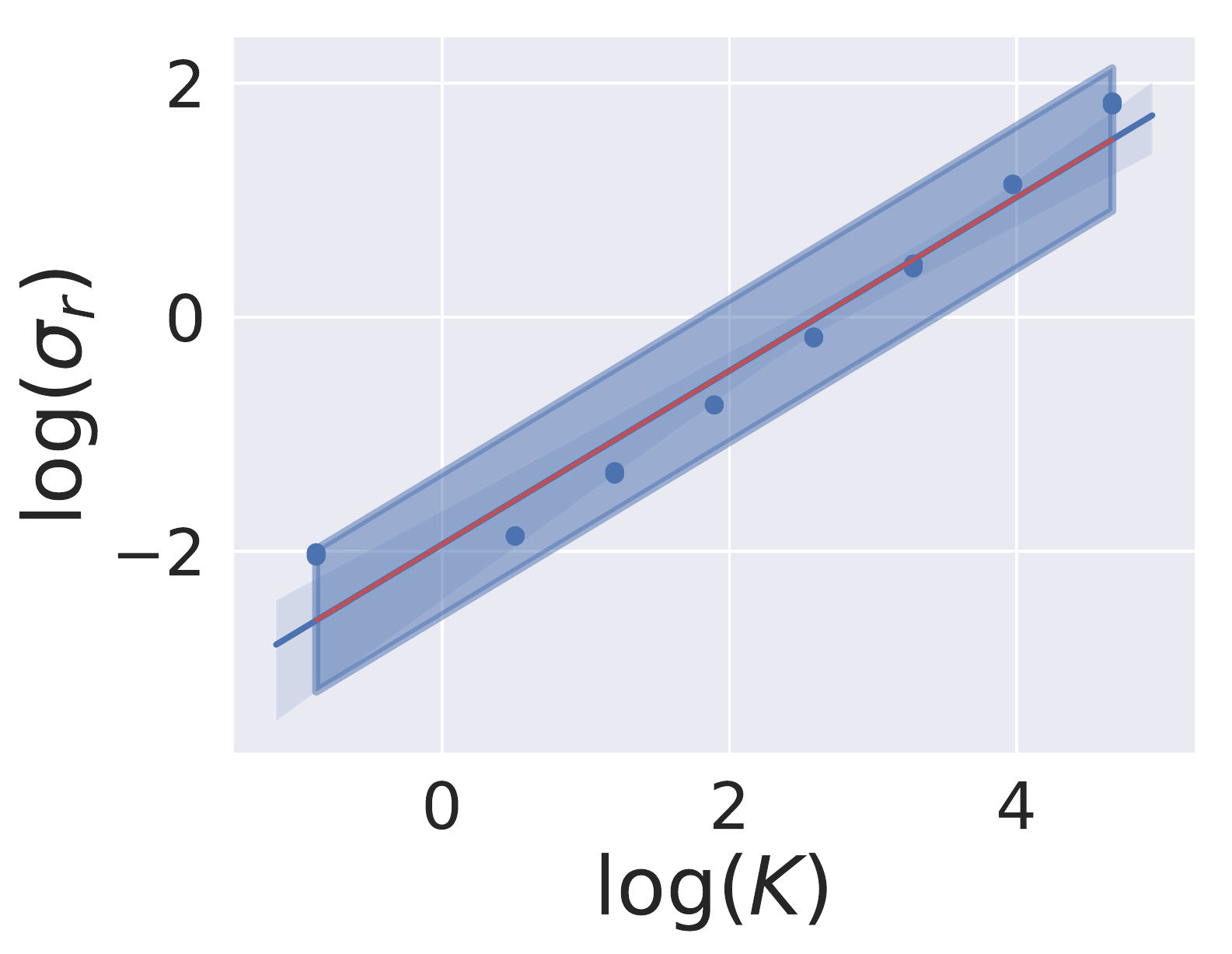} &
		\includegraphics[width=.325\linewidth,clip,keepaspectratio]{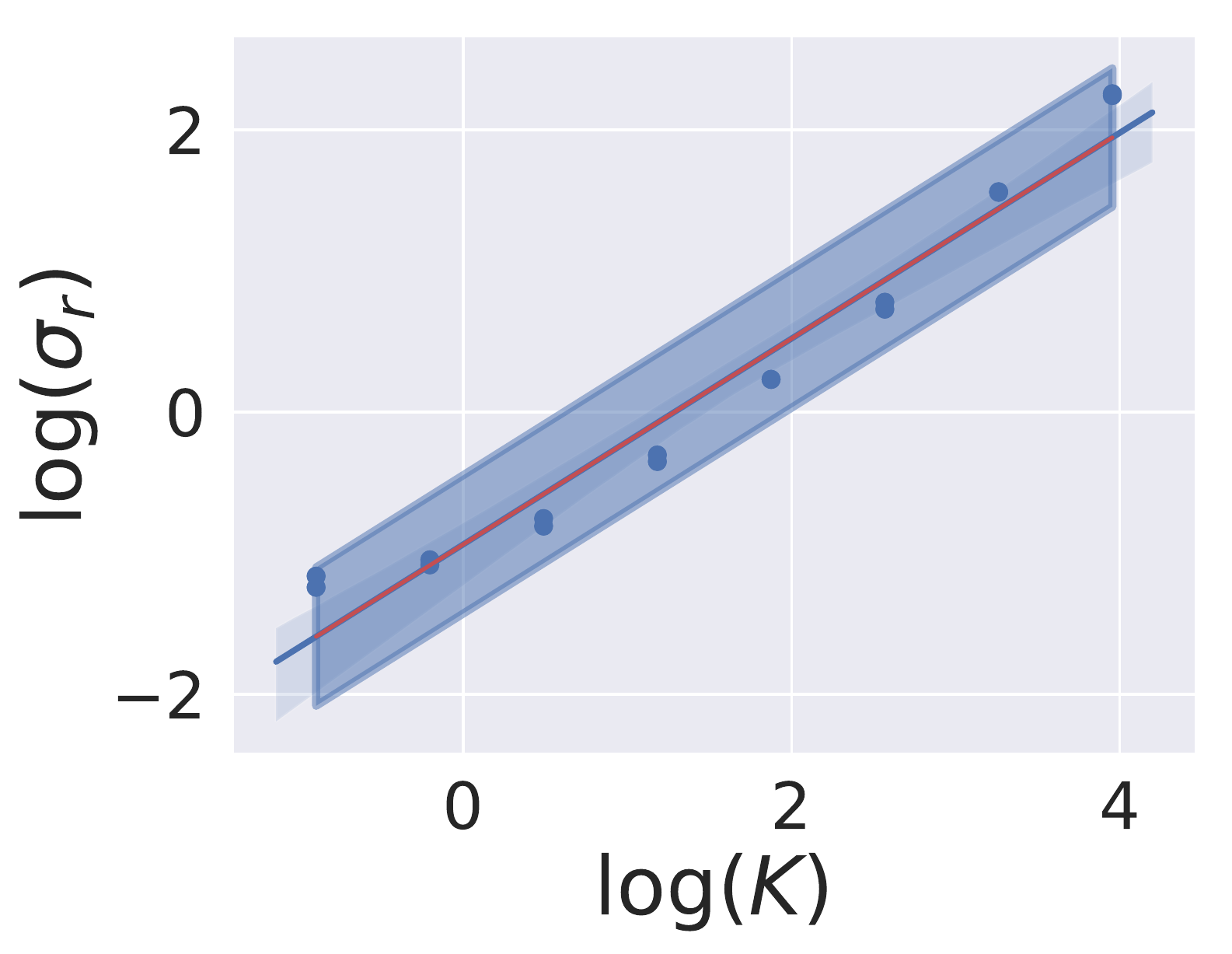} \\
		\footnotesize (a) SonyA7S2 & \footnotesize (b) NikonD850 & \footnotesize (c) CanonEOS70D \\
	\end{tabular}
	\vspace{-4pt}
	\caption{Linear least-square fitting from estimated noise parameter samples (blue dots) from three cameras. Upper and lower figures show the joint distributions of $(K, \sigma_{TL})$ and $(K, \sigma_r)$ respectively, where we sample the noise parameters from the blue shadow regions. }
	\vspace{-4pt}
	\label{fig:joint-dist}
\end{figure}


\vspace{-10pt}
\subsubsection{Modeling Joint Parameter Distributions}
To choose noise parameters for our noise formation model at various ISO settings, we need to model joint parameter distributions such that the noise parameters can be sampled in a coupled way. 
Note the system overall gain $K$ is closely related to the ISO setting, so it's rational to model the joint distributions of $K$ and other noise parameters.  
As shown in Figure~\ref{fig:joint-dist}, we model the joint distributions of ($K$, $\sigma_{TL}$) and ($K$, $\sigma_{r}$) using the linear least squares method\footnote{Other non-linear models (\eg exponential) could yield better fits but not clearly improve the denoising performance. For simplicity, we use the linear model for joint distribution.} to find the line of best fit for two sets of log-scaled measurements. Our noise parameter sampling procedure is
\begin{align}
&\log \left( K \right)  \sim U \left( \log (\hat{K}_{min}), \log (\hat{K}_{max}) \right), \nonumber  \\   \label{eq: sampling}
&\log \left( \sigma_{TL} \right) | \log \left( K \right) \sim \mathcal{N} \left(a_{TL} \log (K)  + b_{TL} ,  \hat{\sigma}_{TL} \right), \\  
&\log \left( \sigma_{r} \right) | \log \left( K \right) \sim \mathcal{N} \left(a_{r} \log (K)  + b_{r} ,  \hat{\sigma}_{r} \right),  \nonumber
\end{align}
where $U(\cdot,\cdot)$ denotes a uniform distribution and $\mathcal{N} (\mu, \sigma)$ denotes a Gaussian distribution with mean
$\mu$ and standard deviation $\sigma$. $\hat{K}_{min}$ and $\hat{K}_{max}$ are
the estimated overall system gains at the minimum and maximum ISO of a camera
respectively. $a$ and $b$ indicate the fitted line's slope and intercept respectively.
$\hat{\sigma}$ is an unbiased
estimator of standard deviation of the linear regression under the Gaussian error
assumption. For shape parameter $\lambda$ and color bias $\mu_c$, 
we simply sample them from the
empirical distribution\footnote{We do not find any analytical distributions that can characterize the distribution of color bias (see Figure~\ref{fig:color-bias}).} of the estimated parameter samples as we do not observe any clear statistical relationship to $K$ (as well as ISO).  

\vspace{-6pt}
\subsubsection{Noisy Image Synthesis}
To synthesize noisy images, clean images are chosen and divided by low light factors sampled uniformly from $[100, 300]$ to simulate low photon count in the dark. Noise is then generated and added to
the scaled clean samples, according to Equation~\eqref{eq: noise-formation}
\eqref{eq: sampling}. The created noisy images are finally normalized by multiplying the same low light factors to expose bright but excessively noisy contents.

\begin{figure}[!t]
	\centering
	\setlength\tabcolsep{0.8pt}
	\begin{tabular}{cc}
		\includegraphics[width=.485\linewidth,clip,keepaspectratio]{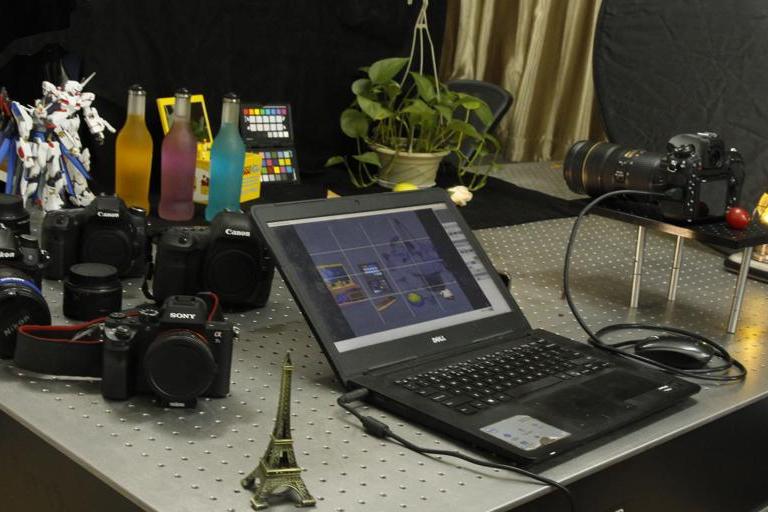} &
		\includegraphics[width=.485\linewidth,clip,keepaspectratio]{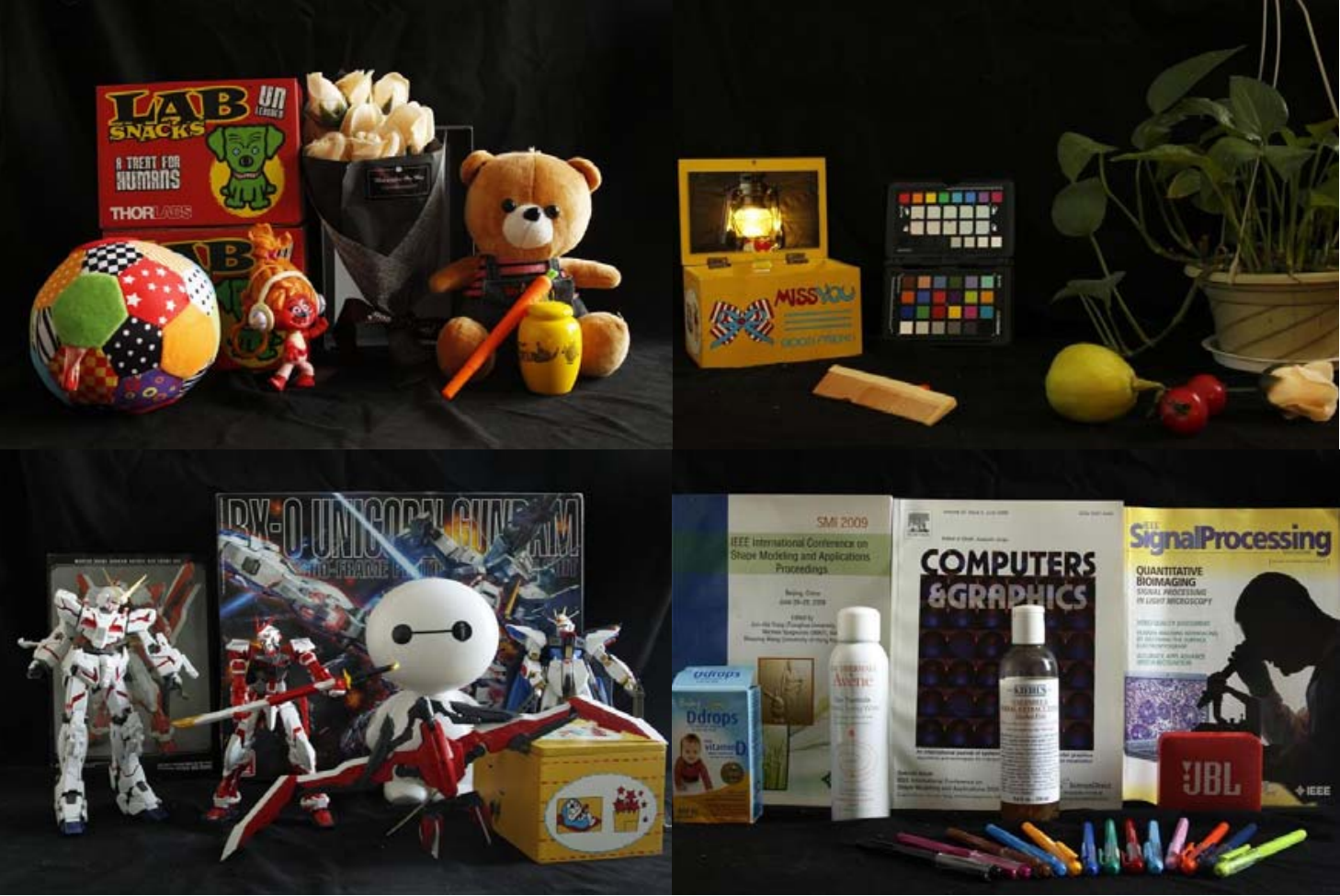} \\
		\footnotesize (a) Image capture setup & \footnotesize (b) Sample images  \\
	\end{tabular}
	\vspace{-4pt}
	\caption{Image capture setup and some sample images from our ELD dataset. }
	\vspace{-4pt}
	\label{fig:dataset}
\end{figure}

\subsection{Extreme Low-light Denoising Dataset (ELD)} \label{sec:eld-dataset}
To systematically study the generality of the proposed noise formation model, 
we collect an extreme low-light dataset that covers 10 indoor
scenes and 4 camera devices from multiple brands (\ie SonyA7S2, NikonD850, CanonEOS70D, CanonEOS700D) for benchmarking.   
We also record bias and flat-field frames for each camera to calibrate our noise model. 
The data capture setup is shown in Figure~\ref{fig:dataset}. 
The camera is mounted on a sturdy optical table and controlled by a remote software to avoid misalignments caused by camera motion, and the scene is illuminated by natural or direct current light sources to avoid flickering effect of alternating current lights \cite{Abdelhamed_2018_CVPR,Sheinin_2017_CVPR}.
For each scene of a given camera, a reference image at the base ISO was firstly taken, followed by noisy images whose exposure time was deliberately decreased by low light factors $f$ to simulate extreme low light conditions.  Another reference image then was taken akin to the first one, to ensure no accidental errors (\eg drastic illumination change or accidental camera/scene motion) occurred.
The detailed procedures are presented in Algorithm~\ref{alg:protocol}.  
We  choose three ISO levels (800,
1600, and 3200)\footnote{Most modern digital cameras are ISO-invariant when ISO is set higher than 3200 \cite{ISOless_Clark}} and two low light factors (100, 200) for noisy images to capture our dataset, resulting in 240 (3$\times$2$\times$10$\times$4) raw image pairs in total. The hardest example in our dataset resembles the image captured at a ``pseudo" ISO up to 640000 (3200$\times$200).

\begin{algorithm}[!t]
	\caption{Image capture protocol}
	\label{alg:protocol}
	\small
	\begin{algorithmic}[1] 
		\REQUIRE 
		$g_b = 100$; $g_h  t^* = g_b t$; 
		\FOR{Each Camera}
		\FOR{Each Scene}
		\STATE Meter the scene to find a exposure time $t$ that well exposes the image at the base ISO $g_b$;
		\FOR{Each ISO $g_h$ for noisy images}
		\STATE Take a reference image at exposure setting ($g_b$, $t$);  
		\FOR{Each low light factor $f$}
		\STATE Capture a noisy image at ($g_h$,  $t^* / f $);
		\ENDFOR
		\STATE Take another reference image at ($g_b$, $t$);
		\ENDFOR
		\ENDFOR
		\ENDFOR
	\end{algorithmic}
\end{algorithm}

\section{Experiments} \label{sec:experiments}

In this section,
we first present the experimental setting including both implementation details and competing methods. 
Then we conduct comprehensive ablation studies for an in-depth analysis of the noise models and compare our method against prior art. 
Both quantitative and visual results on various datasets including the new ELD dataset are presented. Finally, we test our approach on low-light videography and several downstream vision tasks (depth estimation, optical flow estimation, object detection/recognition) in the dark.

\subsection{Experimental Setting} \label{sec:experimental-setting}
\subsubsection{Implementation Details} \label{sec:implementation-details}
\begin{figure}[!t]
	\centering
	\includegraphics[width=.99\linewidth,clip,keepaspectratio]{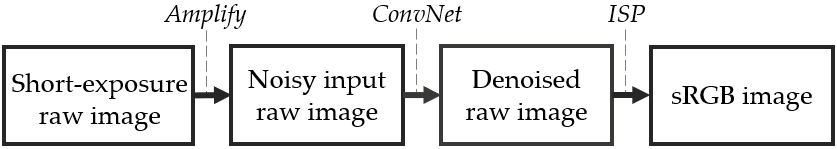} 
	\caption{Our raw image denoising pipeline.}
	\label{fig:denosing_pipeline}
\end{figure}

To substantiate the effectiveness of our noise formation model, a deep neural network scheme is constructed to perform low-light raw denoising. We utilize the same U-Net architecture  \cite{ronneberger2015u} as \cite{Chen_2018_CVPR}.
Raw Bayer images from SID Sony training
set are used to create training data.
We pack the raw Bayer images into four channels, and crop non-overlapped $512\times 512$ regions augmented by random flipping and rotation.
Our approach only uses clean raw images, as the corresponding noisy images are generated by our proposed noise model on-the-fly.
Besides, we also train networks based upon other training schemes as references, including training with paired real data (short exposure and long exposure counterpart) and training with paired real noisy images (\ie Noise2Noise~\cite{pmlr-v80-lehtinen18a}). The whole pipeline of our raw image denoising is depicted in Figure~\ref{fig:denosing_pipeline} (note the RGB2RGB and raw2RGB experiments in Section~\ref{sec:color} will use slightly different pipelines).

Our implementation is based on PyTorch. 
We train the models with 200 epoch using a simple $L_1$ loss and the Adam optimizer~\cite{kingma2014adam} with batch size 1. The base learning rate is set to $10^{-4}$ and halved at epoch 100, then reduced to $10^{-5}$ at epoch 180.

\subsubsection{Competing Methods} \label{sec:competing-method}
To understand how accurate our proposed noise model would be, we compare our method with
\begin{packed_enum}
	
	\item the representative non-deep methods, \ie BM3D \cite{dabov2007BM3D}
	and Anscombe-BM3D (A-BM3D) \cite{makitalo2011optimal}; 

	\item the approaches that use real noisy data for training, \ie Noise2Noise \cite{pmlr-v80-lehtinen18a} and ``paired real data" \cite{Chen_2018_CVPR}\footnote{The work of \cite{Chen_2018_CVPR} used paired real data to train raw-to-sRGB image processing (\ie noisy raw images with clean sRGB counterparts as training labels). Here we adapt its setting to raw-to-raw denoising.};
	
	\item the state-of-the-art noise models, \ie  Noiseflow~\cite{Abdelhamed_2019_ICCV}
	and heteroscedastic Gaussian noise models (G+P)~\cite{Foi2008Practical}.

\end{packed_enum}

\begin{figure*}[t]
	\centering
	\small
	\setlength\tabcolsep{1pt}
	\begin{tabular}{cccccc}
		\includegraphics[width=0.155\linewidth]{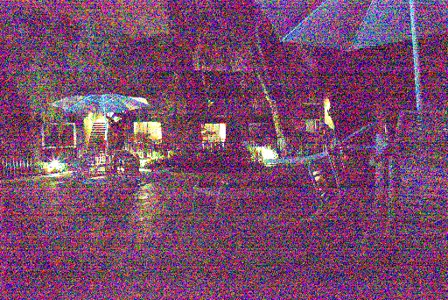}        
		& \includegraphics[width=0.155\linewidth]{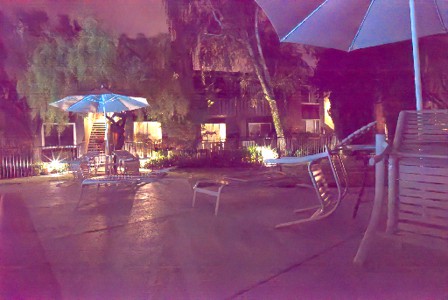}        
		& \includegraphics[width=0.155\linewidth]{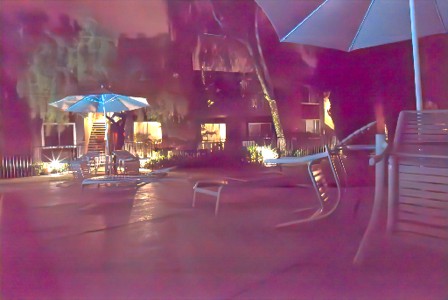} 		
		& \includegraphics[width=0.155\linewidth]{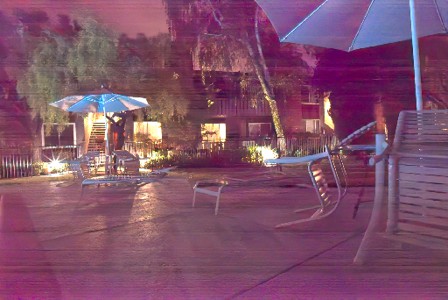} 
		& \includegraphics[width=0.155\linewidth]{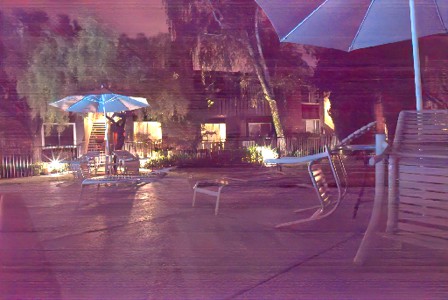} 
		& \includegraphics[width=0.155\linewidth]{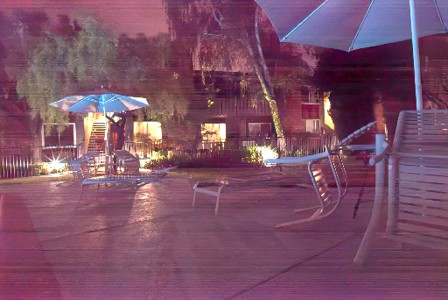} \\
		  (a) Input  &  (b) Noise2Noise &  (c) Noiseflow &  (d) $G$  & (e) $G$+$P$ & (f) $G$+$P^*$  \\
		 (16.95) &  (27.67) &  (27.15)  &  (27.99) &  (28.01)   &  (28.30)  \\		
		 \includegraphics[width=0.155\linewidth]{./figures/images/sid/20107/sony-GP-inc4.jpg}        
		& \includegraphics[width=0.155\linewidth]{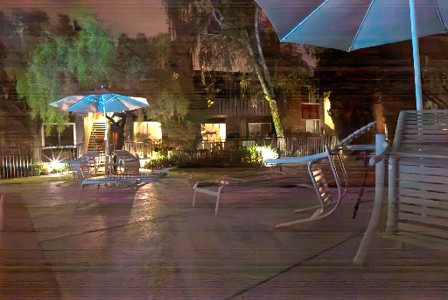} 
		& \includegraphics[width=0.155\linewidth]{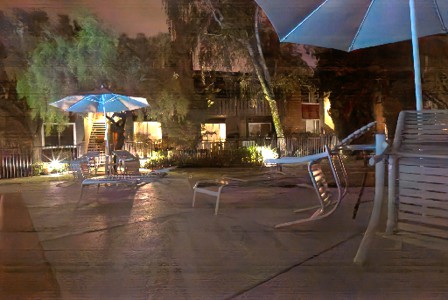} 		
		& \includegraphics[width=0.155\linewidth]{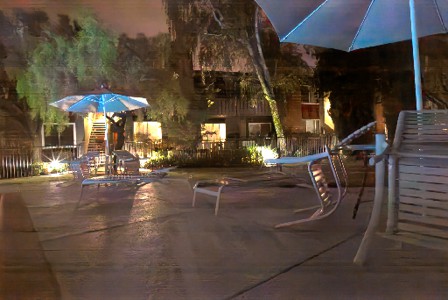} 
		& \includegraphics[width=0.155\linewidth]{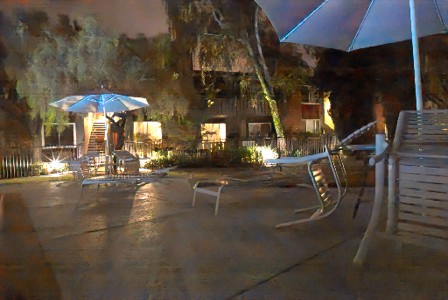}
		& \includegraphics[width=0.155\linewidth]{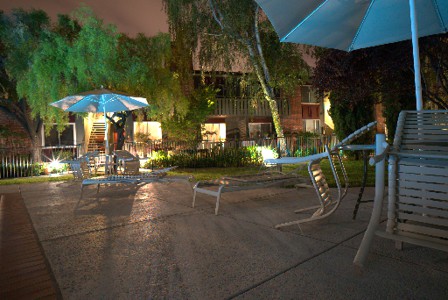} \\
		  (g) $G^*$+$P^*$  & (h) $G^*$+$P^*$+$B$  & (i) $G^*$+$P^*$+$B$+$R$  &  (j) $G^*$+$P^*$+$B$+$R$+$U$ & (k) Paired real data & (l) Reference \\
		(28.41) &  (32.03) &  (32.20)  &  (32.25) &  (31.81)   &  (PSNR)  \\
	\end{tabular}
	\vspace{-3pt}
	\caption{Visual result comparison of different training schemes. Our full noise model ($G^*$+$P^*$+$B$+$R$+$U$) suppresses the color bias, residual bandings and chroma artifacts compared to other baselines. (\textbf{Best viewed with zoom)}} 
	\label{fig:ablation-visual}
	\vspace{-4pt}
\end{figure*}


\begin{figure*}[!t]
	\centering
	\setlength\tabcolsep{1pt}	
	\begin{tabular}{cccccccc}	
		Input & BM3D & A-BM3D & Noise2Noise  & Noiseflow  & G+P  & \fontsize{8pt}{8pt}\selectfont Paired real data & Ours   \\
		\includegraphics[width=0.118\linewidth]{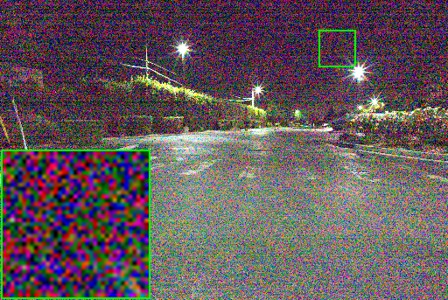}
		&	\includegraphics[width=0.118\linewidth]{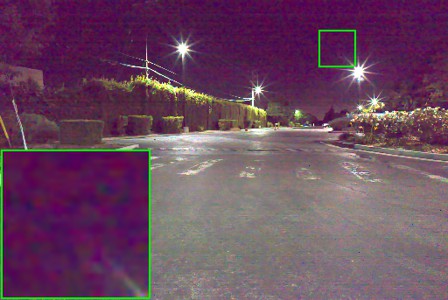}
		&	\includegraphics[width=0.118\linewidth]{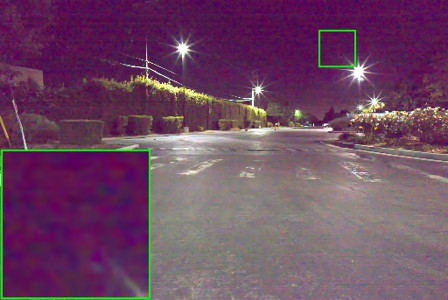}
		&	\includegraphics[width=0.118\linewidth]{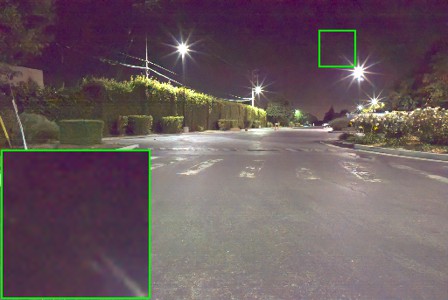}		
		&	\includegraphics[width=0.118\linewidth]{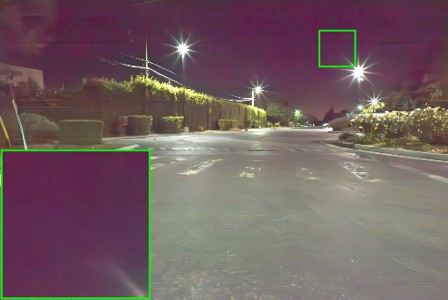}
		&	\includegraphics[width=0.118\linewidth]{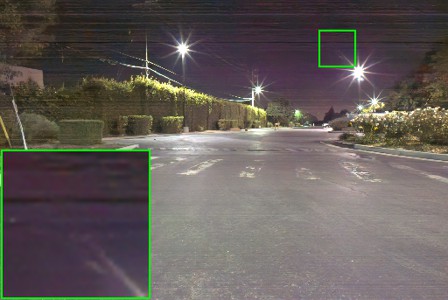}
		&	\includegraphics[width=0.118\linewidth]{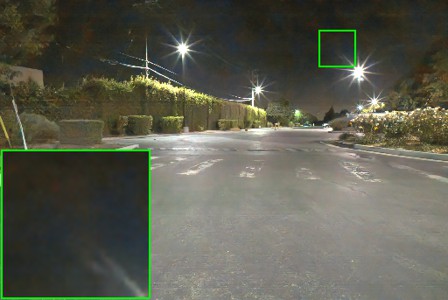}
		&	\includegraphics[width=0.118\linewidth]{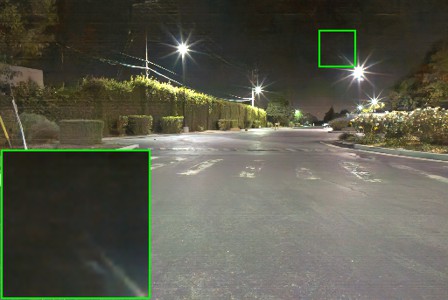} \vspace{-2pt} \\
		23.08 & 31.83 & 31.90 & 34.30  & 33.23 & 34.18 & 34.48 & 34.40		\\
		\includegraphics[width=0.118\linewidth]{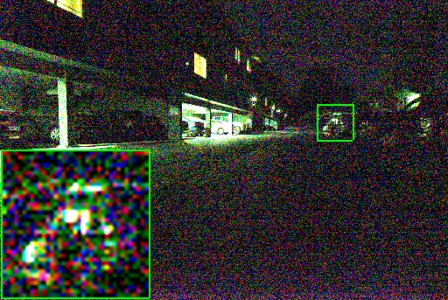}
		&	\includegraphics[width=0.118\linewidth]{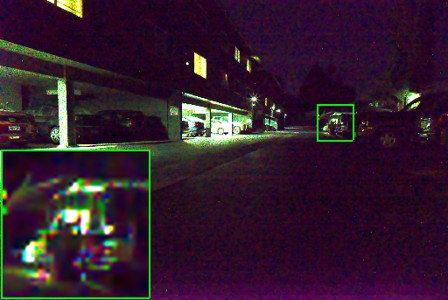}
		&	\includegraphics[width=0.118\linewidth]{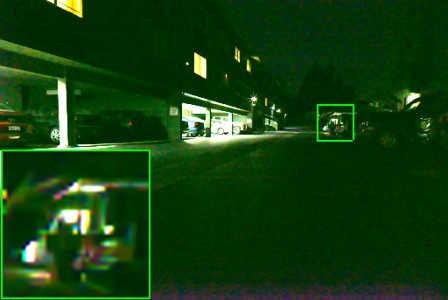}
		&	\includegraphics[width=0.118\linewidth]{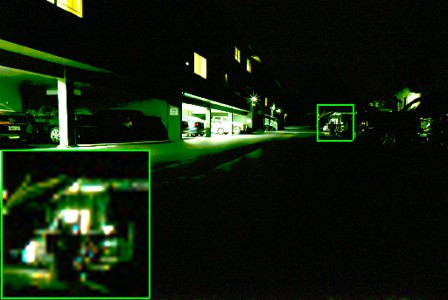}
		&	\includegraphics[width=0.118\linewidth]{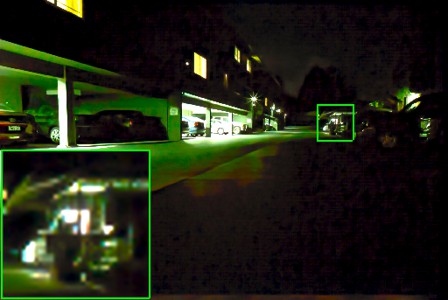}
		&	\includegraphics[width=0.118\linewidth]{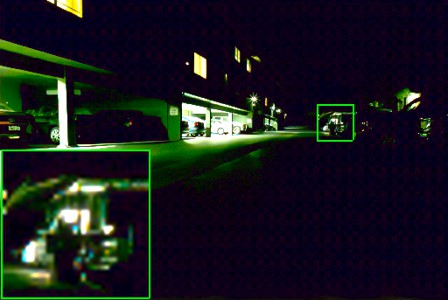}
		&	\includegraphics[width=0.118\linewidth]{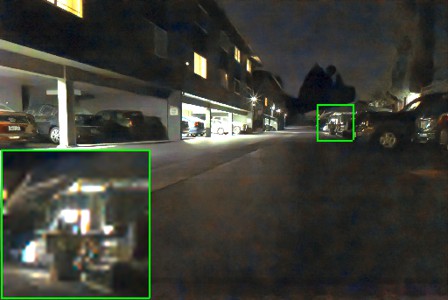}
		&	\includegraphics[width=0.118\linewidth]{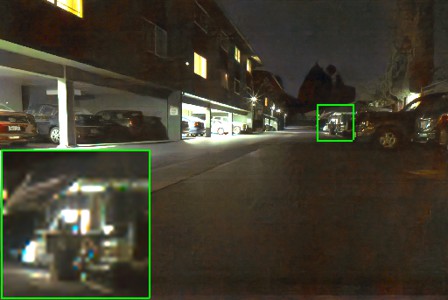} \vspace{-2pt} \\	
		26.82 & 30.66 & 35.58 & 33.57 & 34.54 & 33.67  &  37.98 & 38.67 		\\
		\includegraphics[width=0.118\linewidth]{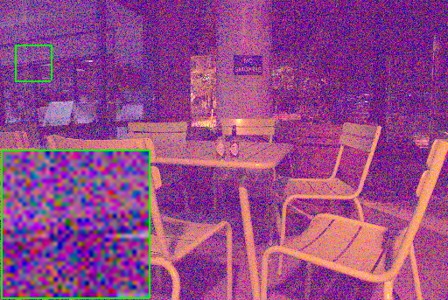}
		&	\includegraphics[width=0.118\linewidth]{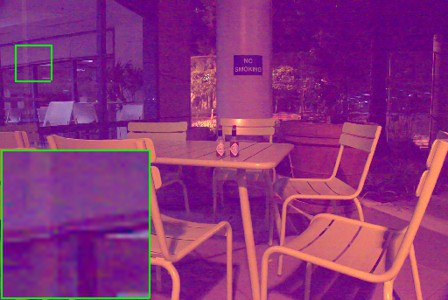}
		&	\includegraphics[width=0.118\linewidth]{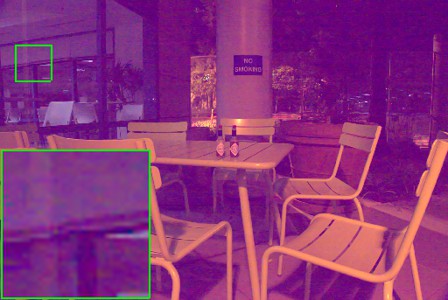}
		&	\includegraphics[width=0.118\linewidth]{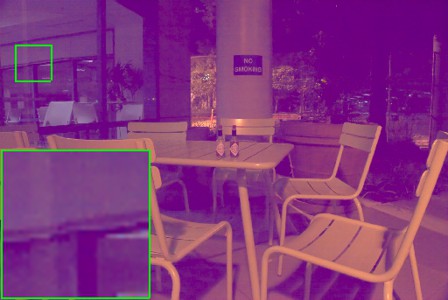}		
		&	\includegraphics[width=0.118\linewidth]{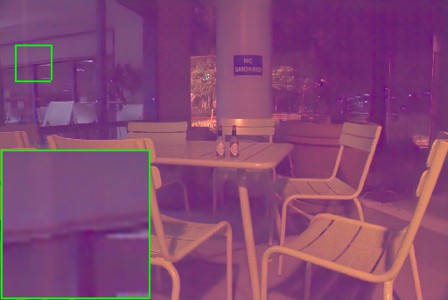}
		&	\includegraphics[width=0.118\linewidth]{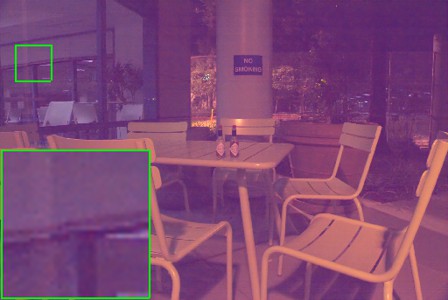}
		&	\includegraphics[width=0.118\linewidth]{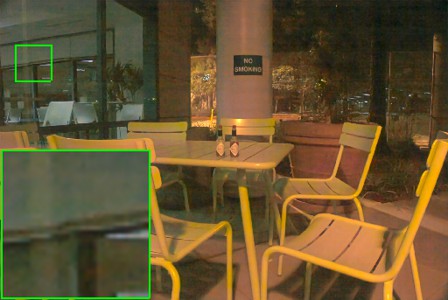}
		&	\includegraphics[width=0.118\linewidth]{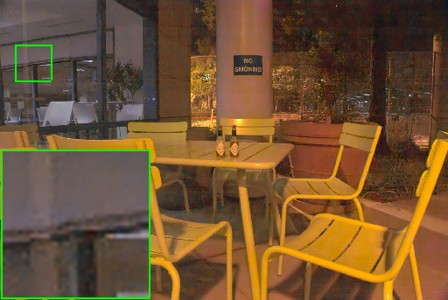} \vspace{-2pt}  \\			 	
		29.25 & 35.82 & 35.81 &  35.54  & 35.26 & 36.10 &  45.97 &  47.14	\\
		\includegraphics[width=0.118\linewidth]{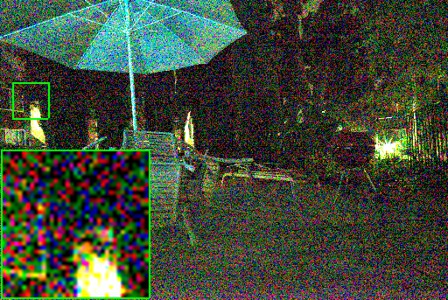}
		&	\includegraphics[width=0.118\linewidth]{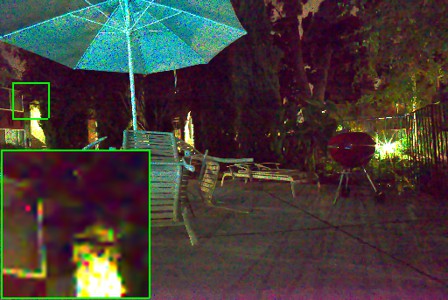}
		&	\includegraphics[width=0.118\linewidth]{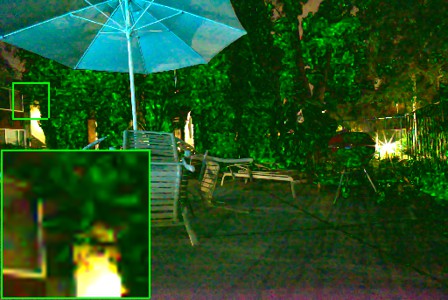}
		&	\includegraphics[width=0.118\linewidth]{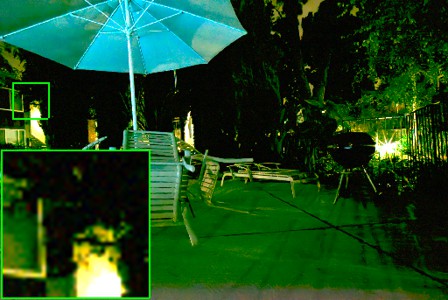}		
		&	\includegraphics[width=0.118\linewidth]{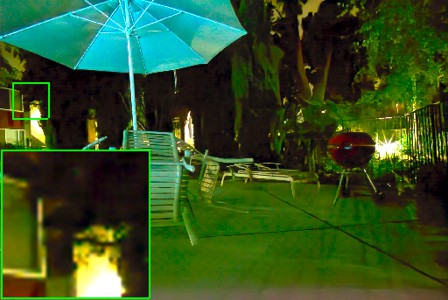}
		&	\includegraphics[width=0.118\linewidth]{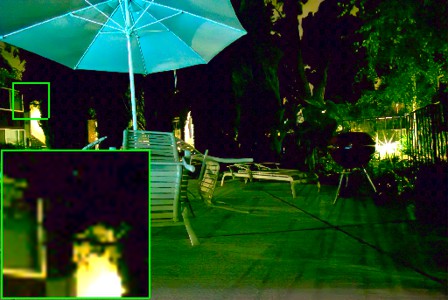}
		&	\includegraphics[width=0.118\linewidth]{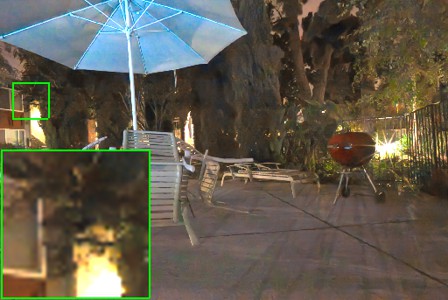}
		&	\includegraphics[width=0.118\linewidth]{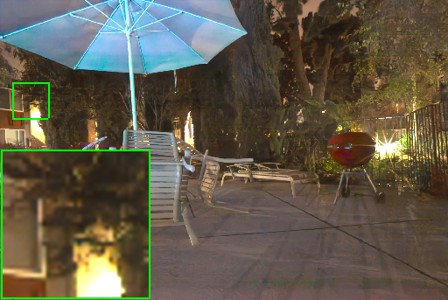} \vspace{-2pt} \\			 
		23.87 & 29.39 & 28.79 &  30.12  & 31.60 & 29.88 &  35.31 & 35.43	\\		 	
	\end{tabular} 
	\vspace{-3pt}
	\caption{Raw image denoising results on both indoor and outdoor scenes from SID Sony dataset. The PSNR value of each resulting image is reported below; similarly hereinafter. (\textbf{Best viewed on screen with zoom})}
	\label{fig:sony-vis}
	\vspace{-3pt}
\end{figure*}

For non-deep methods, the noise level parameters required are provided by the off-the-shelf image noise level estimators \cite{Foi2008Practical,Chen_2015_ICCV}. 
We note though there are other non-deep methods \cite{Gu2014Weighted,Xu_2017_ICCV,xu2018trilateral} that might achieve better results than BM3D, their computation cost is generally unaffordable for large-size images.
As for Noiseflow, it requires paired real data to obtain noise data (by subtracting the ground truth images from the noisy ones) for training. We train the Noiseflow model on the SID Sony training dataset, then use the trained model to synthesize noisy dataset for training denoising networks.

\subsection{Results on SID Sony Dataset} \label{sec:noise-ablation}
Single image raw denoising experiment is firstly conducted on SID Sony dataset. 
The numerical results are reported on whole SID Sony validation and test sets, instead of only indoor scenes in the preliminary version \cite{wei2020physics}.
To account for the imprecisions of shutter speed and analog gain \cite{Abdelhamed_2018_CVPR}, a single scalar
is calculated and multiplied into the reconstructed image to
minimize the mean square error evaluated by the ground
truth.

\vspace{-6pt}
\subsubsection{Ablation Study on Noise Models}
To verify the efficacy of the proposed noise model, we compare the performance of networks trained with different noise models developed in Section \ref{sec: raw formation}.
All noise parameters are calibrated using our ELD dataset, and sampled using Equation~\eqref{eq: sampling}.
The results of the other methods described in Section~\ref{sec:competing-method} are also presented as references. 

\begin{table}[!t]
	\centering
	\caption{Quantitative Results on Sony set of the SID dataset. 
		$G$: the Gaussian model for read noise $N_{read}$; $G^*$: the tukey lambda model for $N_{read}$; $P$: the Gaussian approximation for photon shot noise $N_p$; 
		$P^*$: the true Poisson model for $N_p$; $B$: the color-wise DC model for color bias; 
		$R$: the Gaussian model for row noise $N_r$; $U$: the uniform distribution model for quantization noise $N_q$.
		The best and second-best results are indicated by \textcolor{red}{red} and \textcolor{blue}{blue}, respectively. }
	\vspace{-2pt}
	\footnotesize
	\begin{tabular}{lccc} 
		\toprule
		& $\times 100$ & $\times 250$ & $\times 300$ \\
		Model & \!PSNR / SSIM\! & \!PSNR / SSIM\! & \!PSNR / SSIM\! \\ 
		\midrule
		BM3D   & 38.49 / 0.875 & 34.48 / 0.821 & 30.98 / 0.766 \\ \hline	
		A-BM3D  & 39.24 / 0.869 & 33.74 / 0.742 & 30.29 / 0.729 \\ \hline	
		\midrule
		Noise2Noise  & 40.47 / 0.890 & 35.74 / 0.757 & 32.34 / 0.707 \\ \hline
		Noiseflow & 41.11 / 0.926 & 36.97 / 0.838 & 32.38 / 0.759 \\ \hline	
		Paired real data  & \textcolor{red}{42.76} / \textcolor{blue}{0.948} & \textcolor{blue}{40.59} / \textcolor{red}{0.935} & \textcolor{red}{36.48} / \textcolor{red}{0.919} \\ \hline		
		\midrule
		$G$  & 39.47 / 0.866 & 34.83 / 0.726 & 31.80 / 0.685 \\ \hline
		$G$+$P$  & 40.44 / 0.894 & 35.67 / 0.773 & 32.26 / 0.716 \\ \hline
		$G$+$P^*$  & 41.26 / 0.925 & 37.27 / 0.849 & 33.68 / 0.800 \\ \hline
		$G^*$+$P^*$  & 41.64 / 0.936 & 38.20 / 0.891 & 34.56 / 0.844 \\ \hline
		$G^*$+$P^*$+$B$  & 42.55 / 0.948 & 40.44 / 0.930 & 36.29 / 0.912 \\ \hline
		$G^*$+$P^*$+$B$+$R$  & 42.62 / 0.948 & 40.55 / 0.930 & 36.33 / 0.913 \\ \hline
		$G^*$+$P^*$+$B$+$R$+$U$  & \textcolor{blue}{42.75} / \textcolor{red}{0.949} & \textcolor{red}{40.60} / \textcolor{blue}{0.932} & \textcolor{blue}{36.46} / \textcolor{blue}{0.916} \\ \hline
		\bottomrule
	\end{tabular}
	\label{tb:componet}
	\vspace{-11pt}
\end{table}

\begin{table*}[!t]
	\centering
	\caption{Quantitative results of different methods on our ELD dataset which contains four representative cameras. }
	\footnotesize
	\setlength{\tabcolsep}{1mm}{
		\begin{tabular}{C{.11\linewidth}|C{.05\linewidth}|C{.05\linewidth}|C{.085\linewidth}|C{.09\linewidth}|C{.115\linewidth}|C{.105\linewidth}|C{.1\linewidth}|C{.085\linewidth}|C{.085\linewidth}}
			\hline
			\multirow{2}{*}{\textsc{Camera}} & \multirow{2}{*}{$f$} & \multirow{2}{*}{\textsc{Index}}  & \multicolumn{2}{c|}{\cellcolor[HTML]{EFEFEF} \textsc{Non-deep}} & \multicolumn{2}{c|}{\cellcolor[HTML]{EFEFEF} \textsc{Training with real data}} & \multicolumn{3}{c}{\cellcolor[HTML]{EFEFEF} \textsc{Training with synthetic data}} \\ \cline{4-10}
			& & &  \!\!BM3D \cite{dabov2007BM3D}\!\! & \!\!A-BM3D~\cite{makitalo2011optimal}\!\! & \!\!Noise2Noise~\cite{pmlr-v80-lehtinen18a}\!\!  & \!\!Paired data~\cite{Chen_2018_CVPR}\!\! & \!\!Noiseflow~\cite{Abdelhamed_2019_ICCV}\!\! & \!\!G+P \cite{Foi2008Practical}\!\! & \!\!Ours\!\! \\\hline
			\multirow{4}{*}{SonyA7S2}  & \multirow{2}{*}{$\times 100$} &PSNR&$37.69$&$37.74$&$41.63$&
			\textcolor{blue}{$44.50$}&$40.10$&$42.46$&\textcolor{red}{$45.44$}\\\cline{3-10}
			& &SSIM&$0.803$&$0.776$& $0.856$ &\textcolor{blue}{$0.971$} &$0.831$&$0.889$&\textcolor{red}{$0.975$}\\\cline{2-10}
			& \multirow{2}{*}{$\times 200$} &PSNR&$34.06$&$35.26$& $37.98$ & \textcolor{blue}{$42.45$} &$37.19$&$38.88$&\textcolor{red}{$43.42$}\\\cline{3-10}
			& &SSIM&$0.696$&$0.721$& $0.775$ & \textcolor{blue}{$0.945$} &$0.757$&$0.812$&\textcolor{red}{$0.954$}\\\hline
			\multirow{4}{*}{NikonD850} & \multirow{2}{*}{$\times 100$} &PSNR&$33.97$&$36.60$& $40.47$ & \textcolor{blue}{$41.28$} &$40.62$&$40.29$&\textcolor{red}{$42.27$}\\\cline{3-10}
			& &SSIM&$0.725$&$0.779$& $0.848$ & \textcolor{red}{$0.938$} &$0.870$&$0.845$&\textcolor{blue}{$0.936$}\\\cline{2-10}
			& \multirow{2}{*}{$\times 200$} &PSNR&$31.36$&$32.59$& $37.98$ & \textcolor{blue}{$39.44$} &$37.61$&$37.26$&\textcolor{red}{$40.36$}\\\cline{3-10}
			& &SSIM&$0.618$&$0.723$& $0.820$ & \textcolor{red}{$0.910$} &$0.805$&$0.786$&\textcolor{blue}{$0.908$}\\\hline
			\multirow{4}{*}{CanonEOS70D} 
			& \multirow{2}{*}{$\times 100$} &PSNR&$30.79$&$31.88$&  $38.21$  & $40.10$ &$36.69$&\textcolor{blue}{$40.94$}&\textcolor{red}{$41.20$}\\\cline{3-10}
			& &SSIM&$0.589$&$0.692$& $0.826$ & $0.931$ &$0.787$&\textcolor{blue}{$0.934$}&\textcolor{red}{$0.949$}\\\cline{2-10}
			& \multirow{2}{*}{$\times 200$} &PSNR&$28.06$&$28.66$& $34.33$ & $37.32$ &$34.88$&\textcolor{blue}{$37.64$}&\textcolor{red}{$38.78$}\\\cline{3-10}
			& &SSIM&$0.540$&$0.597$& $0.704$ & $0.867$ &$0.772$&\textcolor{blue}{$0.873$}&\textcolor{red}{$0.908$}\\\hline
			\multirow{4}{*}{CanonEOS700D} 
			&\multirow{2}{*}{$\times 100$} &PSNR&$29.70$&$30.13$& $38.29$ & $39.05$  &$37.79$&\textcolor{blue}{$40.08$}&\textcolor{red}{$40.49$}\\\cline{3-10}
			&&SSIM&$0.556$&$0.640$& $0.859$ & \textcolor{blue}{$0.906$} &$0.861$&$0.897$&\textcolor{red}{$0.938$}\\\cline{2-10}
			&\multirow{2}{*}{$\times 200$} &PSNR&$27.52$&$27.68$& $34.94$ & $36.50$ &$35.36$&\textcolor{blue}{$37.86$}&\textcolor{red}{$38.15$}\\\cline{3-10}
			&&SSIM&$0.537$&$0.579$& $0.766$ & $0.850$ &$0.785$&\textcolor{blue}{$0.879$}&\textcolor{red}{$0.899$}\\\hline
	\end{tabular}}
	\label{tb:ELD}
\end{table*}

\begin{figure*}[!t]
	\centering
	\setlength\tabcolsep{1pt}
	\begin{tabular}{cccccccc}
		Input  & BM3D & A-BM3D &  Noise2Noise  & Noiseflow & G+P &  \fontsize{8pt}{8pt}\selectfont Paired real data & Ours  \\
		\includegraphics[width=0.118\linewidth]{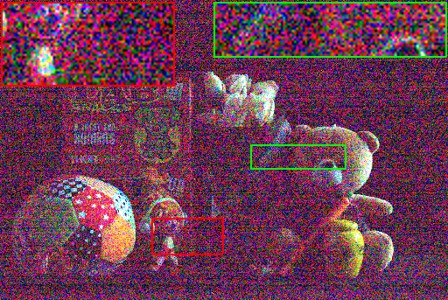}
		&	 \includegraphics[width=0.118\linewidth]{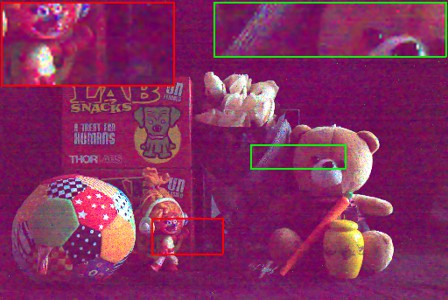}        	 
		&  \includegraphics[width=0.118\linewidth]{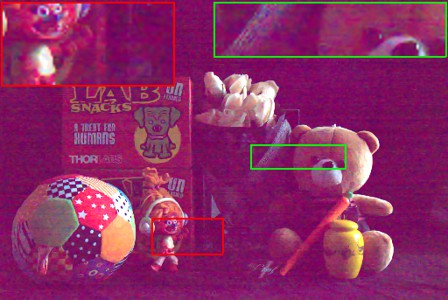}
		&  \includegraphics[width=0.118\linewidth]{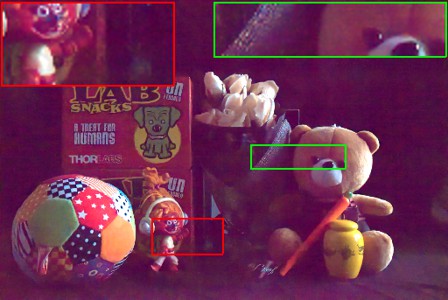}					
		&	 \includegraphics[width=0.118\linewidth]{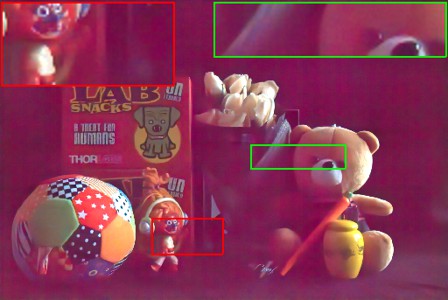}
		&  \includegraphics[width=0.118\linewidth]{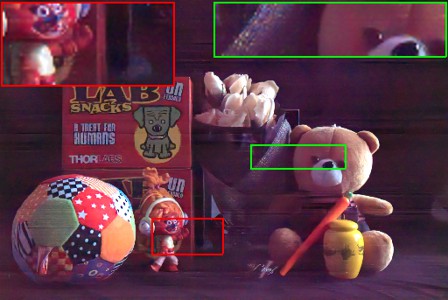}
		&  \includegraphics[width=0.118\linewidth]{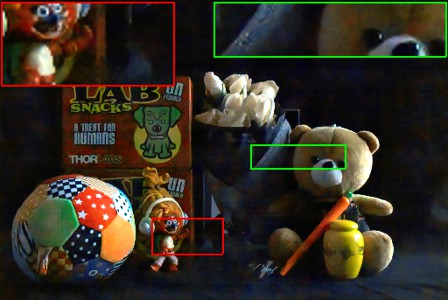}	
		&  \includegraphics[width=0.118\linewidth]{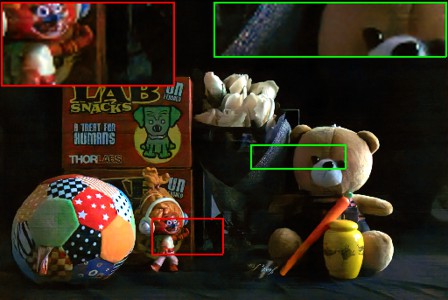} \vspace{-2pt} \\	
		27.95 & 35.45 & 35.46 &  38.81  & 36.12 & 40.25 &  44.53 & 45.58  \\
		\includegraphics[width=0.118\linewidth]{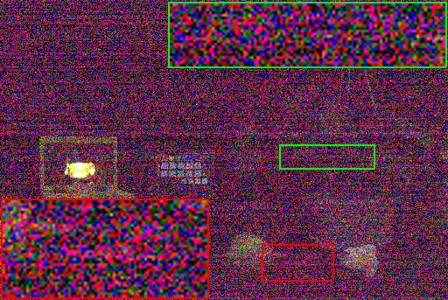}
		&	 \includegraphics[width=0.118\linewidth]{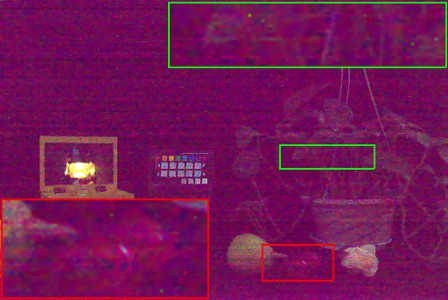}        	 
		&  \includegraphics[width=0.118\linewidth]{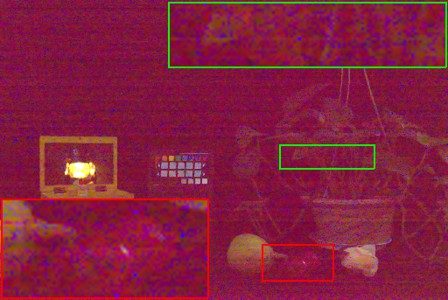}
		&  \includegraphics[width=0.118\linewidth]{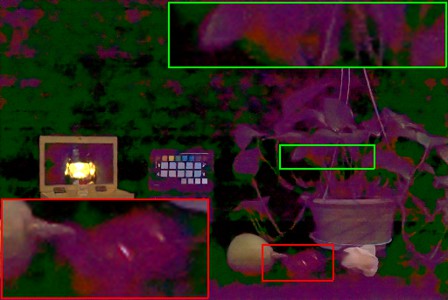}		
		&	 \includegraphics[width=0.118\linewidth]{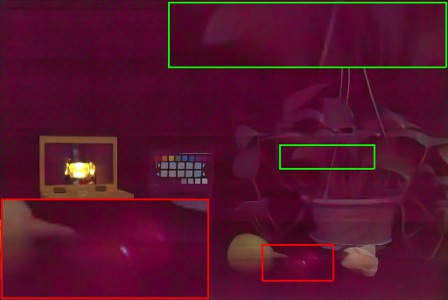}
		&  \includegraphics[width=0.118\linewidth]{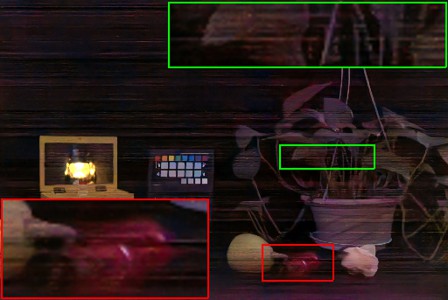}
		&  \includegraphics[width=0.118\linewidth]{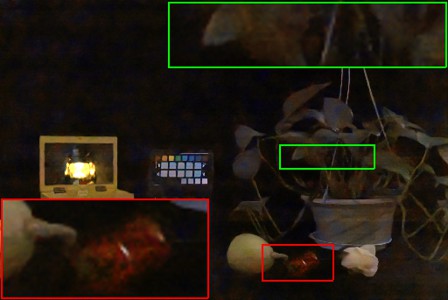}		
		&  \includegraphics[width=0.118\linewidth]{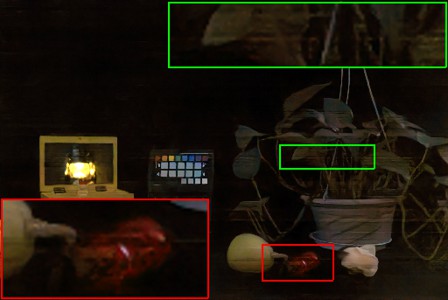} \vspace{-2pt} \\	
		28.48 & 35.73 & 35.02 & 41.71  & 37.40 & 44.49 &  46.57 & 49.51	\\
		\includegraphics[width=0.118\linewidth]{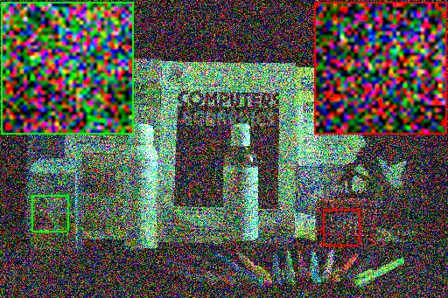}
		&	 \includegraphics[width=0.118\linewidth]{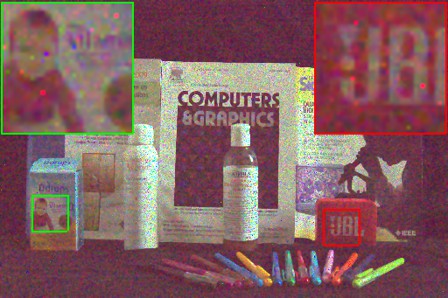}        	 
		&  \includegraphics[width=0.118\linewidth]{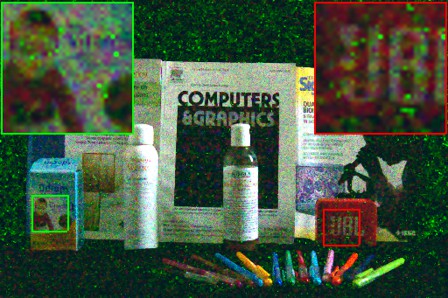}
		&  \includegraphics[width=0.118\linewidth]{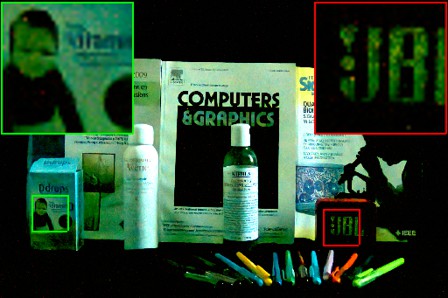}		
		&	 \includegraphics[width=0.118\linewidth]{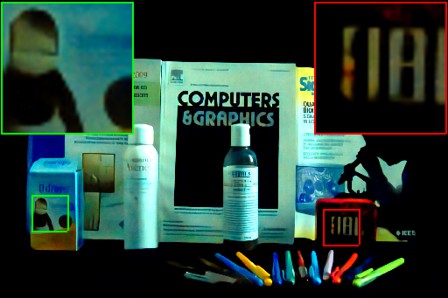}
		&  \includegraphics[width=0.118\linewidth]{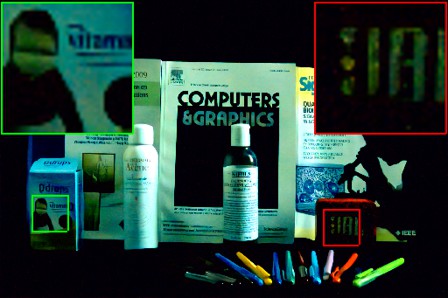}
		&  \includegraphics[width=0.118\linewidth]{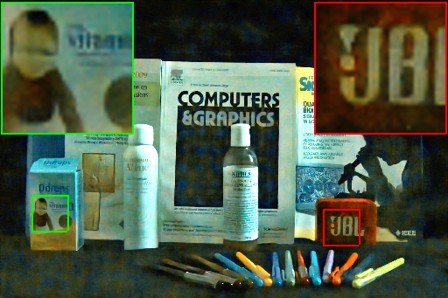}		
		&  \includegraphics[width=0.118\linewidth]{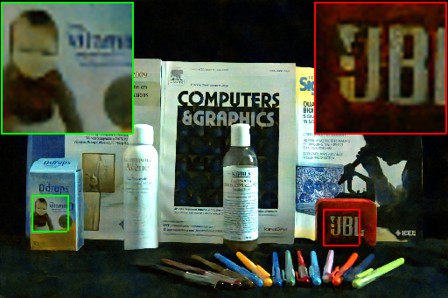} \vspace{-2pt} \\	
		20.71 & 29.07 & 29.02 &  33.44  & 33.66 & 32.93 &  35.82 & 37.07	\\
		\includegraphics[width=0.118\linewidth]{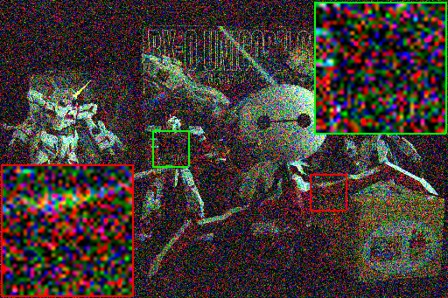}
		&	 \includegraphics[width=0.118\linewidth]{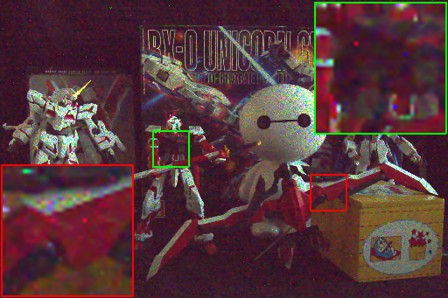}        	 
		&  \includegraphics[width=0.118\linewidth]{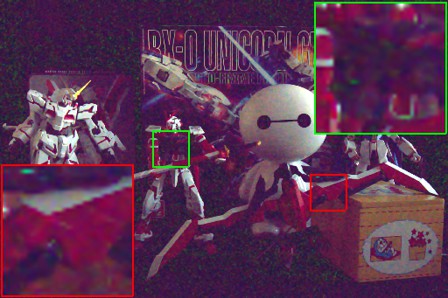}
		&  \includegraphics[width=0.118\linewidth]{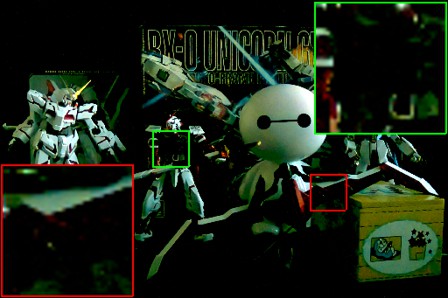}		
		&	 \includegraphics[width=0.118\linewidth]{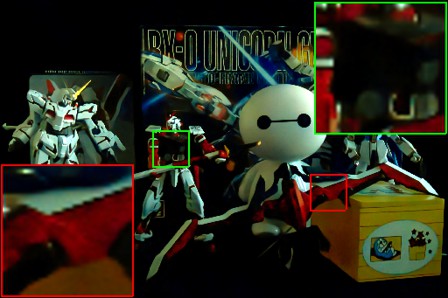}
		&  \includegraphics[width=0.118\linewidth]{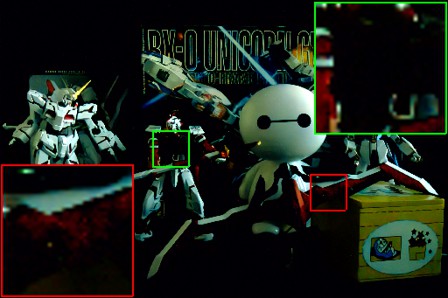}
		&  \includegraphics[width=0.118\linewidth]{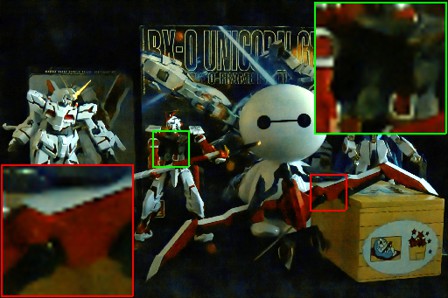}		
		&  \includegraphics[width=0.118\linewidth]{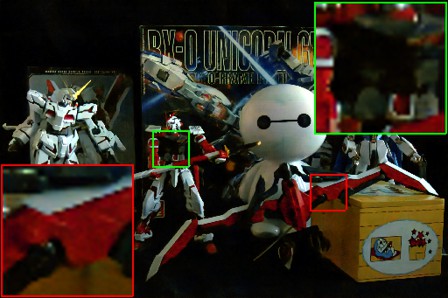} \vspace{-2pt} \\	
		26.57 & 33.55 & 36.31 &  38.31  & 40.13 &  38.14 &  41.89 & 42.65 \\
	\end{tabular} 
	\vspace{-4pt}
	\caption{Raw image denoising results on our ELD dataset, where we show the visual results on four evaluation scenes. (\textbf{Best viewed on screen with zoom})
	}
	\vspace{-4pt}
	\label{fig:method-comparision}
\end{figure*}

As shown in Table~\ref{tb:componet}, the domain gap is significant between the
heteroscedastic Gaussian model ($G$+$P$) and the de facto noise model
(characterized by the model trained with paired real data). 
This can be attributed to 1) the Gaussian approximation of Possion distribution is not
justified under extreme low illuminance; 2) the long-tail nature of read noise is overlooked; 3) the color-wise DC noise component is not  modeled;  4) the horizontal banding noise are not considered in the model.
By taking all these factors into account, our final model, \ie $G^*$+$P^*$+$B$+$R$+$U$ gives
rise to a striking result: the result is comparable to or sometimes even better
than the model trained with paired real data. 
A visual comparison of our final model and other baseline methods is shown in Figure~\ref{fig:ablation-visual}, suggesting the effectiveness of our proposed noise formation model. 

\subsubsection{Method Comparison}

The qualitative results of all competing methods on both indoor and outdoor scenes from SID Sony set are presented in Figure~\ref{fig:sony-vis}. 
It can be seen that the random noise can be suppressed by the model learned with heteroscedastic Gaussian noise (G+P) \cite{Foi2008Practical},  but the resulting colors are distorted,  the banding artifacts become conspicuous, and the image details are barely discernible.  
Training only with real low-light noisy data is not effective enough,
due to the color-bias effect (that violates the zero-mean noise assumption) and
the large variance of corruptions (that leads to a large variance of the
Noise2Noise solution) \cite{pmlr-v80-lehtinen18a}. 
Besides, the learning-based noise model - Noiseflow \cite{Abdelhamed_2019_ICCV} cannot fully capture the low-light real noise structure, leading to color-shifted and over-smoothed results. 
By contrast, our model produces visually appealing results as if it had been trained with paired real data.

\subsection{Results on our ELD Dataset}

\subsubsection{Method Comparisons}
\begin{figure}[!t]
	\centering
	\begin{subfigure}[b]{.32\linewidth}
		\centering
		\includegraphics[width=1\linewidth,clip,keepaspectratio]{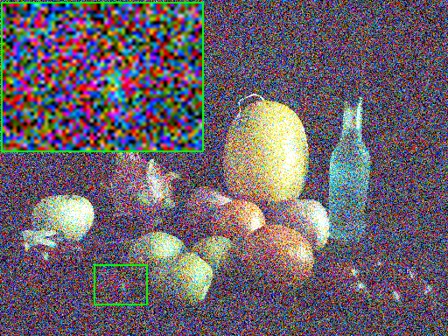}
		\caption{Input}
	\end{subfigure}
	\begin{subfigure}[b]{.32\linewidth}
		\centering
		\includegraphics[width=1\linewidth,clip,keepaspectratio]{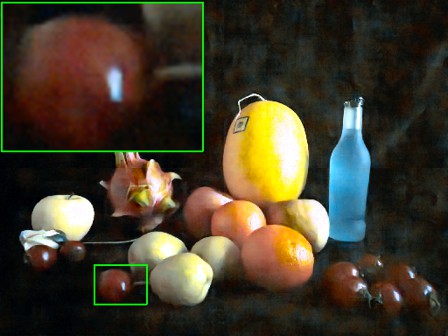}
		\caption{Paired real data}
	\end{subfigure}
	\begin{subfigure}[b]{.32\linewidth}
		\centering
		\includegraphics[width=1\linewidth,clip,keepaspectratio]{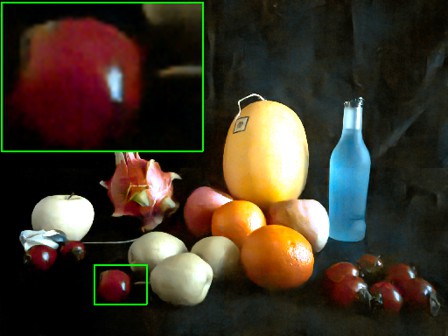}
		\caption{Ours}
	\end{subfigure}
	\vspace{-4pt}
	\caption{The denoising results of a low-light image captured by a Huawei Honor 10 camera.}
	\vspace{-4pt}
	\label{fig:smartphone}
\end{figure}

To see whether our noise model can be applicable to other camera devices as
well, we assess model performance on our ELD dataset.
Table~\ref{tb:ELD} and Figure~\ref{fig:method-comparision} summarize the
results of all competing methods on ELD dataset.  
It can be seen that the non-deep denoising methods, \ie BM3D and A-BM3D, fail to address the banding residuals, the color bias and the extreme values presented in the noisy input,  whereas our model recovers vivid image details, which can be hardly detected from the noisy image by human observers. 
\textit{Moreover, our model trained with synthetic data even often outperforms the model trained with paired real data.} 
We note the finding here conforms with the evaluation of sensor noise presented in Section~\ref{sec:noise-param}, especially in Figure~\ref{fig:TL-PPCC} and \ref{fig:noise_comparision}, where we show the underlying noise distribution varies camera by camera. 
Consequently, training with paired real data from SID Sony camera inevitably overfits to the noise pattern merely existed on the Sony camera, 
leading to suboptimal results on other types of cameras.  In contrast, our model relies on a very flexible noise model and a noise calibration process,  making it adapts to noise characteristics of other (calibrated) camera models as well.  Additional evidence can be found in Figure~\ref{fig:smartphone}, where we apply these two models to an image captured by a smartphone camera. 
Our reconstructed image is clearer and cleaner than what is restored by the model trained with paired real data. 

\begin{figure}[!t]
	\centering
	\begin{subfigure}[b]{.45\linewidth}
		\centering
		\includegraphics[width=1\linewidth,clip,keepaspectratio]{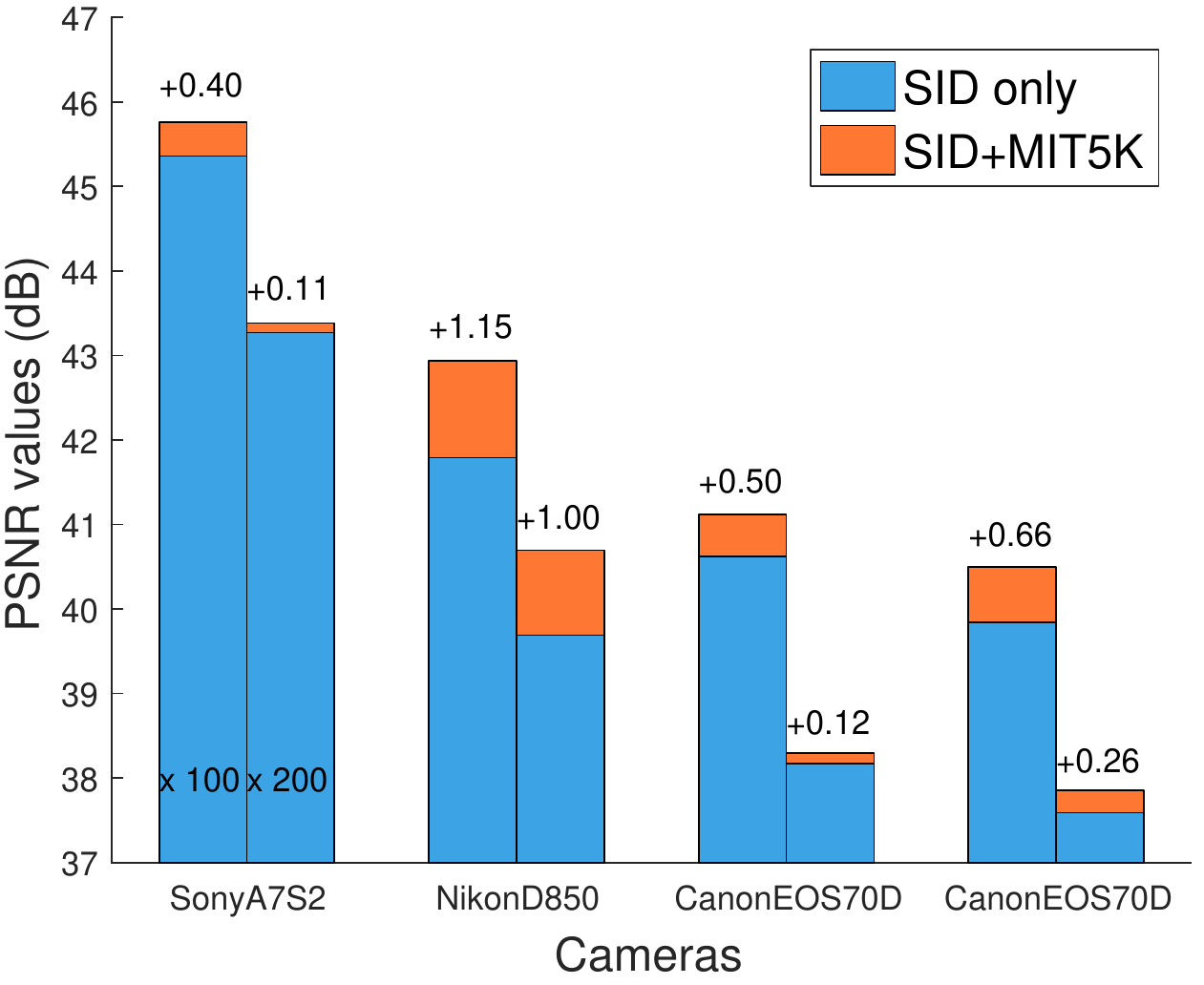}
		\caption{}
		\label{fig:mit5k}
	\end{subfigure}
	\begin{subfigure}[b]{.45\linewidth}
		\centering
		\includegraphics[width=1\linewidth,clip,keepaspectratio]{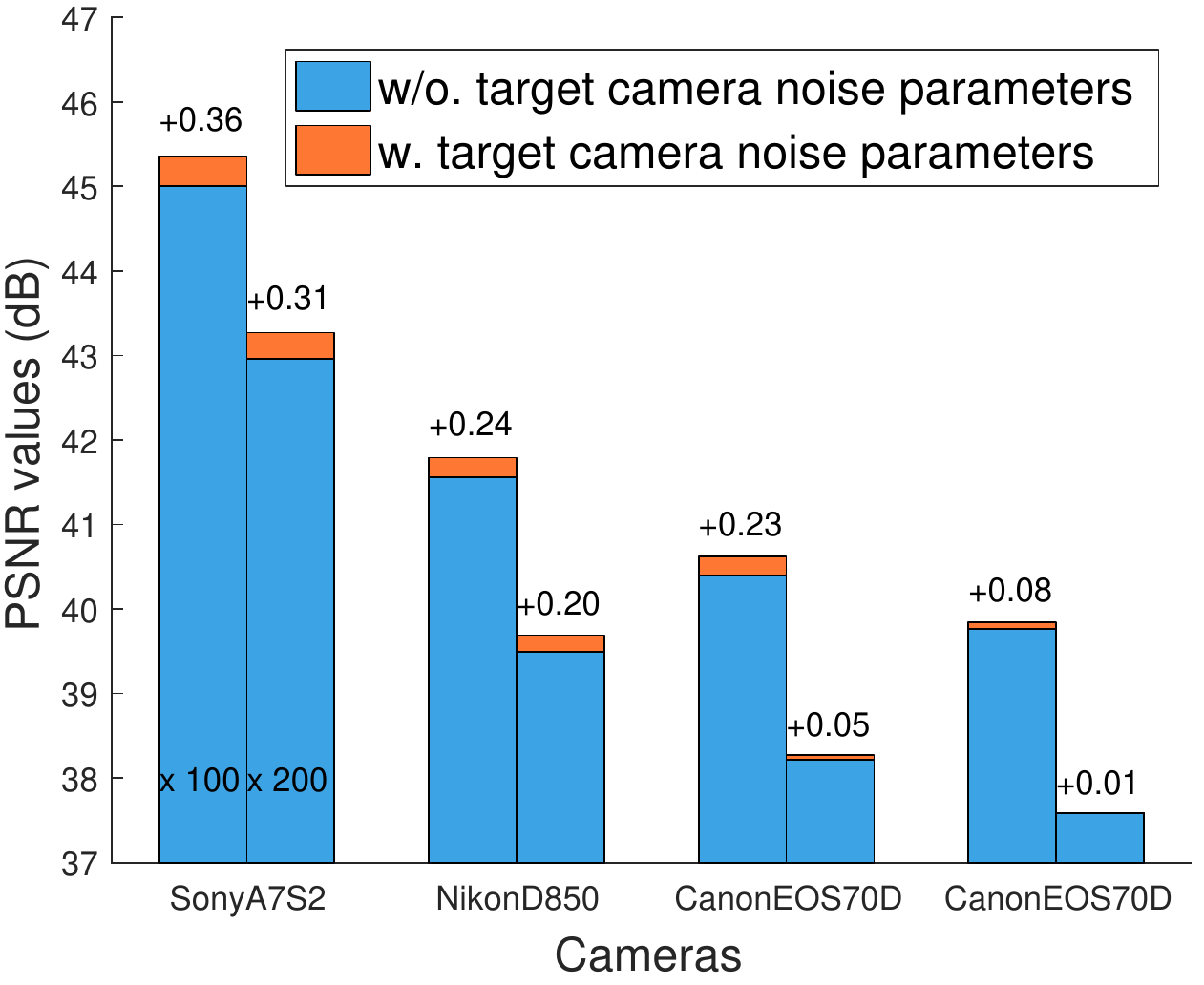}
		\caption{}
		\label{fig: parameter-sensitivity}
	\end{subfigure}
	\caption{(a) Performance boost when training with more synthesized data. (b) Noise parameter sensitivity test. }
\end{figure}
 
\begin{figure}[!t]
	\centering
	\begin{subfigure}[b]{.32\linewidth}
		\centering
		\includegraphics[width=1\linewidth,clip,keepaspectratio]{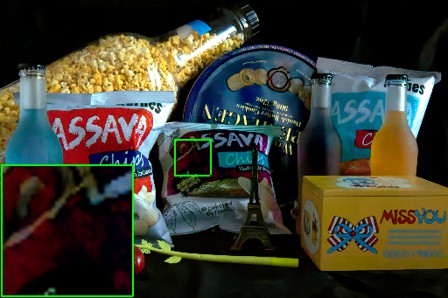}
		\caption{SID only}
	\end{subfigure}
	\begin{subfigure}[b]{.32\linewidth}
		\centering
		\includegraphics[width=1\linewidth,clip,keepaspectratio]{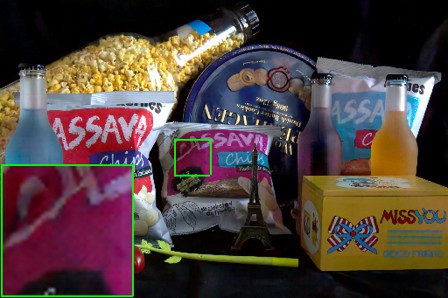}
		\caption{SID + MIT5K}
	\end{subfigure}
	\begin{subfigure}[b]{.32\linewidth}
		\centering
		\includegraphics[width=1\linewidth,clip,keepaspectratio]{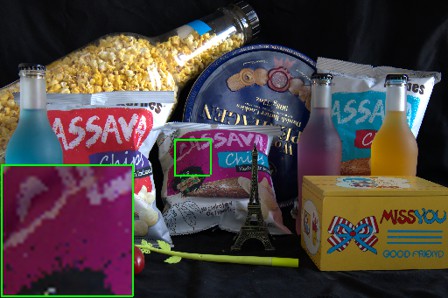}
		\caption{Ground Truth}
	\end{subfigure}
	\caption{The recovery results of a low-light image captured by Nikon D850 camera. }
	\label{fig:mit5k-vis}
\end{figure}


\subsubsection{Training with More Synthesized Data}
A useful merit of our approach compared to training with paired
real data, is that our model can be easily incorporated with more real clean
samples to train. Figure~\ref{fig:mit5k} shows the relative improvements of our
model when training with the dataset synthesized by additional clean raw images from MIT5K dataset~\cite{fivek}. We find the major improvements are owing to the more accurate color and brightness restoration, as shown in Figure~\ref{fig:mit5k-vis}. By training with more raw image samples from diverse cameras, the network learns to infer the appearances of pictures more naturally and precisely. 


\subsubsection{Sensitivity to Noise Calibration}
Another benefit of our approach is we only need clean samples and a noise calibration process to adapt to a new camera, in contrast to capturing real noisy images accompanied with densely-labeled ground truth. 
Besides, the noise calibration process can be simplified once we already have a collection of parameter samples from various cameras. Figure~\ref{fig: parameter-sensitivity} shows models can reach comparable performance on target cameras without noise calibration, by simply sampling parameters from other three calibrated cameras instead. 

\begin{figure*}[!t]
	\centering
	\setlength\tabcolsep{1pt}
	\begin{tabular}{cccccccc}	
		Input & BM3D & A-BM3D  & Noise2Noise &  G+P & \fontsize{8pt}{8pt}\selectfont Paired real data & Ours & Reference  \\
		\includegraphics[width=0.118\linewidth]{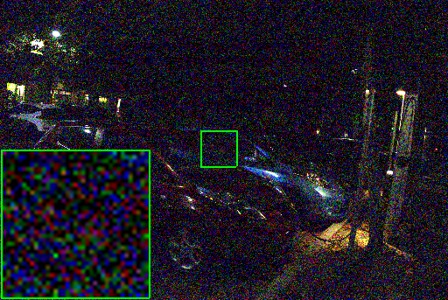}
		&		\includegraphics[width=0.118\linewidth]{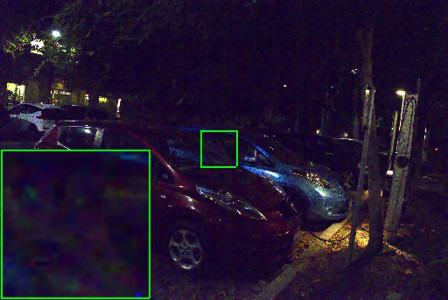}
		&		\includegraphics[width=0.118\linewidth]{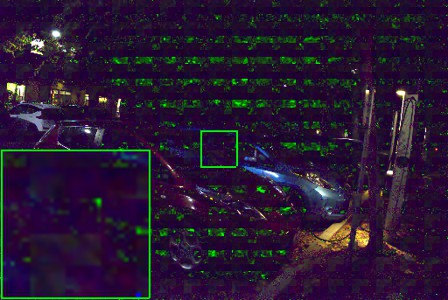}
		&	\includegraphics[width=0.118\linewidth]{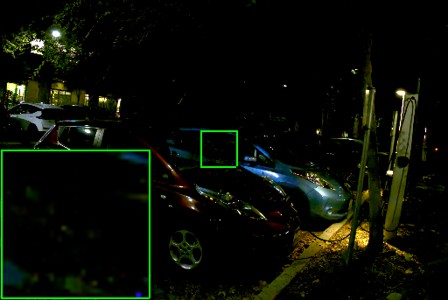}		
		&		\includegraphics[width=0.118\linewidth]{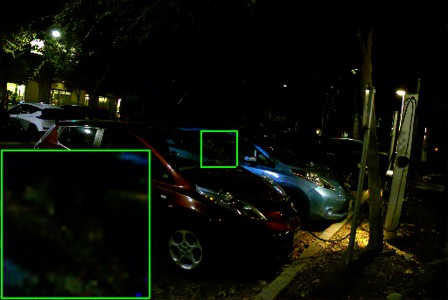}
		&		\includegraphics[width=0.118\linewidth]{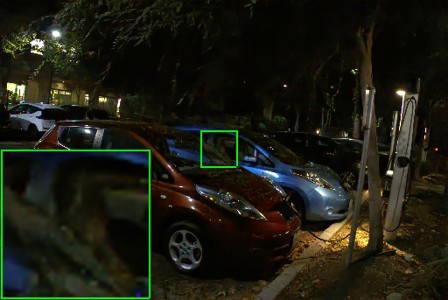}	
		&	\includegraphics[width=0.118\linewidth]{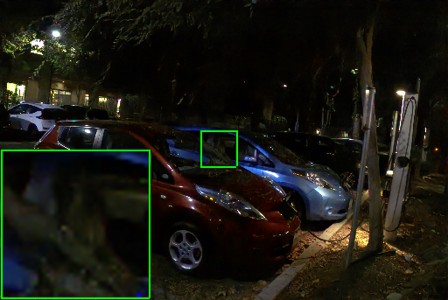}
		&	\includegraphics[width=0.118\linewidth]{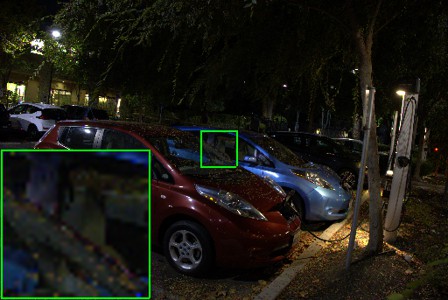} \vspace{-2pt} \\ 
		29.54 & 35.54 & 30.06 &  35.06  & 34.53 & 36.68 & 36.46 & $\infty$ 		\\
		\includegraphics[width=0.118\linewidth]{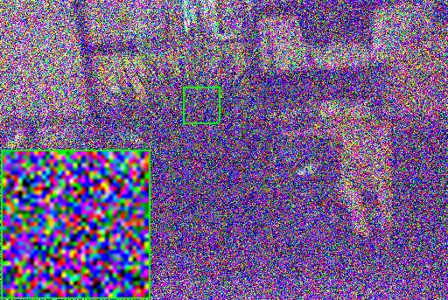}
		&		\includegraphics[width=0.118\linewidth]{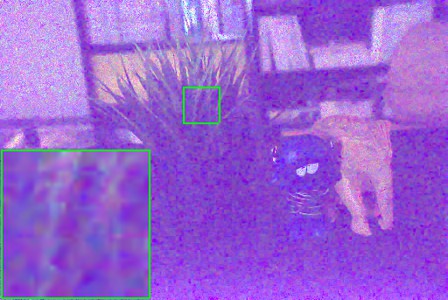}
		&		\includegraphics[width=0.118\linewidth]{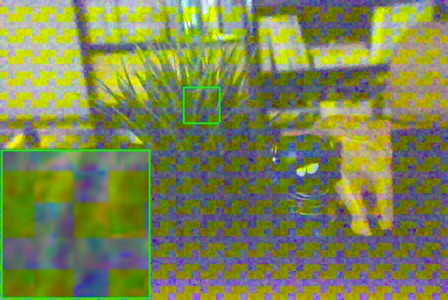}
		&	\includegraphics[width=0.118\linewidth]{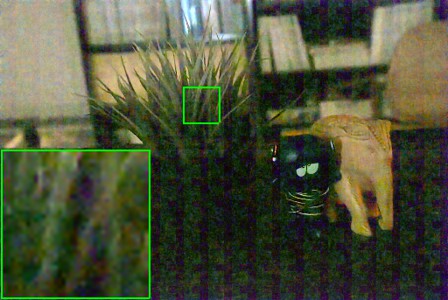}		
		&		\includegraphics[width=0.118\linewidth]{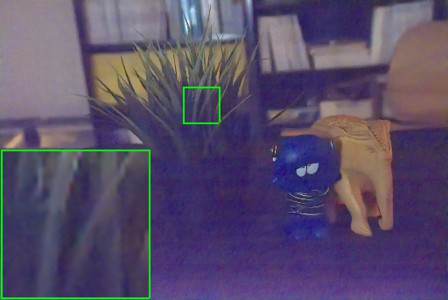}	
		&		\includegraphics[width=0.118\linewidth]{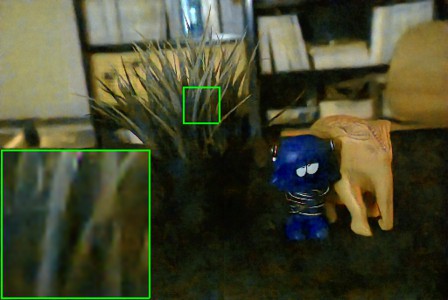}
		&	\includegraphics[width=0.118\linewidth]{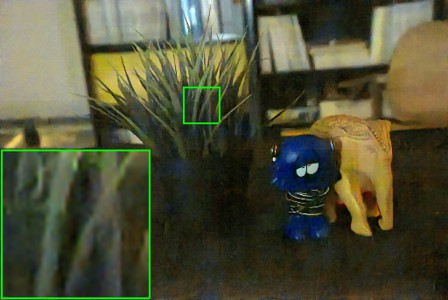}
		&	\includegraphics[width=0.118\linewidth]{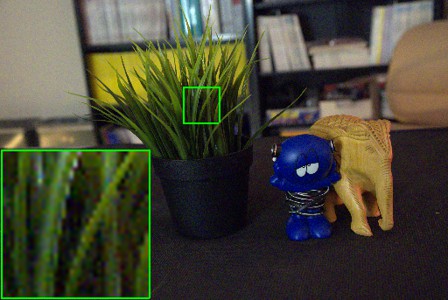} \vspace{-2pt} \\ 
		13.50 & 25.85 & 27.83 &  35.70  &  29.52 & 36.79 & 37.65 &  $\infty$ 		\\
		\includegraphics[width=0.118\linewidth]{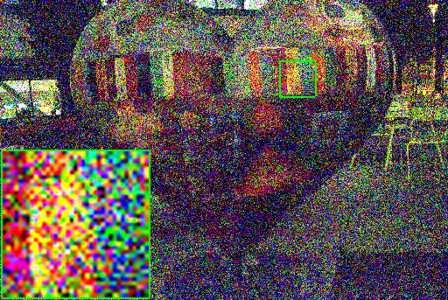}
		&		\includegraphics[width=0.118\linewidth]{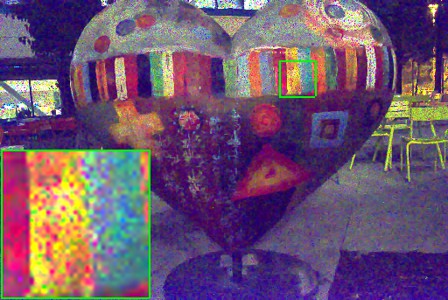}
		&		\includegraphics[width=0.118\linewidth]{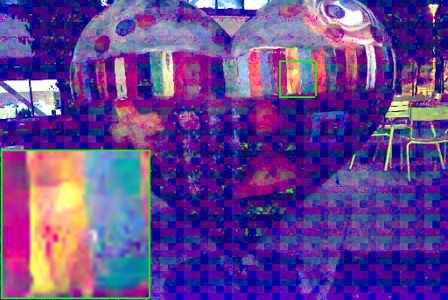}
		&	\includegraphics[width=0.118\linewidth]{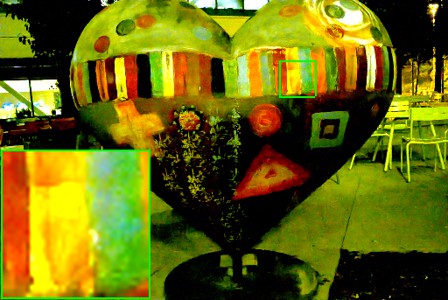}
		&		\includegraphics[width=0.118\linewidth]{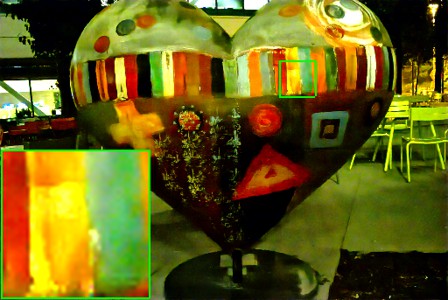}	
		&		\includegraphics[width=0.118\linewidth]{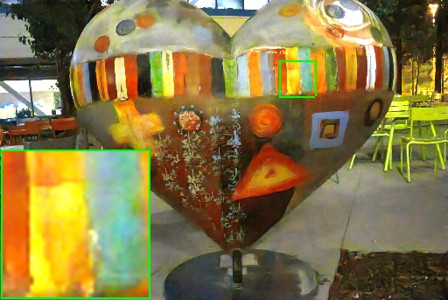}		
		&	\includegraphics[width=0.118\linewidth]{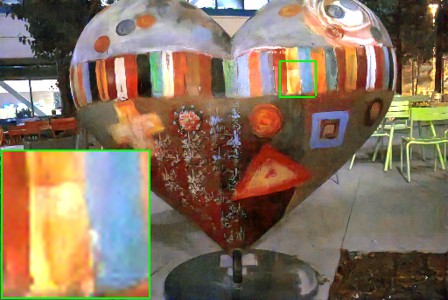}
		&	\includegraphics[width=0.118\linewidth]{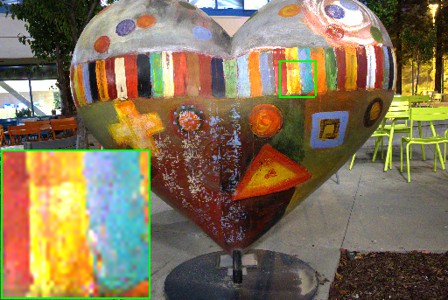} \vspace{-2pt} \\ 
		24.62 & 31.95 & 32.87 &  33.44  & 32.42 & 38.50 & 38.74 & $\infty$ 		\\
		\includegraphics[width=0.118\linewidth]{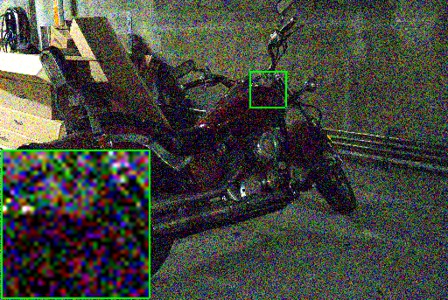}
		&		\includegraphics[width=0.118\linewidth]{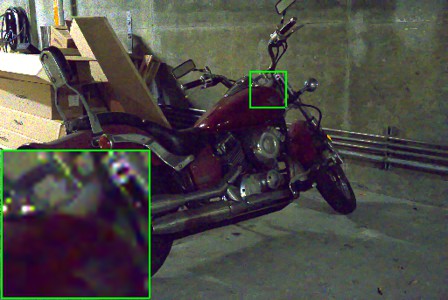}
		&		\includegraphics[width=0.118\linewidth]{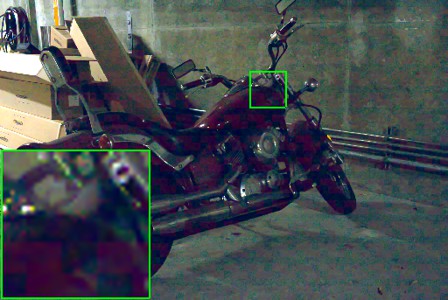}
		&	\includegraphics[width=0.118\linewidth]{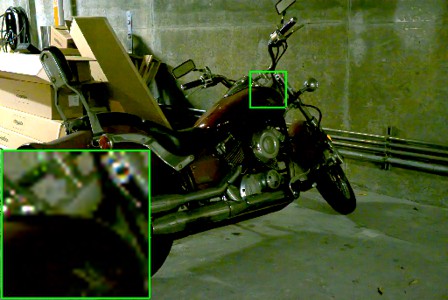}
		&		\includegraphics[width=0.118\linewidth]{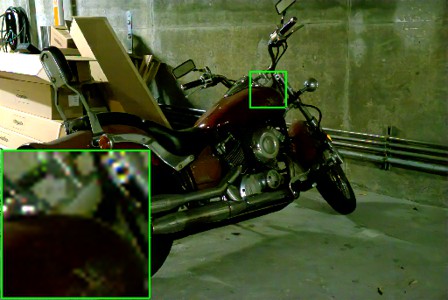}	
		&		\includegraphics[width=0.118\linewidth]{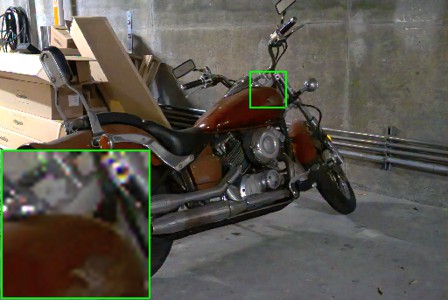}
		&	\includegraphics[width=0.118\linewidth]{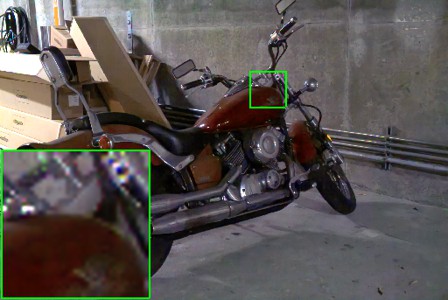}
		&	\includegraphics[width=0.118\linewidth]{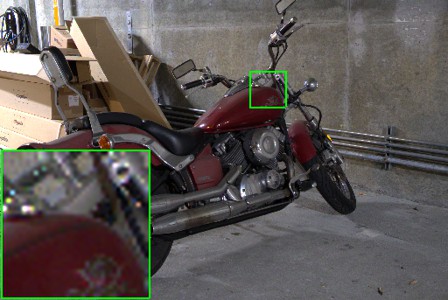} \vspace{-2pt} \\ 
		25.05 & 32.27 & 33.99 & 32.73  &  32.24 & 37.50 & 37.10 & $\infty$ 	
	\end{tabular} 
	\vspace{-3pt}
	\caption{Raw image denoising results on images from SID Fuji dataset. All images are converted from raw X-Trans space to sRGB for visualization. (\textbf{Best viewed on screen with zoom})}
	\label{fig:fuji-vis}
	\vspace{-3pt}
\end{figure*}

\subsection{Applicability to Non-Bayer Raw Data} \label{sec:non-bayer}

In the previous sections, we have evaluated our noise formation model on sensor raw images with Bayer color filter array (CFA). Here, we show our noise formation model is also applicable to photosensors with different CFA, \eg X-Trans used in Fuji cameras. We follow \cite{Chen_2018_CVPR} to preprocess the X-Trans raw data by packing it into 9 channels, and create multiple training datasets for different training schemes, based upon the SID Fuji training dataset.  

Table \ref{tb:fuji} summarizes the quantitative results of different methods on the SID Fuji testing dataset.\footnote{Noiseflow cannot be trained on 9-channel raw images (its normalizing flow implementation requires the number of input channels to be even) thus is not compared here } As we do not have bias/flat-field frames recorded by the Fuji camera, we simply sample the noise parameters required by noise models using the cameras in our ELD dataset. Even without accurate noise calibration, our final model still reaches a performance comparable to the model trained with paired real data (\eg 38.33 v.s. 38.56 in $\times 250$).
Figure~\ref{fig:fuji-vis} visually compares the results of different methods.

\begin{table}[!t]
	\centering
	\caption{Quantitative results on Fuji set of SID dataset. 
		Note our results are obtained \emph{without} noise calibration on the target Fuji camera. 
	}
	\footnotesize
	\begin{tabular}{lccc} 
		\toprule
		& $\times 100$ & $\times 250$ & $\times 300$ \\
		Model & PSNR / SSIM & PSNR / SSIM & PSNR / SSIM \\ 
		\midrule
		BM3D   & 37.26 / 0.889 & 33.16 / 0.804 & 30.59 / 0.739 \\ \hline
		A-BM3D  & 37.19 / 0.879 & 30.34 / 0.746 & 28.92 / 0.678 \\ \hline
		\midrule
		Noise2Noise  & 39.03 / 0.893 & 35.18 / 0.769 & 33.54 / 0.740 \\ \hline
		Paired real data  & \textcolor{red}{40.13} / \textcolor{blue}{0.958} & \textcolor{red}{38.56} / \textcolor{red}{0.941} &  \textcolor{red}{37.44} /  \textcolor{red}{0.928} \\ \hline
		
		\midrule
		$G$+$P$  & 38.84 / 0.923 & 34.86 / 0.815 & 32.83 / 0.753 \\ \hline
		Ours  & \textcolor{blue}{39.88} / \textcolor{red}{0.959} & \textcolor{blue}{38.33} / \textcolor{blue}{0.937} & \textcolor{blue}{37.23} / \textcolor{blue}{0.924} \\ \hline
		\bottomrule
	\end{tabular}
	\label{tb:fuji}
\end{table}

\subsection{Applicability to sRGB Denoising} \label{sec:color}


Given a known image processing pipeline (ISP), our noise formation model allows us to generate downstream noise on sRGB space as well, \emph{i.e.}, noisy raw images are firstly generated according to our noise formation model, then are post-processed by a pre-defined ISP to synthesize the final noisy color images. 
As a proof of concept, here, we assume the color images are finished by a simple ISP with typical functions, including white balance, color correction and a camera response function (CRF).\footnote{We do not perform demosaicking; instead, the Bayer data is split into separate RGB channels to form raw RGB, where the green channel is obtained by averaging the two green pixels in two-by-two block.} Note that the major goal here is to corroborate the applicability of our noise model to color image denoising. In practice, it can be easily extended into real-world cases when more realistic ISP models for targeted camera devices are accessible (\eg for camera vendors and/or ISP designers), but this is beyond the scope of this paper. 

The necessary parameters to specify some of these operations including the per-channel gain for white balance and the camera-to-sRGB conversion matrix, are directly read from the camera raw files\footnote{These parameters are only used in training stage, they can be implicitly inferred by a trained model at inference.}.  
The CRF is a non-linear mapping to compress a wide range of irradiance values within a fixed
range of measurable image intensity values. It often serves as a key feature to distinguish different ISPs \cite{kim2012new,grossberg2004modeling}. Rather than assuming a standard gamma curve as CRF, we adopt a realistic CRF calibrated on a SonyA7S2 camera (the one used in SID dataset) using a PCA-based radiometric calibration method \cite{grossberg2004modeling}. 


\begin{table}[!t]
	\centering
	\caption{Color image (RGB2RGB) denoising  results on SID Sony dataset. All the results are evaluated in sRGB space. 
	}
	\footnotesize
	\begin{tabular}{lccc} 
		\toprule
		& $\times 100$ & $\times 250$ & $\times 300$ \\
		Model & PSNR / SSIM & PSNR / SSIM & PSNR / SSIM \\ 
		\midrule
		Noise2Noise  & 19.67 / 0.479 & 15.11 / 0.332 & 14.48 / 0.321 \\ \hline
		Noiseflow  & 20.77 / 0.512 & 16.66 / 0.380 & 15.57 / 0.352 \\ \hline		
		Paired real data  & \textcolor{red}{24.63} / \textcolor{blue}{0.598} & \textcolor{red}{23.15} / \textcolor{red}{0.560} & \textcolor{red}{22.23} / \textcolor{red}{0.531}  \\ \hline
		\midrule
		$G$+$P$  & 21.08 / 0.529 & 16.64 / 0.390 & 15.38 / 0.354 \\ \hline
		Ours  & \textcolor{blue}{24.45} / \textcolor{red}{0.599} & \textcolor{blue}{22.95} / \textcolor{blue}{0.553} & \textcolor{blue}{21.64} / \textcolor{blue}{0.523} \\ \hline
		\bottomrule
	\end{tabular}
	\label{tb:color}
\end{table}

We retrain the networks for color image (\ie RGB2RGB) denoising based on the same training policy described in Section~\ref{sec:implementation-details}, except the used raw data have been preprocessed by the pre-defined ISP to produce their color counterparts.  
As shown in Table~\ref{tb:color}, the network trained with our proposed noise model still reaches a highly competitive performance compared to the network learned with paired real data, indicating the applicability of our model to color image denoising as well. 

Moreover, by preparing datasets of both raw and RGB formats, we can train networks to play different roles in the whole low-light imaging pipeline, including raw space denoising as a kind of preprocessing (Raw2Raw), color space denoising as a kind of post-processing (RGB2RGB), and joint performing multiple tasks as a kind of ISP (Raw2RGB) -- the networks are learned to do denoising, white balance, color correction and camera response (tone mapping, gamma correction) simultaneously. 
We compare these three approaches on the SID Sony dataset, using either paired real data or synthetic data generated by our noise model.  
The quantitative results are shown in Table~\ref{tb:ISP}. 
Interestingly, the approaches of Raw2Raw+ISP and Raw2RGB achieve comparable performance, and outperform the RGB2RGB approach. 
We note although the Raw2Raw setup is flexible to plug in different ISPs to render different styles of sRGB images, it 
requires non-blind white balance coefficients, which are often difficult to be deduced in very low light. 
By contrast, Raw2RGB can automatically infer color constancy from the noisy raw input itself.

\begin{table}[!t]
	\centering
	\caption{Performance comparison of learning models applied to different stages of the image processing pipeline. 
		We test the same set of noisy images from SID Sony dataset to generate input data of both raw and RGB formats, in which the input RGB images are generated from their corresponding raw counterparts via a pre-defined ISP.  The outputs of Raw2Raw are post-processed by the ISP with oracle white balance.  All the results are evaluated on sRGB space.
	}
	\footnotesize
	\begin{tabular}{lccc} 
		\toprule
		& $\times 100$ & $\times 250$ & $\times 300$ \\
		Model & PSNR / SSIM & PSNR / SSIM & PSNR / SSIM \\ 
		\midrule
		\textsc{Paired data}  &  &  & \\
		Raw2Raw+ISP & \textcolor{red}{25.37} / 0.600  & 23.49 / 0.554 & 22.13 / 0.523 \\ \hline
		RGB2RGB & 24.63 / 0.598 & 23.15 / 0.560 & 22.23 / 0.531  \\ \hline
		Raw2RGB & 24.92 / \textcolor{red}{0.612} & 23.33 / \textcolor{red}{0.571}  & \textcolor{red}{22.69} / \textcolor{red}{0.547} \\ \hline
		\midrule
		\textsc{Ours}  &  &  & \\
		Raw2Raw+ISP & \textcolor{blue}{24.95} / \textcolor{blue}{0.608} & \textcolor{red}{23.60} / 0.560 & 22.13 / 0.533 \\ \hline
		RGB2RGB & 24.45 / 0.599 & 22.95 / 0.553 & 21.64 / 0.523 \\ \hline
		Raw2RGB & 24.65 / 0.605 & \textcolor{blue}{23.50} / \textcolor{blue}{0.561} & \textcolor{blue}{22.53} / \textcolor{blue}{0.542} \\ \hline
		\bottomrule
	\end{tabular}
	\label{tb:ISP}
\end{table}

\subsection{Applicability to Extreme Low-light Videography} \label{sec:video}

\begin{figure*}[!t]
	\centering
	\setlength\tabcolsep{1.5pt}
	\begin{tabular}{cccccccc}		
		& Input & VBM4D &  Noise2Noise  & Noiseflow  & $G$+$P$ & \fontsize{8pt}{8pt}\selectfont Paired real data & Ours \\
		\rotatebox[origin=c]{90}{\textsc{Frame t}} 
		& 		\includegraphics[align=c,width=0.124\linewidth]{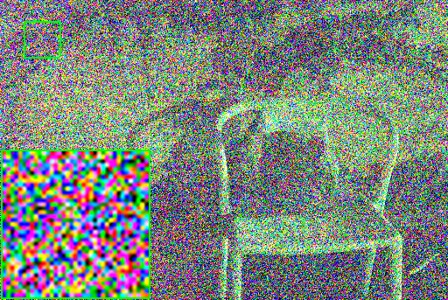}
		&		\includegraphics[align=c,width=0.124\linewidth]{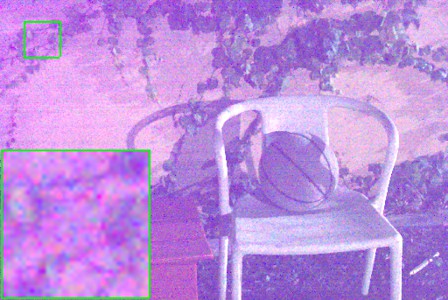}
		&		\includegraphics[align=c,width=0.124\linewidth]{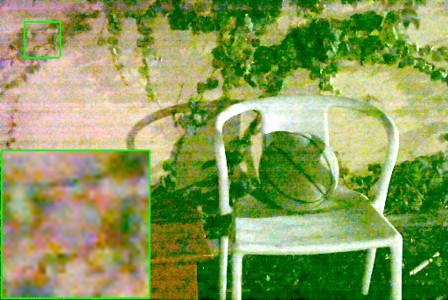}			
		&		\includegraphics[align=c,width=0.124\linewidth]{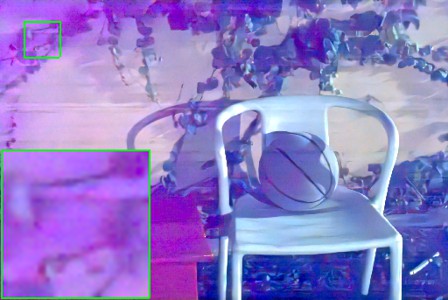}
		&		\includegraphics[align=c,width=0.124\linewidth]{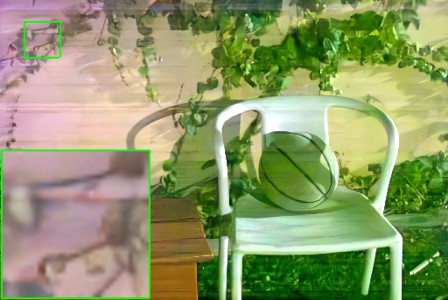}
		&		\includegraphics[align=c,width=0.124\linewidth]{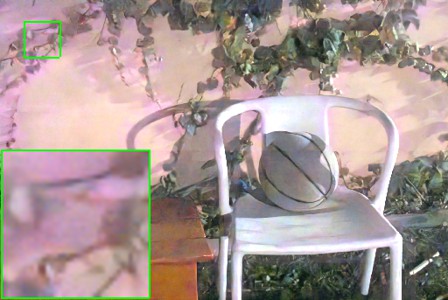}	
		&		\includegraphics[align=c,width=0.124\linewidth]{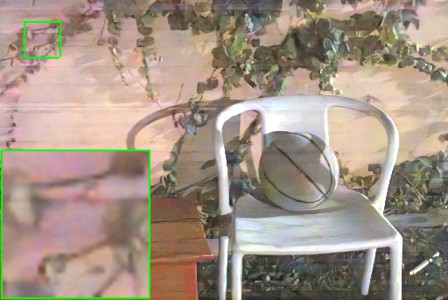} \\
		\addlinespace[1.5pt]
		& 10.84 & 21.26 &  28.11   &   26.69  &  30.08 & 30.97 & 30.95 \\
		
		\rotatebox[origin=c]{90}{\textsc{t+1}} 
		& 		\includegraphics[align=c,width=0.124\linewidth]{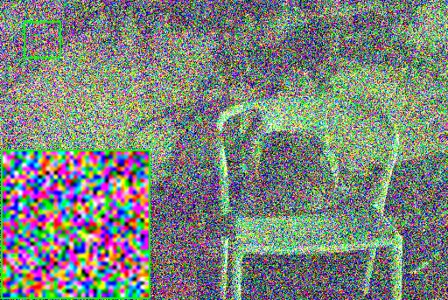}
		&		\includegraphics[align=c,width=0.124\linewidth]{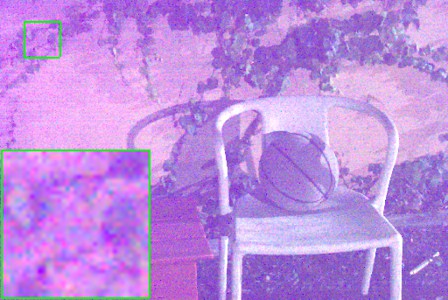}
		&		\includegraphics[align=c,width=0.124\linewidth]{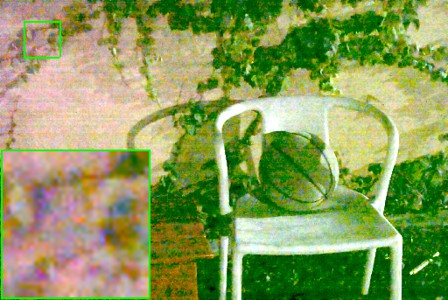}			
		&		\includegraphics[align=c,width=0.124\linewidth]{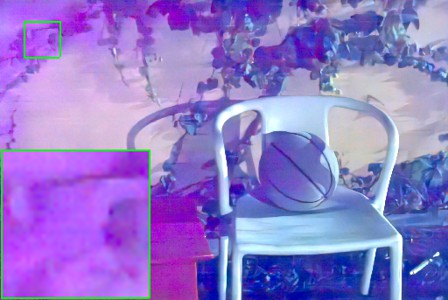}
		&		\includegraphics[align=c,width=0.124\linewidth]{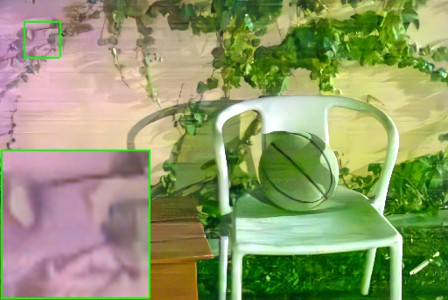}
		&		\includegraphics[align=c,width=0.124\linewidth]{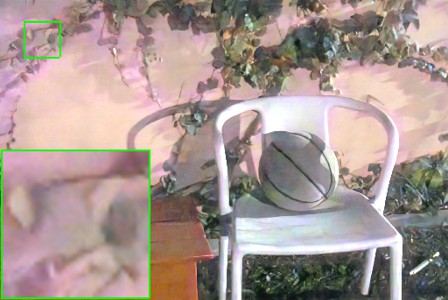}	
		&		\includegraphics[align=c,width=0.124\linewidth]{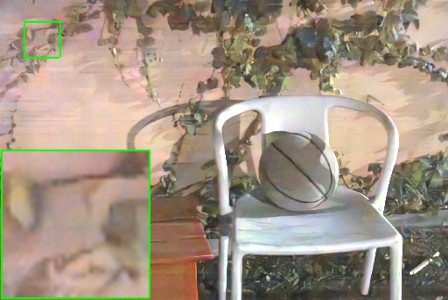} \\
		\addlinespace[1.5pt]
		& 10.86 & 21.24 &  28.20  &   26.67 &   30.23 & 31.20 & 31.17 \\
		
		\rotatebox[origin=c]{90}{\textsc{t+2}}
		& 		\includegraphics[align=c,width=0.124\linewidth]{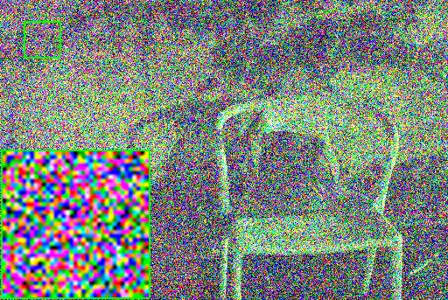}
		&		\includegraphics[align=c,width=0.124\linewidth]{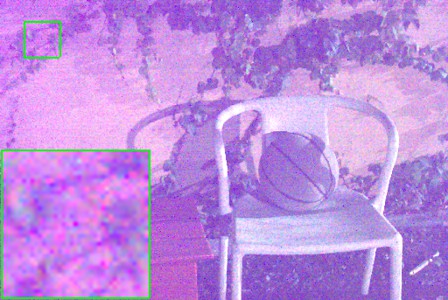}
		&		\includegraphics[align=c,width=0.124\linewidth]{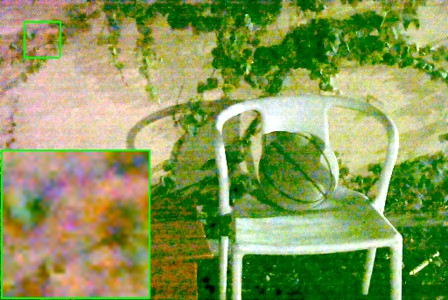}			
		&		\includegraphics[align=c,width=0.124\linewidth]{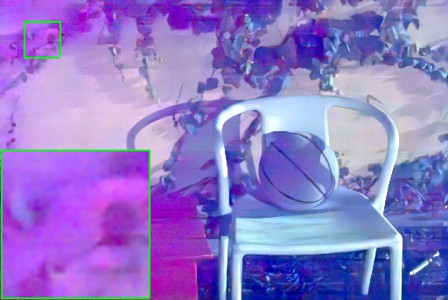}
		&		\includegraphics[align=c,width=0.124\linewidth]{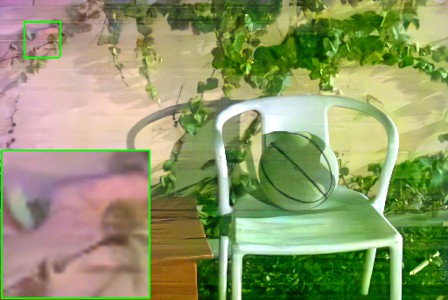}
		&		\includegraphics[align=c,width=0.124\linewidth]{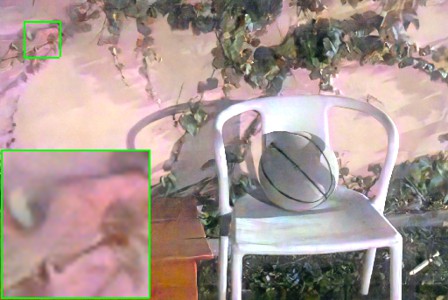}	
		&		\includegraphics[align=c,width=0.124\linewidth]{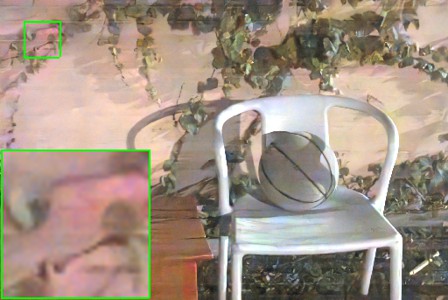} \\
		\addlinespace[1.5pt]
		& 10.85 & 21.20 &  28.14  &  26.77   &   30.20 & 31.01 & 31.13 \\
	\end{tabular} 
	\vspace{-6pt}
	\caption{Raw video denoising results on \textit{static} frames from DRV dataset. (\textbf{Best viewed on screen with zoom}) }
	\vspace{-6pt}
	\label{fig:video-static-vis}
\end{figure*}

\begin{figure*}[!t]
	\centering
	\setlength\tabcolsep{1.5pt}
	\begin{tabular}{ccccc?{1pt}ccc}		
		& & \multicolumn{3}{c?{1pt}}{\textsc{Synthetic dynamic video frames}} &  \multicolumn{3}{c}{\textsc{Real dynamic video frames}} \\
		& & Frame T & Frame T+1 & Frame T+2 & Frame T & Frame T+1 & Frame T+2 \\
		& \rotatebox[origin=c]{90}{\textsc{Input}} 
		& 		\includegraphics[align=c,width=0.132\linewidth]{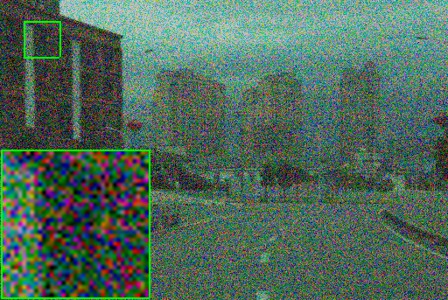}
		&		\includegraphics[align=c,width=0.132\linewidth]{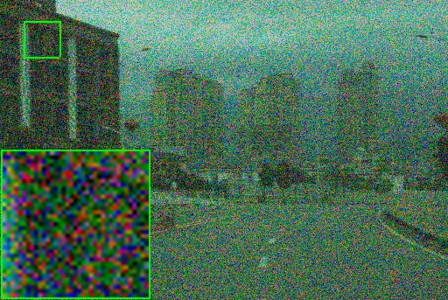}
		&		\includegraphics[align=c,width=0.132\linewidth]{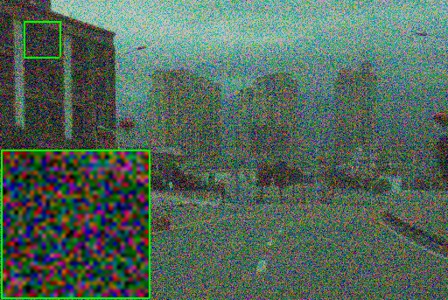}
		&		\includegraphics[align=c,width=0.132\linewidth]{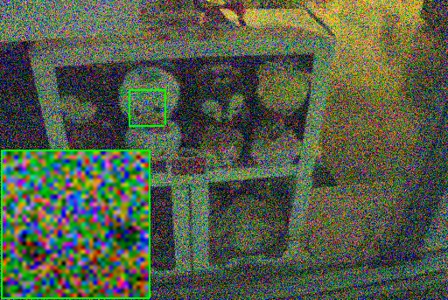}
		&		\includegraphics[align=c,width=0.132\linewidth]{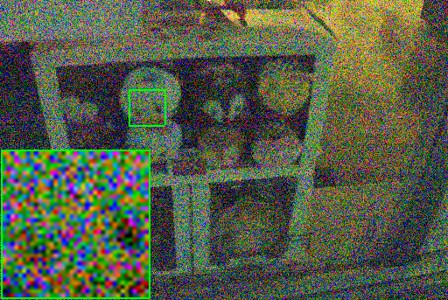}
		&		\includegraphics[align=c,width=0.132\linewidth]{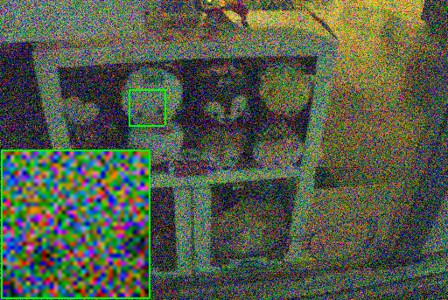} \\
		\specialrule{1pt}{3pt}{4pt}
		\raisebox{2.4\width}{\multirow{2}{*}{\rotatebox[origin=c]{90}{\textsc{UNet (single-frame)}}}}
		& \rotatebox[origin=c]{90}{Paired  data} 
		& 		\includegraphics[align=c,width=0.132\linewidth]{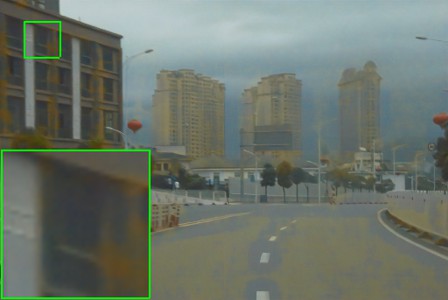}
		&		\includegraphics[align=c,width=0.132\linewidth]{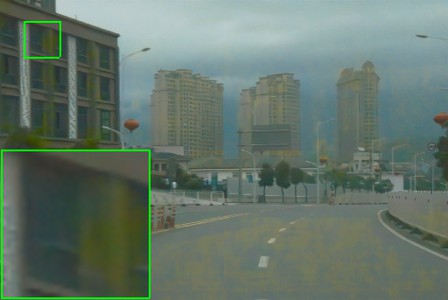}
		&		\includegraphics[align=c,width=0.132\linewidth]{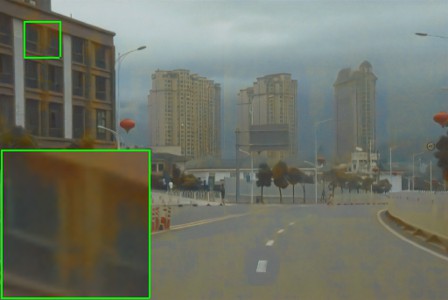}
		&		\includegraphics[align=c,width=0.132\linewidth]{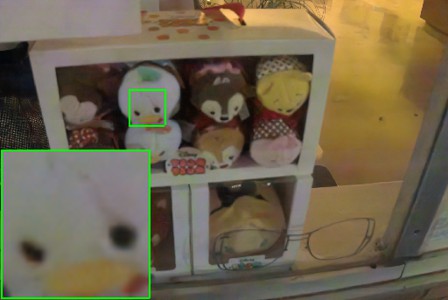}
		&		\includegraphics[align=c,width=0.132\linewidth]{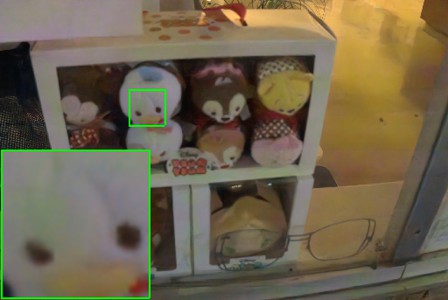}
		&		\includegraphics[align=c,width=0.132\linewidth]{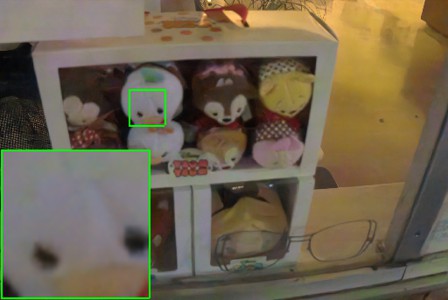} \\
		\addlinespace[1.5pt]
		& \rotatebox[origin=c]{90}{Ours}
		& 		\includegraphics[align=c,width=0.132\linewidth]{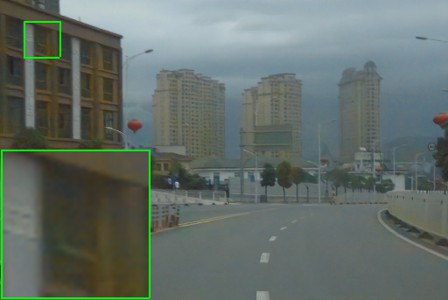}
		&		\includegraphics[align=c,width=0.132\linewidth]{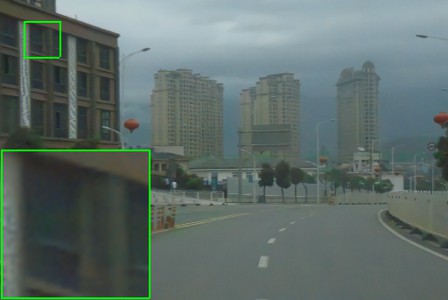}
		&		\includegraphics[align=c,width=0.132\linewidth]{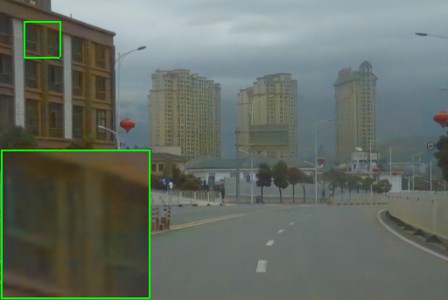}
		&		\includegraphics[align=c,width=0.132\linewidth]{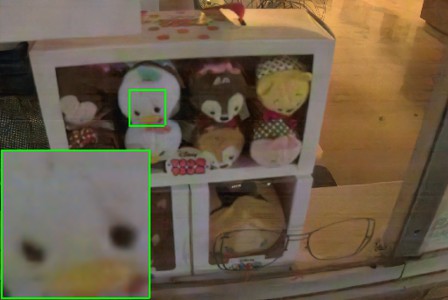}
		&		\includegraphics[align=c,width=0.132\linewidth]{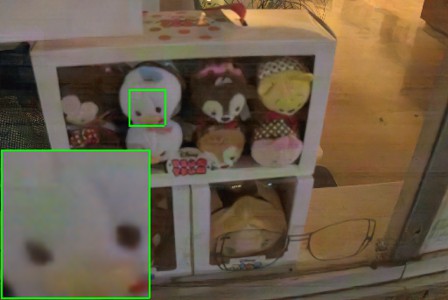}
		&		\includegraphics[align=c,width=0.132\linewidth]{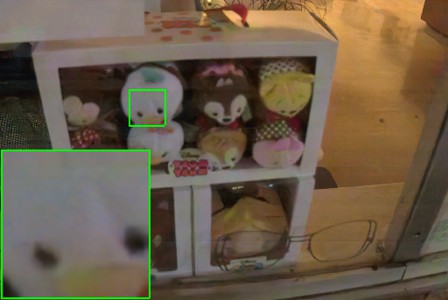} \\
		\specialrule{1pt}{3pt}{4pt}
		\raisebox{0\width}{\multirow{3}{*}{\rotatebox[origin=c]{90}{\textsc{fastDVD (multi-frame)}}}}
		& \rotatebox[origin=c]{90}{\footnotesize Paired data (S)} 
		& 		\includegraphics[align=c,width=0.132\linewidth]{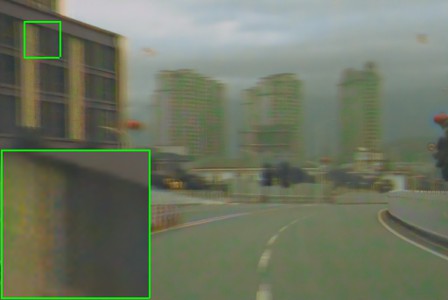}
		&		\includegraphics[align=c,width=0.132\linewidth]{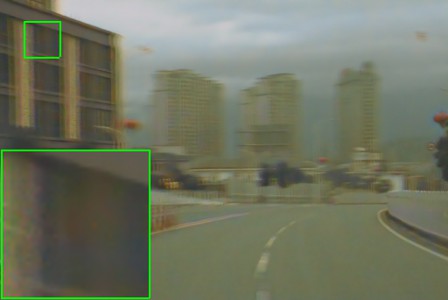}
		&		\includegraphics[align=c,width=0.132\linewidth]{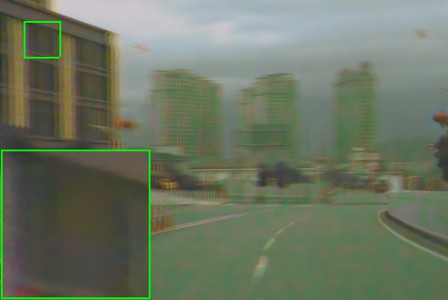}
		&		\includegraphics[align=c,width=0.132\linewidth]{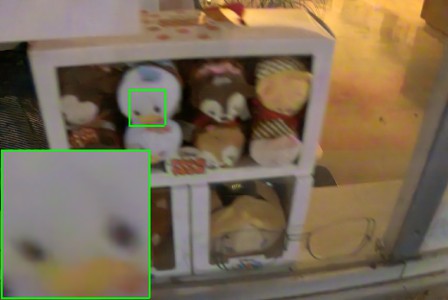}
		&		\includegraphics[align=c,width=0.132\linewidth]{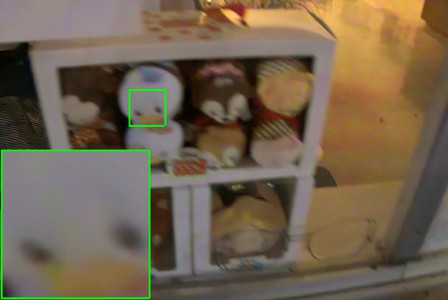}
		&		\includegraphics[align=c,width=0.132\linewidth]{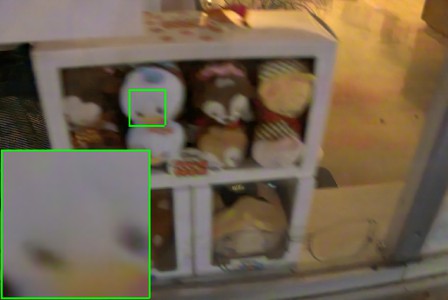} \\
		\addlinespace[1.5pt]
		& \rotatebox[origin=c]{90}{Ours (S)}
		& 		\includegraphics[align=c,width=0.132\linewidth]{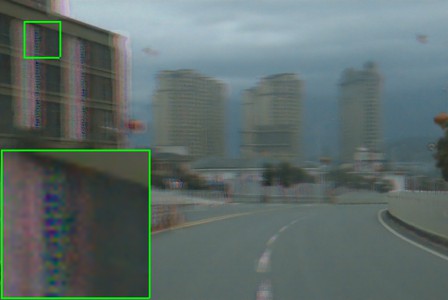}
		&		\includegraphics[align=c,width=0.132\linewidth]{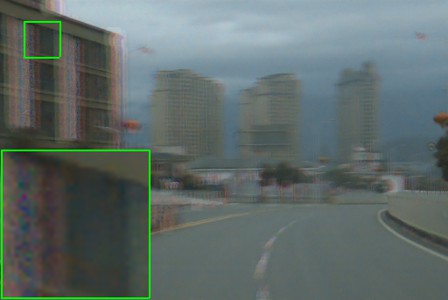}
		&		\includegraphics[align=c,width=0.132\linewidth]{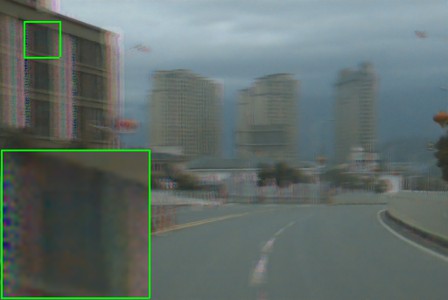}
		&		\includegraphics[align=c,width=0.132\linewidth]{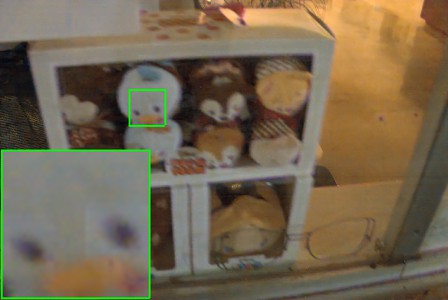}
		&		\includegraphics[align=c,width=0.132\linewidth]{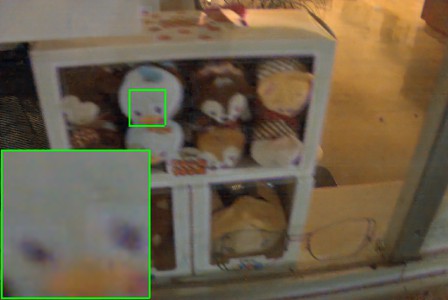}
		&		\includegraphics[align=c,width=0.132\linewidth]{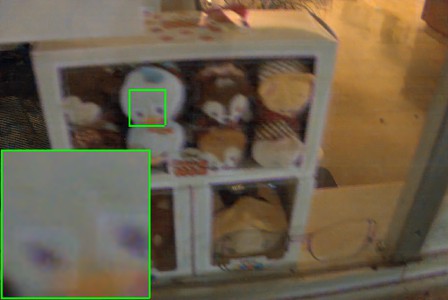} \\
		\addlinespace[1.5pt]
		& \rotatebox[origin=c]{90}{Ours (D)}
		& 		\includegraphics[align=c,width=0.132\linewidth]{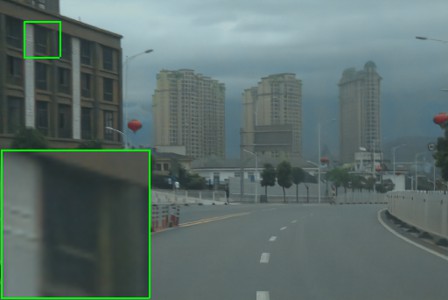}
		&		\includegraphics[align=c,width=0.132\linewidth]{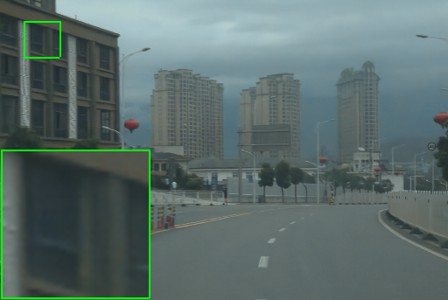}
		&		\includegraphics[align=c,width=0.132\linewidth]{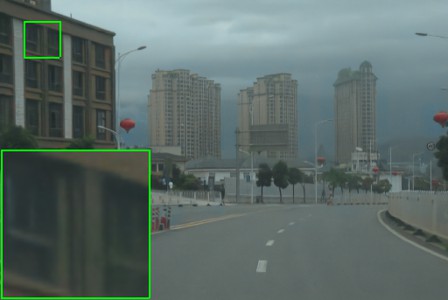}
		&		\includegraphics[align=c,width=0.132\linewidth]{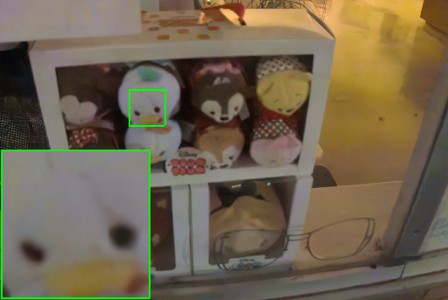}
		&		\includegraphics[align=c,width=0.132\linewidth]{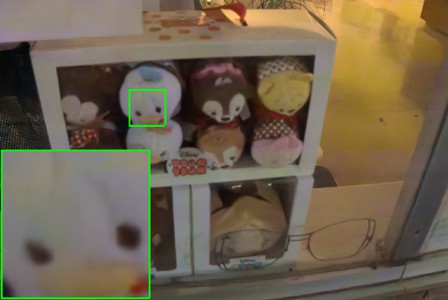}
		&		\includegraphics[align=c,width=0.132\linewidth]{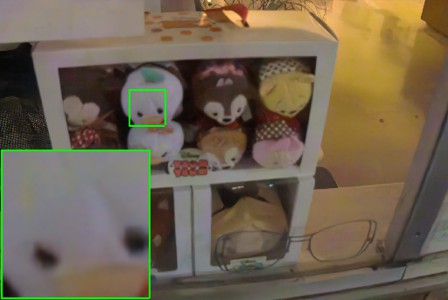} \\
	\end{tabular} 
	\vspace{-3pt}
	\caption{Raw video denoising results on synthetic and real \textit{dynamic} frames from our dynamic video dataset and DRV dynamic video dataset respectively. 
		``S" and ``D" indicate the multi-frame networks are trained on static and dynamic video datasets respectively. (\textbf{Best viewed on screen with zoom})}
	\vspace{-3pt}
	\label{fig:video-dynamic-vis}
\end{figure*}

\begin{table}[!t]
	\centering
	\caption{Quantitative results on \textit{static} video clips from DRV dataset. 
	}
	\footnotesize
	\begin{tabular}{lccc} 
		\toprule
		Model & PSNR$\uparrow$ & SSIM$\uparrow$ & ST-RRED$\downarrow$  \\ 
		\midrule
		VBM4D  \cite{maggioni2012video}  & 33.29 & 0.804 & 1.568 \\ \hline
		\midrule
		Noise2Noise \cite{pmlr-v80-lehtinen18a}  & 32.73 & 0.712 & 0.773 \\ \hline		
		Noiseflow \cite{Abdelhamed_2019_ICCV} & 35.14 & 0.856  & 0.441 \\ \hline		
		Paired real data \cite{Chen_2019_ICCV} & \textcolor{red}{37.24} & \textcolor{red}{0.903}  & \textcolor{red}{0.369} \\ \hline
		\midrule
		$G$+$P$ \cite{Foi2008Practical} & 34.38 & 0.816 & 0.467 \\ \hline
		Ours  & \textcolor{blue}{36.61} & \textcolor{blue}{0.896}  & \textcolor{blue}{0.407} \\ \hline
		\bottomrule
	\end{tabular}
	\label{tb:static-video}
\end{table}

\begin{table}[!t]
	\centering
	\caption{Quantitative results on synthetic \textit{dynamic} video clips from our collected dynamic video dataset. ``S" and ``D" indicate the multi-frame networks are trained on static and dynamic video clips respectively. 
	}
	\footnotesize
	\begin{tabular}{lccc} 
		\toprule
		Model & PSNR$\uparrow$ & SSIM$\uparrow$ & ST-RRED$\downarrow$ \\ 
		\midrule
		\textsc{UNet (single-frame)}  &  & &  \\
		Paired real data    & 33.66  & 0.925  & 0.482 \\ \hline
		Ours   & \textcolor{blue}{36.40} & \textcolor{blue}{0.930}  & \textcolor{blue}{0.230} \\ \hline
		\midrule
		\textsc{fastDVD (multi-frame)}  &  &  & \\
		Paired real data (S)  &  28.66 & 0.870 & 5.746 \\ \hline
		Ours (S)    & 28.47 & 0.860 & 5.373 \\ \hline
		Ours (D)    & \textcolor{red}{37.48} & \textcolor{red}{0.940} & \textcolor{red}{0.224} \\ \hline
		\bottomrule
	\end{tabular}
	\label{tb:dynamic-video}
\end{table}

Our approach can be naturally extended into the extreme low-light videography. 
In this section, we conduct raw video denoising experiment on the Dark Raw Video (DRV) dataset \cite{Chen_2019_ICCV}. 
The DRV training set only contains static videos with corresponding long-exposure ground truth, since 
collecting paired noisy and noise-free dynamic videos simultaneously is usually impossible. 
The learning-based video processing pipeline built on \cite{Chen_2019_ICCV} uses a sophisticated preprocessing (\eg $2\times 2$ binning, VBM4D \cite{maggioni2012video} denoising) followed by a U-Net to postprocess videos frame-by-frame. 
The network is trained on both the recovery loss to encourage the output to be close to the ground truth, as well as the self-consistency loss \cite{Chen_2019_ICCV} to enforce the two output frames of the same scene to be similar. 

To keep our setting simple, we discard the VBM4D preprocessing, and only use the U-Net with the same loss function as that in  \cite{Chen_2019_ICCV}\footnote{In \cite{Chen_2019_ICCV}, the loss function is defined on the VGG \cite{simonyan2014very} feature space. As we focus on raw-to-raw denoising, we use the (raw) image space instead.} to perform raw-to-raw video denoising in a frame-wise manner. 
The full experimental setting resembles the one used in raw-to-raw image denoising (See Section \ref{sec:experimental-setting}) except the extra self-consistency loss  for temporal stability. 
We calibrate our noise model using a Sony RX100 VI camera, the same camera model capturing the DRV dataset. 
During training, our approach still only need clean raw frames, as we can synthesize arbitrary number of  noisy frames to construct paired data. 
We also implement other training schemes (\ie Noise2Noise \cite{pmlr-v80-lehtinen18a}, Noiseflow \cite{Abdelhamed_2019_ICCV}, paired real data \cite{Chen_2019_ICCV}\footnote{We adapt the video processing pipeline of \cite{Chen_2019_ICCV} to raw-to-raw video denoising without sophisticated preprocessing.}, heteroscedastic Gaussian noise model $G$+$P$ \cite{Foi2008Practical}) as baselines based on DRV training dataset, and adopt the same U-Net architecture and loss function for learning.

For each static video, we assess the denoising performance by the averaged PSNR/SSIM over frames (5 continuous frames per video) as well as the Spatio-Temporal Reduced Reference Entropic Differences (ST-RRED) \cite{soundararajan2012video}. The ST-RRED is a high performing video quality assessment metrics, which takes both the spatial and temporal distortions into account. Lower ST-RRED score indicates better restoration accuracy. 
Table~\ref{tb:static-video} and Figure~\ref{fig:video-static-vis} display raw video denoising results on DRV testing dataset. It can be seen that our method still arrives at the performance comparable to the ``paired real data" model, and produces temporal stable results with less chroma artifacts. 

\begin{figure*}[!t]
	\centering
	\setlength\tabcolsep{1pt}
	\begin{tabular}{cccccccc}
	Input  & BM3D  &  Noise2Noise  & Noiseflow & G+P &  \fontsize{8pt}{8pt}\selectfont Paired real data & Ours & Reference \\
	\includegraphics[width=0.118\linewidth]{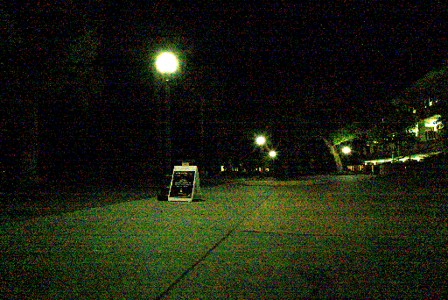}
	&\includegraphics[width=0.118\linewidth]{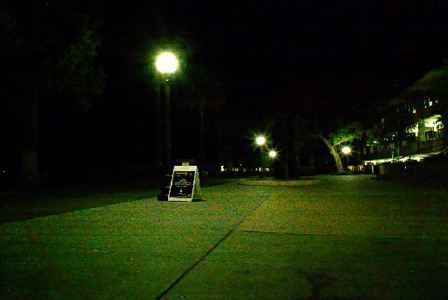}
	&\includegraphics[width=0.118\linewidth]{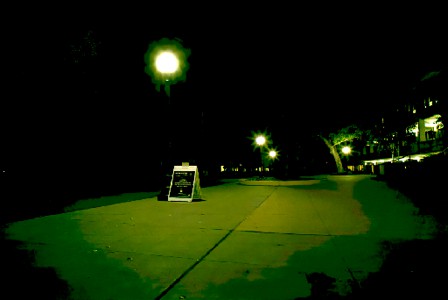}
	&\includegraphics[width=0.118\linewidth]{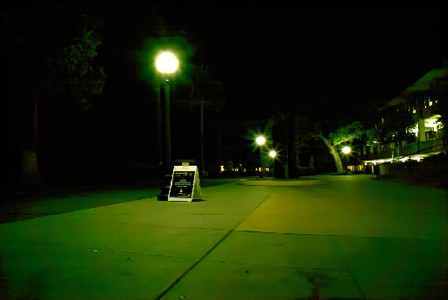}
	&\includegraphics[width=0.118\linewidth]{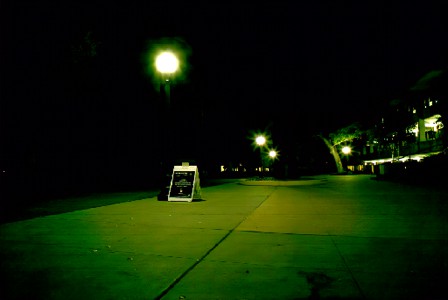}
	&\includegraphics[width=0.118\linewidth]{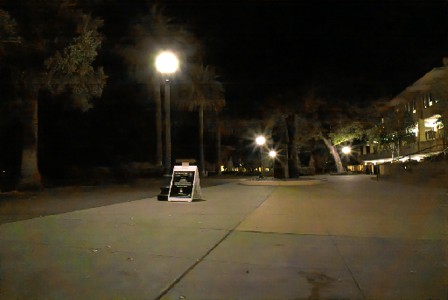}
	&\includegraphics[width=0.118\linewidth]{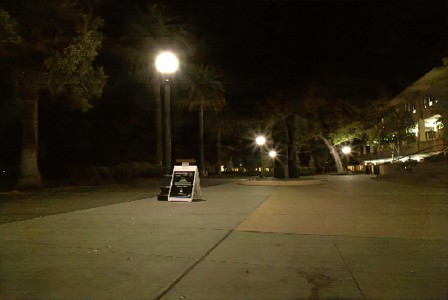}
	&\includegraphics[width=0.118\linewidth]{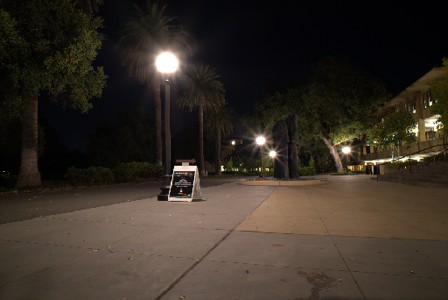} \vspace{-2pt} \\ 
	\includegraphics[width=0.118\linewidth]{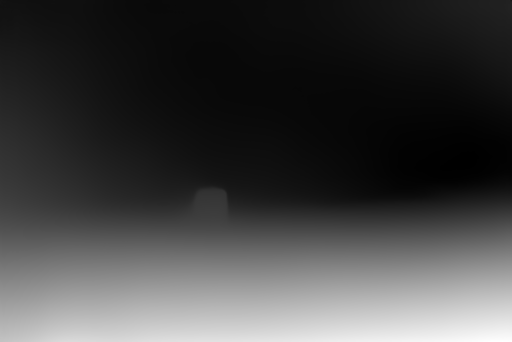}
	&\includegraphics[width=0.118\linewidth]{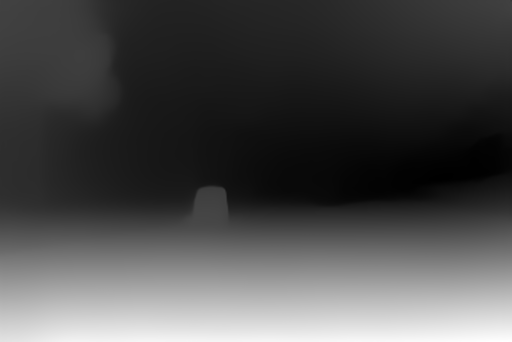}
	&\includegraphics[width=0.118\linewidth]{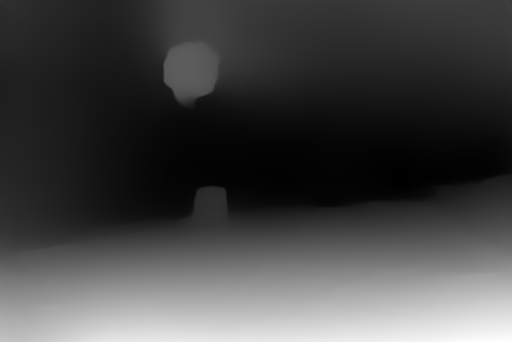}
	&\includegraphics[width=0.118\linewidth]{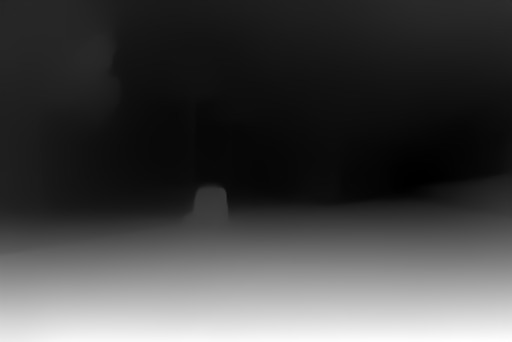}
	&\includegraphics[width=0.118\linewidth]{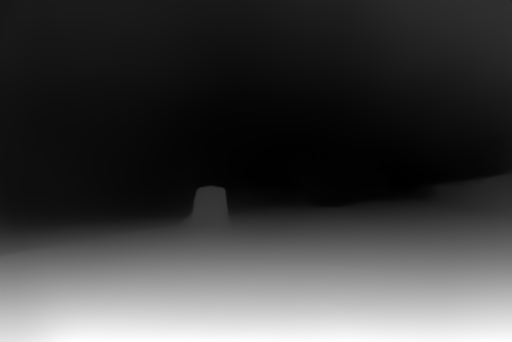}
	&\includegraphics[width=0.118\linewidth]{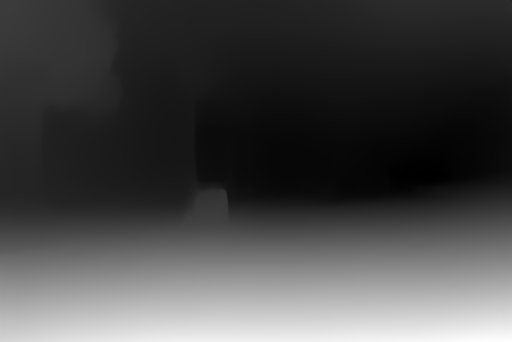}
	&\includegraphics[width=0.118\linewidth]{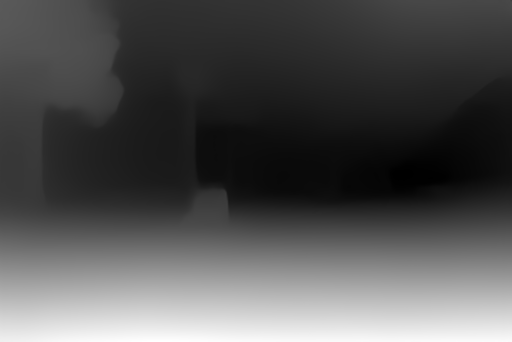}
	&\includegraphics[width=0.118\linewidth]{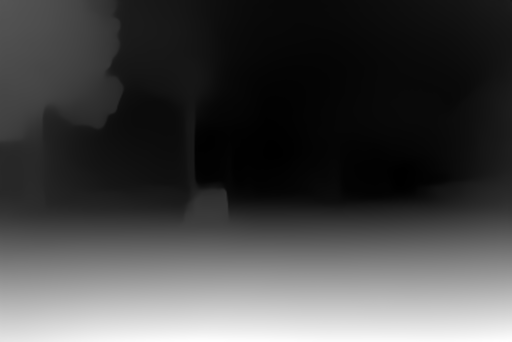} \vspace{-2pt} \\ 
	\end{tabular} 
	\vspace{-3pt}
	\caption{Monocular depth estimation in the dark. (\textbf{Best viewed on screen with zoom}) }
	\vspace{-3pt}
	\label{fig:depth-estimation}
\end{figure*}

\begin{figure*}[!t]
	\centering
	\setlength\tabcolsep{1pt}
	\begin{tabular}{cccccccc}
	Input  & BM3D  &  Noise2Noise  & Noiseflow & G+P &  \fontsize{8pt}{8pt}\selectfont Paired real data & Ours & Reference \\
	\includegraphics[width=0.118\linewidth]{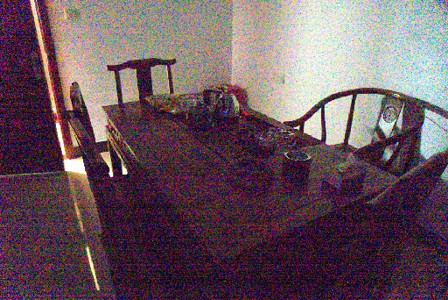}
	&\includegraphics[width=0.118\linewidth]{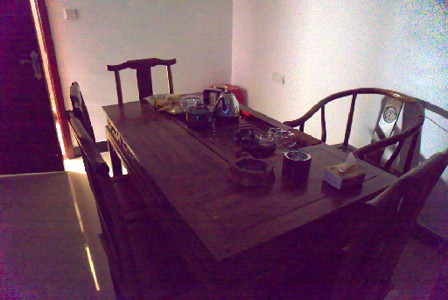}
	&\includegraphics[width=0.118\linewidth]{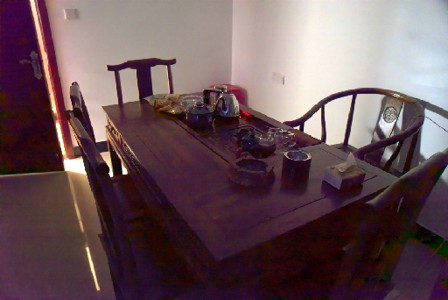}
	&\includegraphics[width=0.118\linewidth]{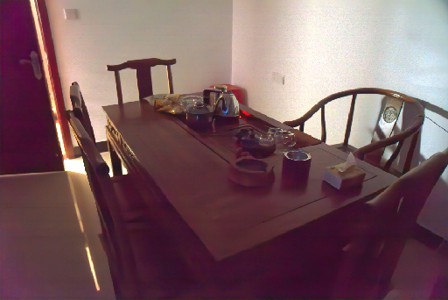}
	&\includegraphics[width=0.118\linewidth]{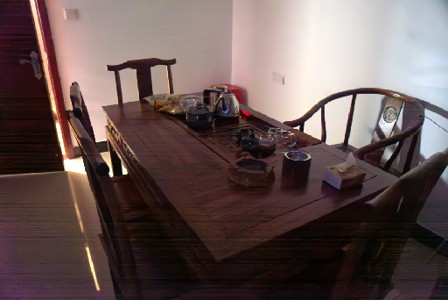}
	&\includegraphics[width=0.118\linewidth]{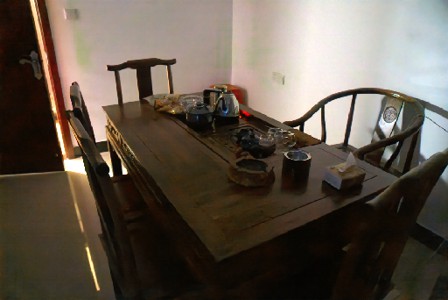}
	&\includegraphics[width=0.118\linewidth]{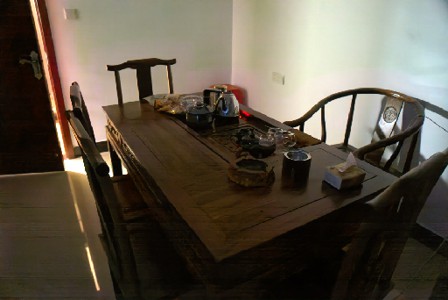}
	&\includegraphics[width=0.118\linewidth]{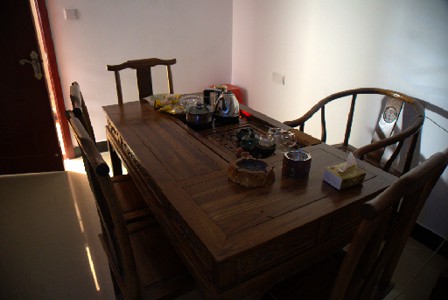}\vspace{-2pt} \\ 
	\includegraphics[width=0.118\linewidth]{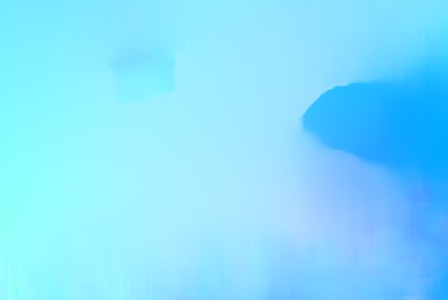}
	&\includegraphics[width=0.118\linewidth]{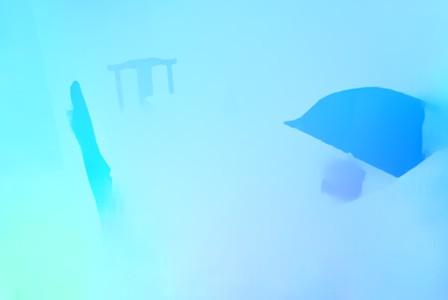}
	&\includegraphics[width=0.118\linewidth]{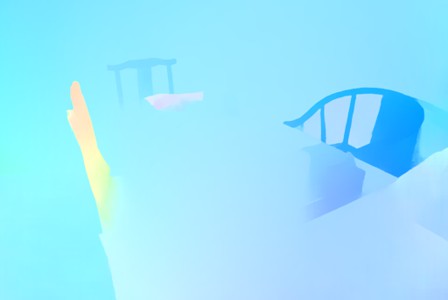}
	&\includegraphics[width=0.118\linewidth]{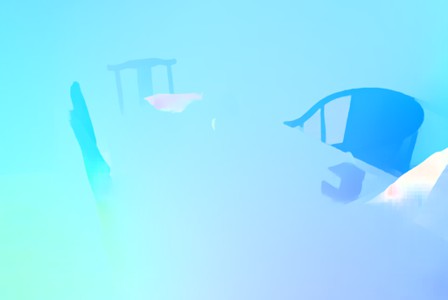}
	&\includegraphics[width=0.118\linewidth]{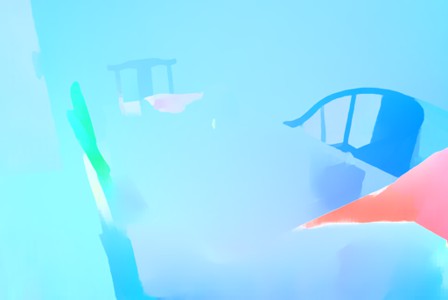}
	&\includegraphics[width=0.118\linewidth]{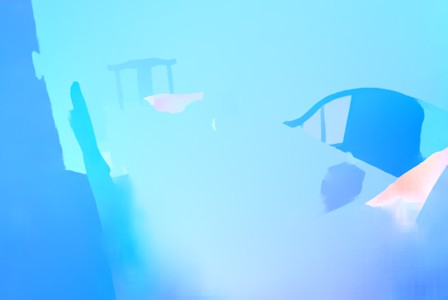}
	&\includegraphics[width=0.118\linewidth]{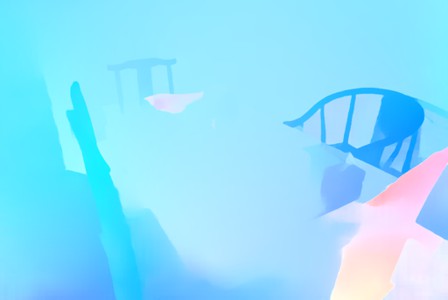}
	&\includegraphics[width=0.118\linewidth]{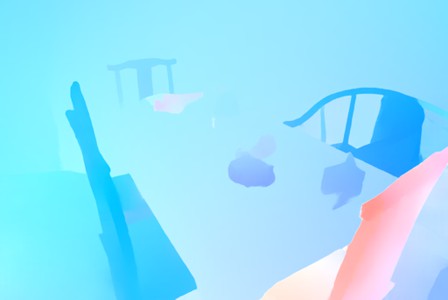}\vspace{-2pt} \\ 
	\end{tabular} 
	\vspace{-3pt}
	\caption{Optical flow in the dark. (\textbf{Best viewed on screen with zoom}) }
	\vspace{-3pt}
	\label{fig:optical-flow}
\end{figure*}

\begin{figure*}[!t]
	\centering
	\setlength\tabcolsep{1pt}
	\begin{tabular}{cccccccc}
	Input  & BM3D  &  Noise2Noise  & Noiseflow & G+P &  \fontsize{8pt}{8pt}\selectfont Paired real data & Ours & Reference \\
	\includegraphics[width=0.118\linewidth]{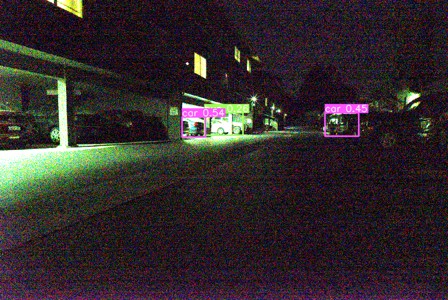}
	&\includegraphics[width=0.118\linewidth]{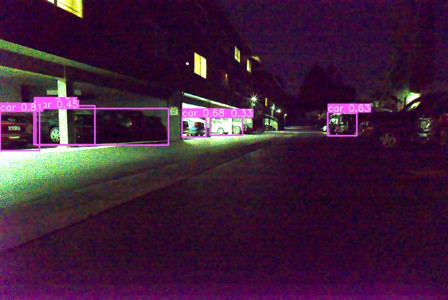}
	&\includegraphics[width=0.118\linewidth]{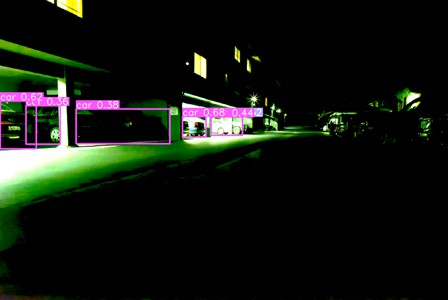}
	&\includegraphics[width=0.118\linewidth]{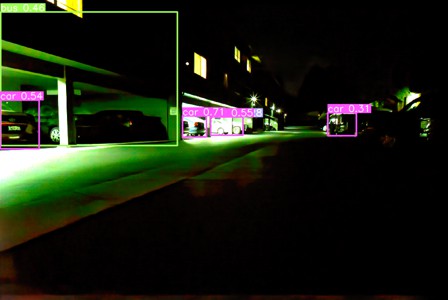}
	&\includegraphics[width=0.118\linewidth]{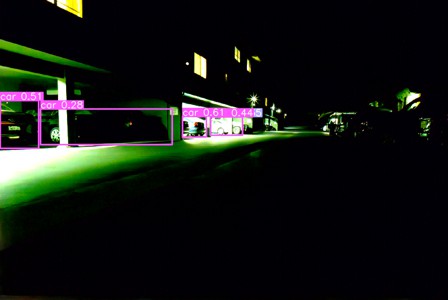}
	&\includegraphics[width=0.118\linewidth]{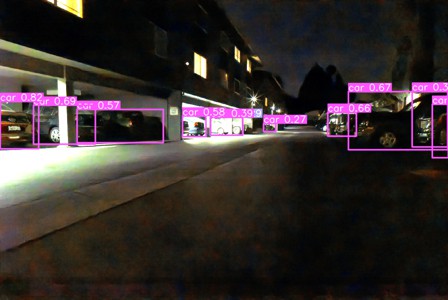}
	&\includegraphics[width=0.118\linewidth]{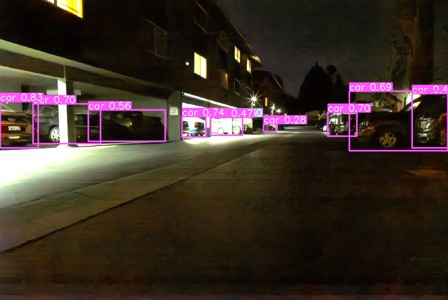}
	&\includegraphics[width=0.118\linewidth]{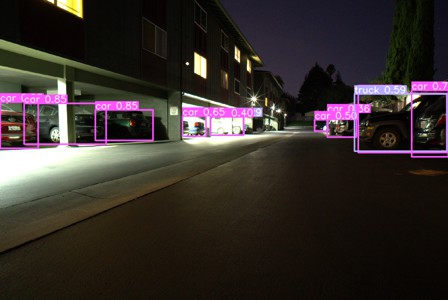} \vspace{-2pt} \\ 
	\includegraphics[width=0.118\linewidth]{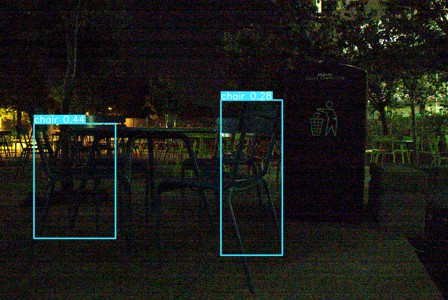}
	&\includegraphics[width=0.118\linewidth]{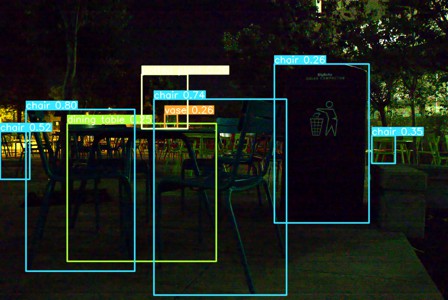}
	&\includegraphics[width=0.118\linewidth]{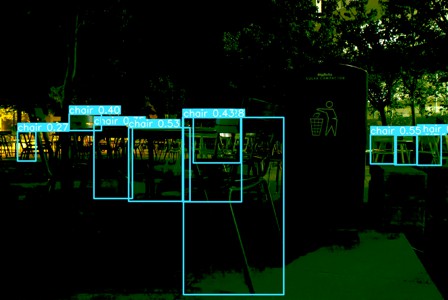}
	&\includegraphics[width=0.118\linewidth]{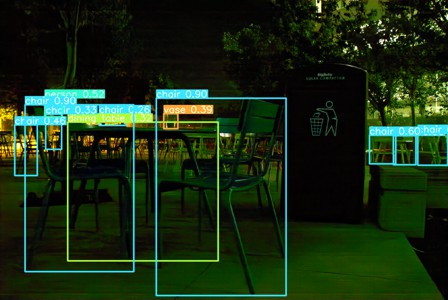}
	&\includegraphics[width=0.118\linewidth]{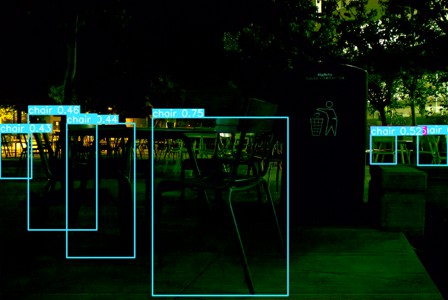}
	&\includegraphics[width=0.118\linewidth]{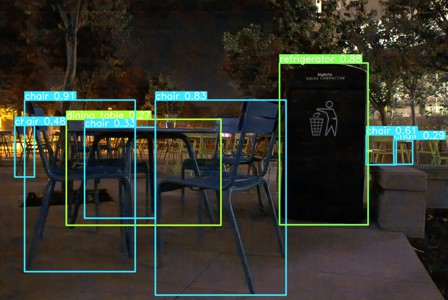}
	&\includegraphics[width=0.118\linewidth]{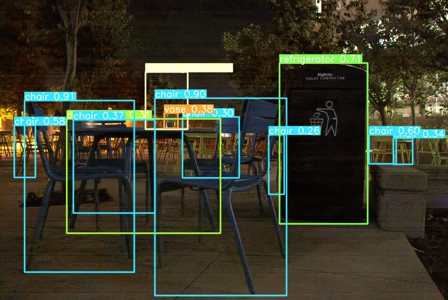}
	&\includegraphics[width=0.118\linewidth]{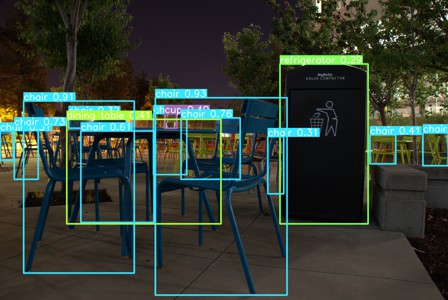} \vspace{-2pt} \\ 
	\end{tabular} 
	\vspace{-3pt}
	\caption{Object detection in the dark. (\textbf{Best viewed on screen with zoom}) }
	\vspace{-3pt}
	\label{fig:object-detection}
\end{figure*}


\begin{figure*}[!t]
	\centering
	\begin{tabular}{cccc}		
		\includegraphics[width=.221\linewidth,clip,keepaspectratio]{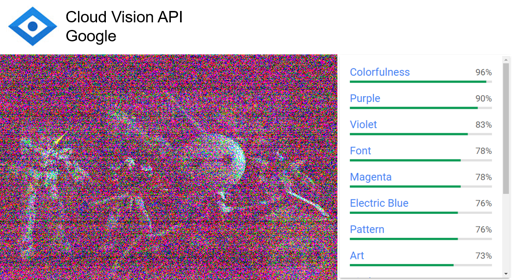} 
&		\includegraphics[width=.221\linewidth,clip,keepaspectratio]{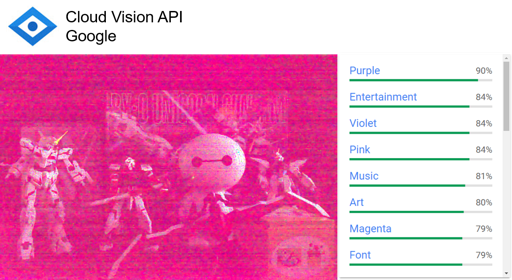} 
&		\includegraphics[width=.221\linewidth,clip,keepaspectratio]{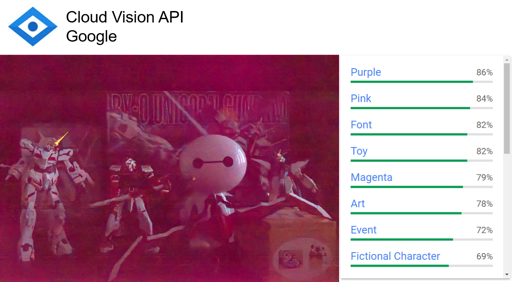} 
&		\includegraphics[width=.221\linewidth,clip,keepaspectratio]{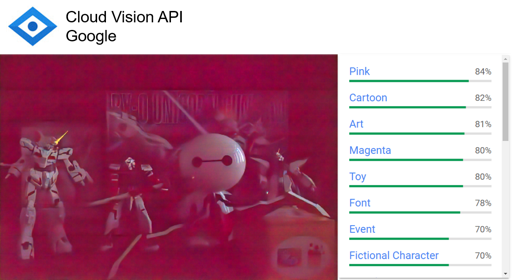}  \\
(a) Input & (b) BM3D & (c) Noise2Noise & (d) Noiseflow \\
		\includegraphics[width=.221\linewidth,clip,keepaspectratio]{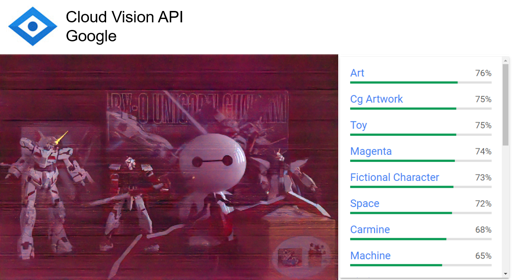}  
&		\includegraphics[width=.221\linewidth,clip,keepaspectratio]{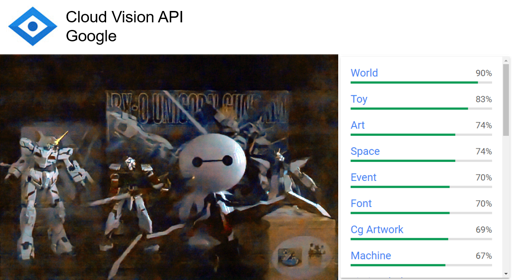} 
&		\includegraphics[width=.221\linewidth,clip,keepaspectratio]{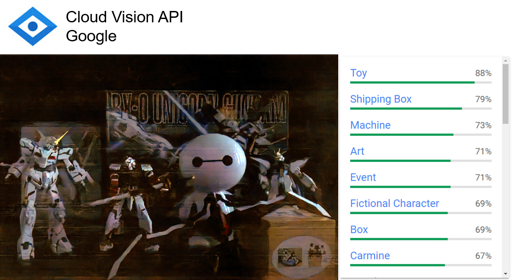} 
&		\includegraphics[width=.221\linewidth,clip,keepaspectratio]{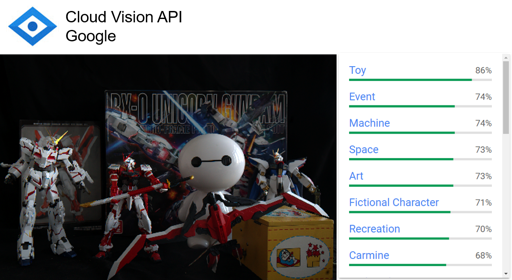} \\
(e) G+P  & (f) Paired real data & (g) Ours & (h) Reference \\
	\end{tabular}
	\vspace{-3pt}
	\caption{Low-light object recognition results on Google Vision API. (\textbf{Best viewed on screen with zoom})}
	\vspace{-3pt}
	\label{fig:google_vision}
\end{figure*}


It is worthwhile to highlight that our synthetic pipeline allows us to create a paired dynamic video dataset via adding synthetic noise to clean dynamic video frames\footnote{To enforce each synthesized frame at a video clip simulates a real one captured at the same low-light setting, we sample noise parameters once, and use the same parameters to synthesize noise instances for each frame.}.
This circumvents the difficulty of paired real dynamic video data collection, and makes learning multi-frame denoising networks feasible. To support this argument, we collect 100 clean raw dynamic video clips (5 to 10 frames per clip) using the Sony RX100 VI camera and randomly divide them into 95 clips for training and 5 clips for testing. 
Both single-frame (U-Net) and multi-frame approaches (FastDVDnet \cite{Tassano_2020_CVPR}) are compared on dynamic videos. These networks are trained either on paired real data or our synthetic data. 
Besides, either static video or dynamic video datasets are provided for training FastDVDnet.  

Table~\ref{tb:dynamic-video} presents the numerical results on synthetic dynamic video clips,
and the visual results on both synthetic and real low-light video frames can be found in Figure~\ref{fig:video-dynamic-vis}. 
It can be seen that the multi-frame network fastDVDnet trained on the static video data cannot generalize to dynamic video frames, since it is impossible to learn motion correspondence based upon only static videos. As a result, it only averages and merges frames, which leads to highly blurry results on dynamic videos. 
In contrast, the fastDVDnet trained on our synthetic dynamic video dataset performs quite well on dynamic videos, which surpasses the single-frame approaches owing to its additional capability to exploit the temporal correlation over frames.

\subsection{Application to Downstream Vision Tasks} \label{sec:high-level}


We have demonstrated the advantages of our noise model on both low-light photography and videography. 
Here, we show our approach could also be useful for downstream computer vision applications, \eg depth estimation in the dark, optical flow in the dark as well as object detection/recognition in the dark. 
For all these tasks, we follow the same pipeline (Figure~\ref{fig:denosing_pipeline}) to process images, i.e., 1) acquiring an image on RAW format; 2) performing denoising on it; 3) converting it into sRGB domain; 4) executing off-the-shelf vision algorithms on the processed sRGB image.  We note in practice, this whole process can be seamlessly integrated into a hardware ISP system executed silently on a chip, such that users do not need to directly interact with RAW formatted images. 

State-of-the-art/representative methods are adopted to carry out vision tasks under very limited luminance, including MiDaS \cite{Ranftl2020} for depth estimation, RAFT \cite{teed2020raft} for optical flow estimation as well as YOLOv5 \cite{glenn_jocher_2021_4418161} for object detection. We also test a commercial black-box object recognition solution provided by Google Vision API, which represents a wild uncontrolled task setting. 
 Both visual and numerical results are shown in Figure~\ref{fig:depth-estimation}-\ref{fig:google_vision} and Table~\ref{tb:vision-applications} respectively, which collectively demonstrate that our low-light denoising algorithm significantly improves the performance of downstream vision applications under extreme low-light settings.


\begin{table}[!t]
	\centering
	\caption{Quantitative results of downstream vision applications on SID Sony set (depth/detection) and our collected moved scenes (flow). 
	We report the root-mean-square error (RMSE), endpoint error (EPE) and the average number of objects recognized per image (\# of Objects) for these three tasks respectively, using the results of long-exposure references as ground truth.}
	\footnotesize
	\begin{tabular}{lccc} 
		\toprule
		\!Model\!   & \!Depth\! & \!Flow\! & \!Detection\! \\ 		
		& (RMSE) & (EPE) & (\# of Objects) \\
		\midrule
		Input & 0.175 & 22.44 & 1.550 \\ \hline
		BM3D & 0.121 & 15.52 & 2.496 \\ \hline
		Noise2Noise & 0.120 & 11.97 & 3.047 \\ \hline
		Noiseflow & 0.124 & 11.32 & 2.837 \\ \hline
		$G$+$P$ & 0.122 & 14.60 & 3.171 \\ \hline		
		Paired real data  & \textcolor{blue}{0.091} & \textcolor{blue}{10.36} & \textcolor{blue}{3.233} \\ \hline
		Ours  & \textcolor{red}{0.086} & \textcolor{red}{8.49} & \textcolor{red}{3.434} \\ \hline
		Reference  & 0.000 & 0.00 & 4.225 \\ \hline
		\bottomrule
	\end{tabular}
	\label{tb:vision-applications}
\end{table}

\section{Discussion of Applicability Scope} \label{sec:discussion-scope}
Our principal approach can be applied on a variety of real-world application scenarios, ranging from extreme low-light image/video processing to downstream computer vision applications at night. However, it should be clarified the applicability scope here is largely restricted to cases where the RAW images are available. As a result, our method is unfortunately unable to handle some practical situations, \eg video surveillance from closed-circuit television data, or processing images from internet. 
Nevertheless, we note RAW image is accessible in many practical and important scenarios from both scientific and industrial domains, where our method is applicable and more desirable. For example, in many scientific imaging applications (\eg microscopy \cite{joens2013helium}, astronomy \cite{mclean2008electronic}, remote sensing \cite{levin2020remote}), what we actually care is to faithfully record the physical world (\ie the scene irradiance) to analyze and induce physical properties. In such cases, the ISP functionalities in those scientific imagers are either simplified (typically with demosaicking only) or fully disabled, resulting in RAW images that are linearly response to irradiance. Besides, from industrial perspective, our method is also useful for camera and ISP designers with full accessibility to camera RAW and ISP. In this regard, our method could not only facilitate developing scientific imaging solutions, but also help camera/smartphone vendors to enhance their ISP system by inserting our sensor-specific denoising algorithm. It has the potential to unlock the powerful low-light imaging capability on commodity/scientific imagers, which in turn has significant impacts on downstream consumers. Moreover, recent years have witnessed significant advancements on smartphone cameras. Nowadays, almost every mainstream smartphone camera (\eg iPhone, Huawei, Samsung, etc.) has provided user interface of RAW image acquisition for non-professional users. We envision acquiring RAW images would be more and more convenient in future.

\section{Conclusion and Future Work} \label{sec:conclusion}
We have presented a physics-based noise formation model together with a noise parameter calibration method to help resolve the difficulty of extreme low-light imaging.  
We revisit the electronic imaging pipeline and investigate the influential noise sources overlooked by existing noise models.  
This enables us to synthesize realistic noisy raw data that match the underlying physical process of noise formation better. 
We systematically study the efficacy of our noise formation model, 
by introducing a new dataset that covers four representative camera devices. 
By training only with our synthetic data, we demonstrate a convolutional neural network can compete with or sometimes even outperform the network trained with paired real data on real-world benchmarks. 

Our approach opens a new door to real-world computational low-light imaging, which significantly alleviates the burden of capturing paired real training data from diverse camera devices.  It's highly flexible and enables rapid adaptation and deployment on a variety of consumer-level cameras. 
Nevertheless, to build a whole computational low-light imaging system, several challenges remain unresolved.
For example, both auto-focus and auto-exposure modules would fail under very low light. Their failure will render user experience unfriendly, as taking desired photos with sharp appearance would be excessively difficult. Furthermore, preserving high dynamic range is very appealing for low-light photography. 
Finally, in many commodity cameras, \eg smartphone cameras, the computation resource is often very limited. Thus it's worth investigating light-weight deep architectures specially tailored for low-light image processing as well.   
We foresee future works to further tackle these challenges and push the boundaries of computational low-light photography.

\ifCLASSOPTIONcaptionsoff
\newpage
\fi

{

	\begin{IEEEbiography}
		[{\includegraphics[width=1in,height=1.25in,clip,keepaspectratio]{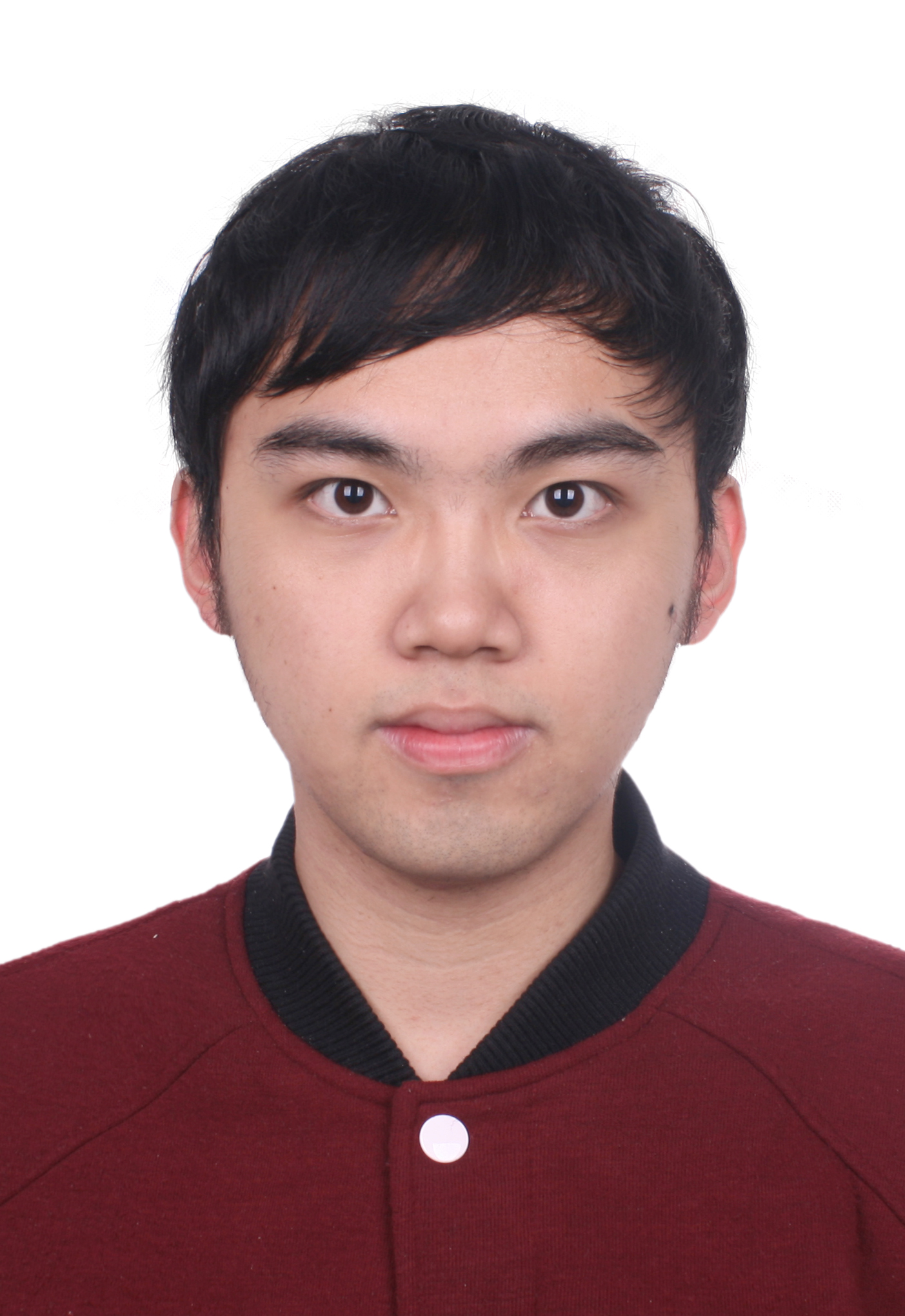}}]
		{Kaixuan Wei} received the B.S. in Electronic Engineering and the M.S. degree in Computer Science from Beijing Institute of Technology in 2018 and 2021, respectively. 
		His research interests include computer vision, computational photography and computational imaging. 
		He serves as reviewer for major computer vision and machine learning conferences and journals including CVPR/ ICCV/ECCV/AAAI/NeurIPS/ICLR/TPAMI/TIP.
		He received the Outstanding Paper Award from ICML'20.
	\end{IEEEbiography}

	\begin{IEEEbiography}
		[{\includegraphics[width=1in,height=1.25in,clip,keepaspectratio]{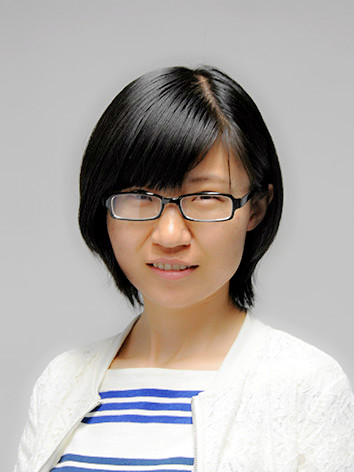}}]
		{Ying Fu} received the B.S. degree in Electronic Engineering from Xidian
		University in 2009, the M.S. degree in Automation from Tsinghua University in
		2012, and the Ph.D. degree in information science and technology from the
		University of Tokyo in 2015. She is currently a professor with the
		School of Computer Science and Technology, Beijing Institute of Technology.
		Her research interests include physics-based vision, image and video processing, and	computational photography. She received the Outstanding Paper Award from ICML'20.
	\end{IEEEbiography}

	\begin{IEEEbiography}
		[{\includegraphics[width=1in,height=1.25in,clip,keepaspectratio]{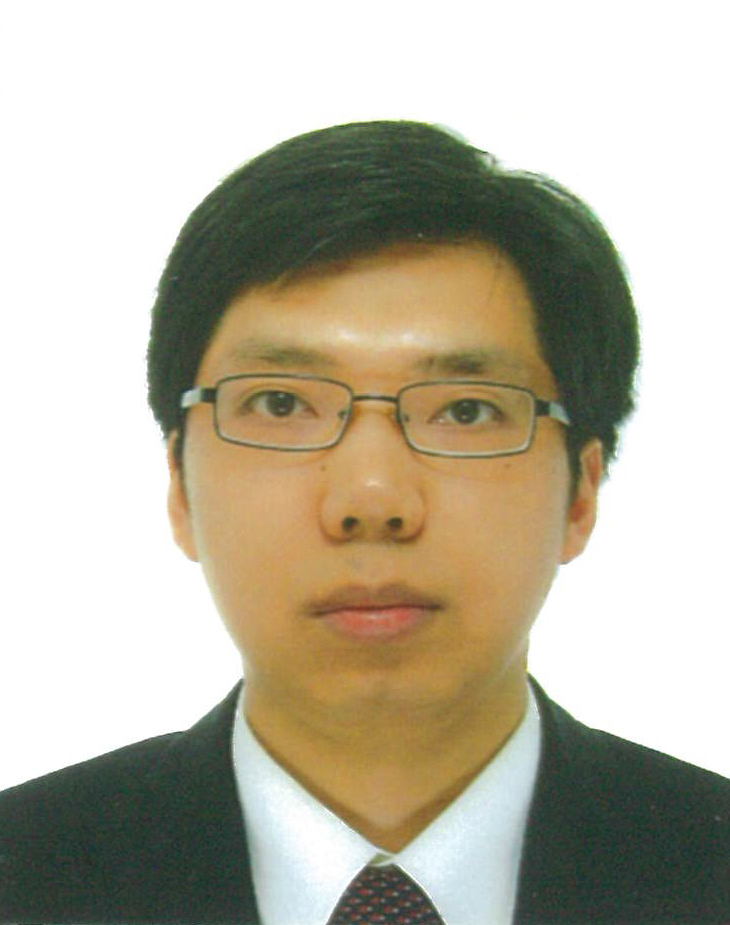}}]
		{Yinqiang Zheng} received the bachelor’s degree from the Department of Automation, Tianjin University, Tianjin, China, in 2006, the master’s degree in engineering from Shanghai Jiao Tong University, Shanghai, China, in 2009, and the doctoral degree in engineering from the Department of Mechanical and Control Engineering, Tokyo Institute of Technology, Tokyo, Japan, in 2013. After working at National Institute of Informatics, Japan, for more than seven years, he is currently an Associate Professor with The University of Tokyo, Japan. His research interests include image processing, computer vision, and mathematical optimization.
    \end{IEEEbiography}			

	\begin{IEEEbiography}
		[{\includegraphics[width=1in,height=1.25in,clip,keepaspectratio]{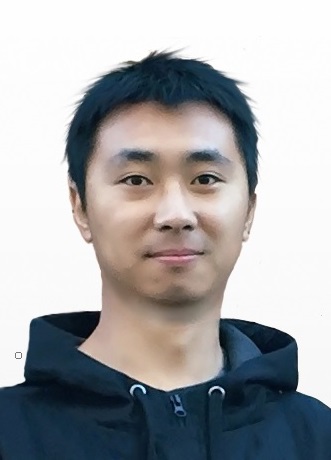}}]
		{Jiaolong Yang} received the B.S. degree in Computer Science from Beijing Institute of Technology in 2010, and the dual Ph.D. degrees in Computer Science and Engineering from the
		Australian National University and Beijing Institute of Technology in 2016. He is currently a Senior Researcher in the Visual Computing Group at Microsoft Research Asia, Beijing.
		His research interests include 3D and low-level computer vision. He serves regularly as program committee member/reviewer for major computer vision conferences and journals including CVPR/ ICCV/ECCV/TPAMI/IJCV, and is an Area Chair for CVPR'21, ICCV'21 and CVPR'22. He received the outstanding PhD thesis award from China Society of Image and Graphics in 2017.
	\end{IEEEbiography}

}

\end{document}